# Presolar grains from meteorites: Remnants from the early times of the solar system

Katharina Lodders [a,*] and Sachiko Amari [b]
[a] Planetary Chemistry Laboratory, Department of Earth and Planetary Sciences and McDonnell Center for the Space Sciences, Washington University, Campus Box 1169, One Brookings Drive, St. Louis, MO 63130, USA
[b] Department of Physics and McDonnell Center for the Space Sciences, Washington University, Campus Box 1105, One Brookings Drive, St. Louis, MO 63130, USA

## Abstract

This review provides an introduction to presolar grains – preserved stardust from the interstellar molecular cloud from which our solar system formed – found in primitive meteorites. We describe the search for the presolar components, the currently known presolar mineral populations, and the chemical and isotopic characteristics of the grains and dust-forming stars to identify the grains' most probable stellar sources.

## 1. Introduction

The history of our solar system started with the gravitational collapse of an interstellar molecular cloud laden with gas and dust supplied from dying stars. The dust from this cloud is the topic of this review. A small fraction of this dust escaped destruction during the many processes that occurred after molecular cloud collapse about 4.55 Ga ago. We define presolar grains as stardust that formed in stellar outflows or ejecta and remained intact throughout its journey into the solar system where it was preserved in meteorites.

The survival and presence of genuine stardust in meteorites was not expected in the early years of meteorite studies. In the 1950s and 1960s, models of solar system formation assumed that the matter from the presolar molecular cloud was processed and homogenized (e.g., Suess 1965, see also Fegley 1993). Most of this matter accreted to the Sun and less than about one percent remained to form the planets, their satellites, and other small objects (asteroids, Kuiper-Edgeworth-belt objects). During collapse and accretion of matter towards the cloud center, gravitational heating vaporized presolar solids, and it was generally assumed that this process resulted in a relatively homogeneous solar nebular gas made of evaporated presolar solids and presolar gas. Upon cooling of the solar nebula, new condensates appeared which accumulated to form the solid bodies in the solar system (Fig. 1). In this very simplified picture, all matter from the presolar cloud would be chemically and isotopically homogenized, and no record about the mineralogy of presolar solids would remain.

*Corresponding author.
e-mail address: lodders@wustl.edu

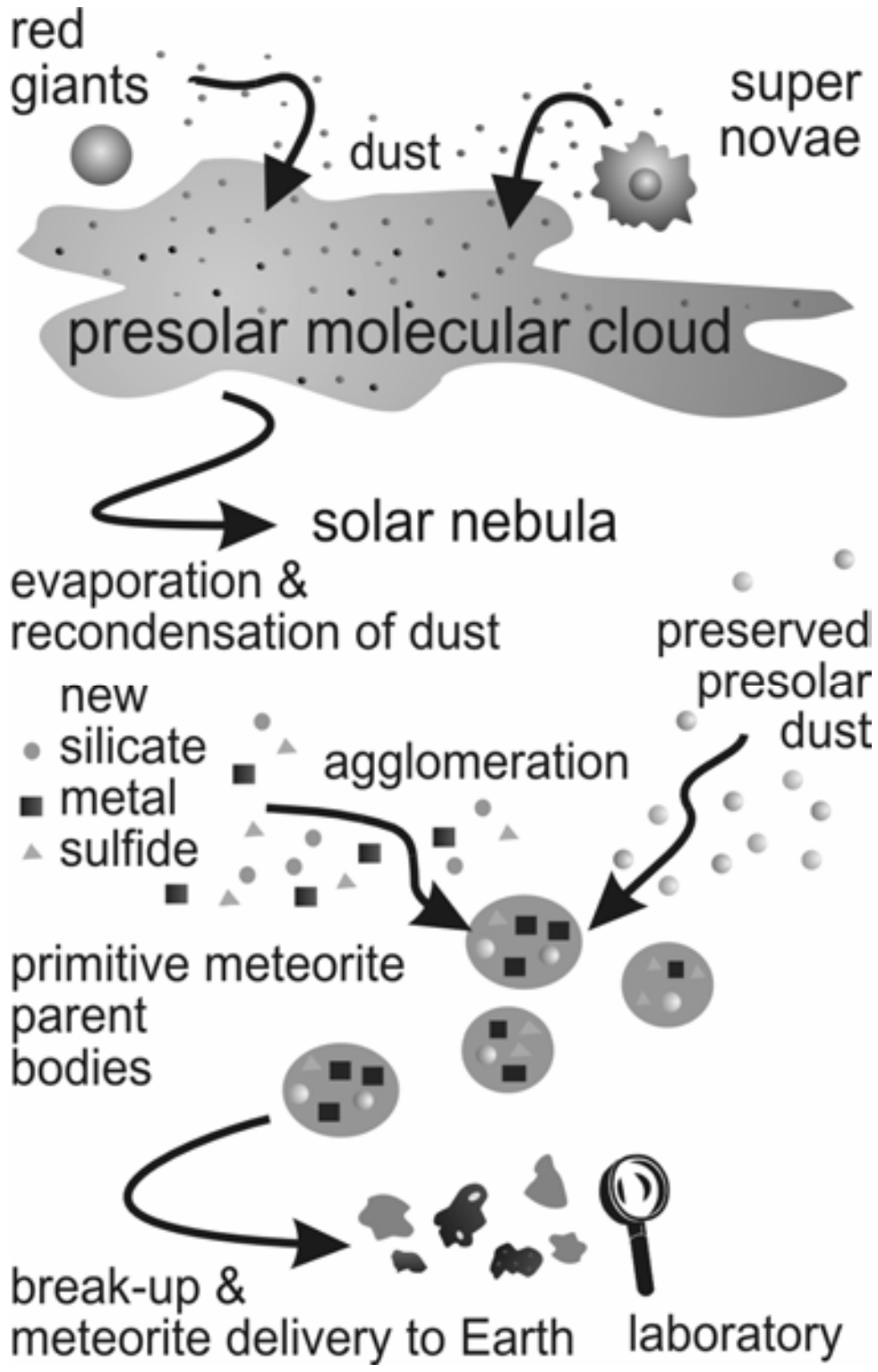

**Fig. 1.** The end of red giant stars and supernovae is accompanied by production of dust grains that also entered the molecular cloud from which our solar system formed. A tiny portion of original stardust survived the passage through the ISM and the events during solar system formation. After extraction of presolar dust from meteorites and examination by various types of instruments, we obtain information about the grains' parent stars

Long before presolar grains were discovered, Cameron (1973) speculated on the question "Interstellar grains in museums?" and concluded that primitive carbonaceous chondrites may harbor presolar grains. Indeed, presolar grains were incorporated into meteorite parent bodies (small asteroid-size objects). Chemical and metamorphic processes on the least metamorphosed meteorite parent bodies were apparently mild and more or less non-destructive to the grains, so upon meteorite delivery from a parent body to Earth, the presolar grains awaited discovery.

There are several motivations to study presolar dust. Fresh dust and gas are continuously supplied to the interstellar medium (ISM), mainly by mass-loss from red giant stars and by exploding novae and supernovae, which is the way such stars enrich the ISM with their nucleosynthetic products over time. Hence the presolar grains from meteorites could provide a glimpse into the dust population that accumulated in the presolar molecular cloud. However, the known presolar minerals are very likely a biased sample of solids from the presolar molecular cloud because of their different stabilities against physical and chemical processing in the ISM, in the solar nebula, on meteorite parent bodies, and in the laboratory during grain isolation procedures.

Another reason to study presolar grains is that their physical, chemical and isotopic signatures record nucleosynthetic and grain formation processes of a variety of stars. Grain morphologies and compositions may reflect formation conditions in stellar environments, and the grains' chemical and isotopic compositions are firm tests of stellar evolution and nucleosynthesis models. Hence the microscopic grains reveal more details about these processes than currently possible by astronomical observations with even the best telescopes.

One way to find the nm- to μm-size presolar grains in meteorites is to search for variations in isotopic compositions because several elements in certain meteoritic components are clearly different from the terrestrial isotopic reference compositions. Many of these "isotope anomalies" cannot be easily explained by mass-fractionation processes or by decay of longer-lived radioactive isotopes and therefore suggest the presence of exotic, possibly presolar, phases.

Even in the 1950s and 1960s there were indications that the solar nebula was not homogeneous. For example, Boato (1953, 1954) and Briggs (1963) found deuterium enrichments in carbonaceous chondrites that were not easy to explain by chemical mass-fractionation processes known at that time. Clayton (1963) encountered a similar problem in trying to explain large $^{13}C$ enrichments in carbonate from the Orgueil meteorite and thought that incomplete homogenization of matter from different nucleosynthetic sources could be responsible. Unusual isotopic compositions in noble gases from meteorites were discovered for Xe (Reynolds and Turner 1964), and Ne (Black and Pepin 1969).

However, the discovery of widespread isotopic variations in O, the major rock-forming element, by Clayton et al. (1973) provided the strongest evidence of incomplete homogenization of matter in the solar nebula. In particular, the so-called calcium aluminum-rich inclusions (CAIs) found in chondritic meteorites show characteristic enrichments of around 5% in $^{16}O$ relative to terrestrial rocks and other major meteorite components. These enrichments cannot be derived by normal mass-dependent fractionation processes in the solar nebula and require another explanation. Clayton et al. (1973) suggested that $^{16}O$, a major product from supernovae, was present in a reservoir in the solar nebula (see, however, alternative explanations of the $^{16}O$ enrichments by mass- independent fractionation processes by Thiemens and Heidenreich 1983, Clayton 2002).

Further support that supernova and other stellar debris may have been present in the presolar molecular cloud came from observations of isotopic variations in other abundant rock-forming elements such as Ca, Ti, and Cr in CAIs (Clayton et al. 1988). The variations in Ca and Ti isotopic compositions in CAIs are typically observed for neutron-rich isotopes such as $^{48}Ca$ and $^{50}Ti$, which are signatures from nucleosynthesis operating in supernovae, but the size of these anomalies (0.01% to 0.1%) is quite small (Clayton et al. 1988). Thus, although CAIs often carry isotopic anomalies, many researchers do not regard them as pure presolar phases, and the origin of isotopic anomalies in CAIs is rather vaguely assigned to a possible incorporation of some presolar components.

In contrast, bona-fide presolar grains are identified by their often huge, orders-of-magnitude-ranging excesses or deficits in isotopic compositions (relative to normal terrestrial isotopic abundances). A presolar phase may have an outstanding isotopic anomaly in at least one element, but often isotope anomalies are observed simultaneously in several major elements such as C, N, O, and Si within a *single* presolar grain. In essentially all cases, the observed isotopic anomalies can only be explained by nucleosynthetic processes, which means that these grains formed near the sites of nucleosynthesis. This leaves the stars to imprint their chemical and isotopic signatures onto the grains. The principal environments for grain formation are the circumstellar shells of red giant stars and asymptotic giant branch (AGB) stars, and supernova ejecta.

The major and trace element chemistry can also be used to identify presolar grains. For example, the occurrence of reduced grains such as graphite and silicon carbide in meteorites consisting mainly of oxidized rock and hydrous silicates (e.g., CI and CM chondrites) is a rather unusual paragenesis, and is most easily explained by an external source of reduced grains. However, the presolar grains are rather small, and carbonaceous dust also may have formed in the solar nebula, so major element chemistry alone cannot provide a strong constraint on the origin of the reduced dust. On the other hand, trace element abundances in some reduced grains cannot be explained by chemical fractionation during condensation from a gas of solar composition. Instead, trace elements in non-solar proportions are required at the grain sources, and several types of stars show non-solar trace element abundances due to nucleosynthesis.

Here we review some of the history of the search and discovery of presolar dust, the minerals already identified, and what other minerals are to be expected. We then describe dust-producing stars and how certain grain types can be related to them. Since the discovery and isolation of presolar grains by Lewis et al. (1987), much work on presolar grains has followed (see reviews by e.g., Anders and Zinner 1993, Bernatowicz and Zinner 1997, Ott 1993, Zinner 1998, 2004).

## 2. The search for presolar grains

Historically, noble gas studies played an important role in cosmochemistry. The noble – or rare –gases are literally rare in many meteorites but even a small addition of a trapped or adsorbed component whose isotopic composition is vastly different from “normal” is easily detected. Because most of the noble gases have more than two stable isotopes and an anomalous composition may not be restricted to a single isotopic ratio, the term “component” is used to describe an mixture with a certain characteristic isotopic composition. In noble gas studies, "stepwise heating" is applied to distinguish the various components. A sample is heated in incremental temperature steps and the isotopic composition of the noble gases released during each step is measured. If the components reside in different phases and/or have been trapped in different sites by different mechanisms, they may become released at different temperatures.

In their studies, Reynolds and Turner (1964) and Black and Pepin (1969) discovered some unusual Ne and Xe components with extreme isotopic compositions. Subsequent studies showed them more clearly after other components that partially masked the unusual ones were removed by chemical treatments. This led to the questions: Is it possible to isolate the pure unusual gas components and to identify the carrier phases?

### 2.1. Presolar signatures in neon isotopes: Ne-E

Until the study by Black and Pepin (1969), the Ne-isotopic composition in meteorites was explained by mixing of three major components with characteristic $^{21}Ne/^{22}Ne$ and $^{20}Ne/^{22}Ne$ ratios, historically called Ne-A, Ne-B, and Ne-S. Neon-A, or "planetary Ne", is observed in primitive meteorites. The planetary noble gas components are found in carbonaceous chondrites which display elemental noble gas abundance patterns strongly enriched in heavy noble gases (= planetary pattern) relative to solar abundances. (It turned out, however, that the major portion of Ne-A is carried by presolar diamond). Neon-B is observed in particularly gas-rich meteorites and is thought to be implanted solar wind or solar energetic particles. Neon-S, or cosmogenic Ne, comes from cosmic ray spallation, which produces all Ne isotopes in about the same amounts.

For a three component mixture, the observed Ne compositions should plot within a triangle spanned by the components Ne-A, Ne-B, and Ne-S (e.g., Anders 1988). However, Black and Pepin (1969) found a Ne component released between 900°C and 1000°C from six carbonaceous chondrites that plotted outside the triangle spanned by the known end-member compositions. This component was poor in $^{21}Ne$, and had $^{20}Ne/^{22}Ne$ <3.4, much below the $^{20}Ne/^{22}Ne$ of ~8 and ~13 for Ne-A and Ne-B, respectively. This previously unknown component very rich in $^{22}Ne$ was named Ne-E by Black (1972), as the letters C and D were already taken for other minor Ne components. Subsequent studies (Eberhardt et al. 1979, Lewis et al. 1979, Jungck 1982) identified two kinds of Ne-E in different density fractions of meteorites. One kind, Ne-E(L), was released at *l*ow temperatures (500-800°C) and its carrier phase was concentrated in *l*ow-density fractions (2.2-2.5g/cm$^3$). The other kind, Ne-E(H), was released at *h*igh temperatures (1100-1500°C) and its carrier was concentrated in *h*igh-density fractions (2.5-3.1g/cm$^3$). The upper limits to $^{20}Ne/^{22}Ne$ dropped further (0.008 for Ne-E(L), 0.2 for Ne-E(H)) when Ne was measured in the Orgueil meteorite (Jungck 1982). This indicated the presence of almost pure $^{22}$Ne in some phase of the separates and supported the idea of a direct nucleosynthetic origin fore Ne-E, as suggested by Black (1972). A possible source of "pure" $^{22}$Ne is from decay of $^{22}$Na ($t_{1/2}$= 2.6 a), and with such a short half-life, $^{22}$Na must have been incorporated into a carrier near its stellar source to allow in-situ decay (Black 1972, Clayton 1975, Eberhardt et al. 1981, Jungck 1982).

### 2.2. Presolar signatures in Xe isotopes: Xe-HL, Xe-S

With nine stable isotopes, the Xe systematics in meteorites and planetary atmospheres is quite complex so that Reynolds (1963) coined Xe studies "Xenology". The Xe components include a radiogenic component ($^{129}$Xe from decay of $^{129}$I with $t_{1/2}$ =17 Ma), a spontaneous fission component from $^{235,238}$U and $^{244}$Pu, a cosmic-ray spallation component, and so-called "trapped" components. Reynolds and Turner (1964) observed that the abundances of the heavy isotopes $^{131\text{-}136}$Xe in the 700-1000°C fractions released from the Renazzo CR-chondrite were akin to those produced from spontaneous fission of heavy actinides. Rowe and Kuroda (1965) reported similar observations. This component was dubbed CCF-Xe (carbonaceous chondrite fission), but the ensuing debate in the late 1960s about the possible presence of super-heavy elements in meteorites implied by CCF-Xe did not last (see Anders 1988). One objection was that enrichments of heavy Xe isotopes were accompanied by enrichments in the lightest ones, $^{124}$Xe and $^{126}$Xe, which is inconsistent with a fission origin (Manuel et al. 1972).

In their search for the carrier of the unusual Xe component in the Allende meteorite, Lewis et al. (1975) found that, after HF-HCl treatment, essentially all trapped Xe of the meteorite was in the residue, which comprised 0.5 mass% of the bulk meteorite. Processing with $HNO_3$ removed only 6% of the residue's mass but also all trapped Xe, and a Xe component (Xe-HL), enriched in both and *h*eavy ($^{134}$Xe and $^{136}$Xe) and *l*ight ($^{124}$Xe and $^{126}$Xe) isotopes, emerged. In Xe-HL, the isotopic anomalies in light and heavy isotopes are always correlated.

Another component with high $^{130}$Xe/$^{132}$Xe was accidentally found when Srinivasan and Anders (1978) tried to characterize Xe-HL in the Murchison meteorite (see Anders 1988). The new component was enriched in the even-numbered isotopes ($^{128}$Xe, $^{130}$Xe, and $^{132}$Xe), and a comparison to nucleosynthesis theory indicated a signature of the slow neutron capture process (*s*-process), thus it was named Xe-S. Typically, Xe-S is accompanied by a Kr component also rich in *s*-process isotopes (Kr-S).

## 3. Isolation and discovery of presolar grains

Abundances of the exotic noble gas components compared to the total noble gas content are relatively low, so the abundances of the carrier minerals (what we now call presolar grains) of the anomalous noble gases were also expected to be quite low. For example, Lewis et al. (1975) demonstrated that the carrier of Xe-HL comprises less than 0.5% by mass of the whole Allende meteorite.

At the time the anomalous noble gas components were discovered, the carrier minerals were not known, but further studies suggested that they are carbonaceous (Table 1; e.g., Eberhardt et al. 1981, Jungck 1982, see summary by Anders 1987, 1988). Lewis et al. (1987) succeeded in isolating diamond, the carrier of Xe-HL. This was followed by the isolation of SiC, the carrier of Kr-S, Xe-S, and Ne-E(H) (Bernatowicz et al. 1987, Tang and Anders 1988), and graphite, the carrier of Ne-E(L) (Amari et al. 1990).

The procedure to separate presolar grains developed by Amari et al. (1994) is shown in Fig. 2. First, silicates, which comprise a major part of stony meteorites (~96%), are removed with HF. Part of the organic matter - aromatic polymers collectively called reactive kerogen - is destroyed by oxidants ($H_2O_2$, $Cr_2O_7^{2-}$). At this point, nearly 99% of the meteorite is dissolved. From the remaining residue, diamond is collected by so-called colloidal separation. In a basic solution diamond particles are negatively charged, repulse each other and stay in solution, while in acidic solution they become neutral and coagulate (Fig. 3), and settle down by centrifugation.

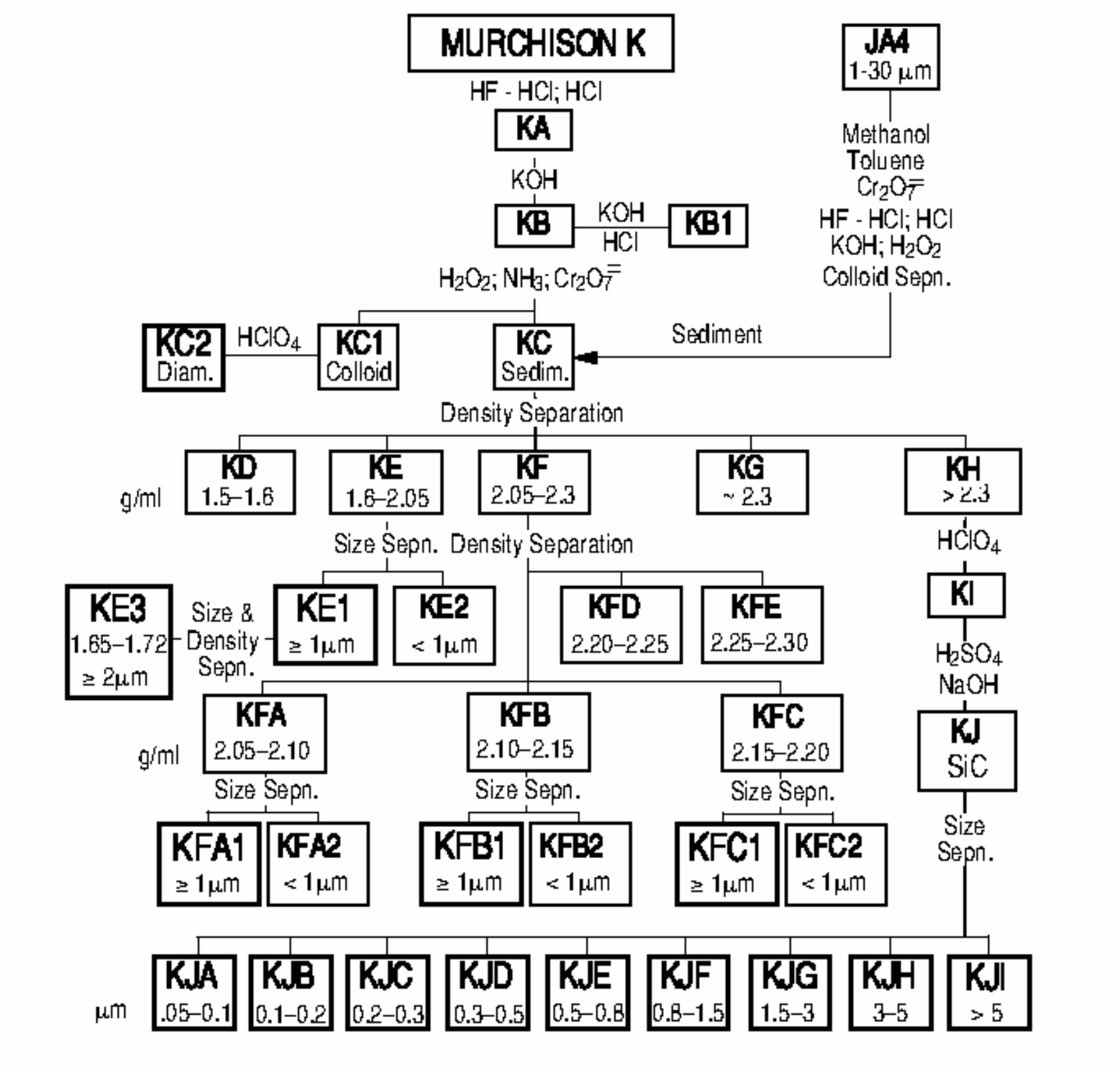

Fig. 2. The steps of the chemical isolation procedure and the resulting presolar grain size- and density-fractions from the Murchison meteorite by Amari et al. (1994).

Presolar graphite has similar chemical properties as other carbonaceous matter that remains in the meteoritic residue, and a density separation is applied to extract presolar graphite from the remaining carbonaceous matter and other minerals with higher densities. The separate with density >2.2g/cm$^3$ are purified with $HClO_4$, which leaves minerals such as SiC and aluminous oxides.

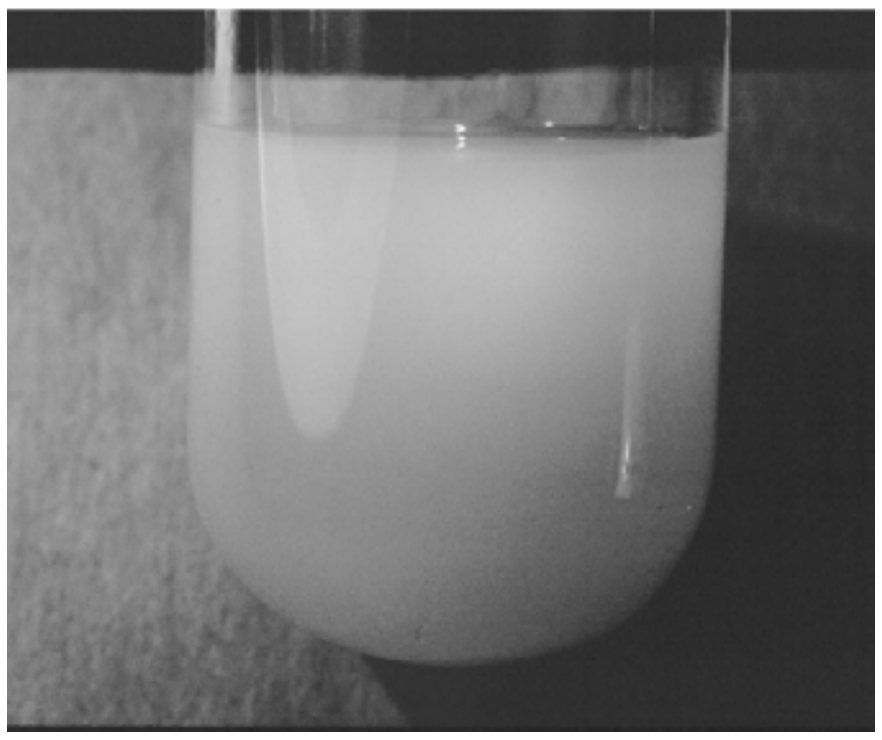

**Fig.3**. Nano-diamonds precipitate as a cloudy white gel from acidic solution but they completely "dissolve" in basic solution. Photo courtesy of Roy S. Lewis.

**Table 1**. Early results on presolar grain identifications ca. 1988

| Working designation | C-α | C-β [a] | C-ε [a] | C-δ | C-θ |
|---|---|---|---|---|---|
| Identified phase | amorphous carbon | SiC | SiC | diamond | amorphous carbon |
| Noble gas identifier | Ne-E(L) [b] | Xe-S | Ne-E(H) | Xe-H(L) | no noble gases |
| Grain size μm | 0.8 – 20 [c] | 0.03 – 20 [c] | | 0.001-0.0025 | 0.2 - 2 |
| Average density (g/cm$^3$) | 1.6–2.2 | 3.22 | 3.22 | 2.22 to 2.33 | <2.2 |
| Abundance[d] (ppm) | < 2 – < 5 | 3 – 7 | | 400 | 105 |
| $\delta$ $^{13}$C (per-mil)[e] | +340 | +1000 | +1500 | ~ -38 | -50 |
| $\delta^{15}$N (per-mil)[e] | ≥ +252 | ~ -500 | | ~ -375 | |
| Possible stellar source [f] | novae, SN, AGB? | AGB | AGB | SN | SN? |

*Sources*: Lewis et al. 1987, Anders 1988, Tang and Anders 1988, Tang et al. 1988, Anders and Zinner 1993

[a] Originally it was thought that C-β and C-ε are two different carriers. It was believed that C-β carries Xe-S, and that C-ε , associated with spinel because of similar density, carries Ne-E(H)., e.g., Anders (1988).

[b] Ne-E(L) = "pure" $^{22}$Ne is mainly from $^{22}$Na decay but may have some $^{22}$Ne from AGB stars

[c] Large grains are extremely rare.

[d] Abundance (ppm by mass) in CM2 chondrites

[e] Values are for carrier-enriched fractions, not for the pure carrier phase

[f] Stellar sources. AGB = asymptotic giant branch star (section 6), SN = Supernova (section 7)

Most of the grains were isolated from meteorites by progressively harsher acid dissolution of their meteorite hosts. The advantage of this method is that individual grains can be studied for their overall morphology and crystal structure. The disadvantage is that other potential presolar minerals are destroyed. Another desired piece of information is how presolar grains interact with the meteorite host during metamorphism, because decreases in presolar grain abundances with metamorphic type are observed. Locating the grains *in situ* is necessary to study grain destruction in meteorite parent bodies and to identify their reaction products. The first *in-situ* detection of presolar grains in meteorites was reported by Alexander et al. (1990).

# 4. Characterization techniques

The techniques used to analyze presolar grains are listed in Table 2. Grain morphology and compositions are examined by scanning electron microscopy, SEM. Structural information is obtained by transmission electron microscopy, TEM, and Raman spectroscopy. Thin-sections of graphite and SiC grains for TEM studies are prepared by two methods. One is to embed grains in resin and to slice them into sections with a nominal 70 nm thickness using an ultramicrotome equipped with a diamond knife (e.g., Bernatowicz et al. 1991). Another method is "focused ion beam lift-out", where a ~3µm-thick Pt-layer is deposited onto a grain to protect it from subsequent sputtering by a high-energy Ga beam that "cuts" parallel trenches to make a ~100 nm thin section (e.g., Stroud et al. 2002).

Most light element isotopic data have been obtained by ion probe. Presolar oxides are a minor oxide population in meteorites, and the majority of meteoritic oxides is isotopically normal. Ion imaging techniques, pioneered by P. Hoppe and further developed by Nittler et al. (1997) are used to efficiently locate presolar oxide grains. The ratio of the signal strengths of $^{16}O$ and $^{18}O$ is used to produce isotopic "maps" of grains on a sample mount (~100×100 µm/image). Any grains with isotope ratios significantly different than the standard then can be located from the isotope map of the mount for further analyses.

Abundance and isotopic measurements of reasonable precision for several elements by conventional ion probe require grains of ≥1µm and relatively high elemental abundances. Analyses of sub-micron grains with much higher precision are possible with the NanoSIMS. Its $Cs^+$ primary beam size can be as small as 30 nm (as opposed to a few µm for the CAMECA IMS-3f) and achieves high spatial resolution. The NanoSIMS also has higher sensitivity at high-mass resolution (e.g., 40 × higher than the IMS-3f for Si isotopic analysis) and a multi-detection system with four mobile and one fixed electron multipliers.

Very high precision data for the isotopic composition of heavy elements (Sr and Ba) in aggregate grain samples have been obtained by conventional thermal ionization mass spectroscopy, TIMS. The sample (e.g., SiC grains) is directly loaded onto a filament and gradually heated, which ionizes different elements at different temperatures, and thus separates elements within similar mass ranges.

**Table 2.** Analytical methods applied in presolar grain studies

| Analytical quantity [a] | Method [b] | References [c] |
|---|---|---|
| *Elemental abundances* | | |
| SiC (single/bulk): Mg, Al, Ca, Ti, V, Fe, Sr, Y, Zr, Ba, Ce | SIMS | Amari et al. 1995c |
| graphite (single): H, N, O Al, Si | SIMS | Hoppe et al. 1995 |
| diamond (bulk): Sc, Cr, Fe, Co, Ni, Ru, Os, Ir | INAA | Lewis et al. 1991b |
| *Isotopes of light elements* | | |
| SiC (single/bulk): C, N, O, Al-Mg, Si, Ca, Ti, Fe | SIMS | Hoppe et al. 1994, Huss et al. 1997, Amari et al. 2000a |
| graphite (single/bulk): C, N, O, Al-Mg, Si, Ca, Ti | SIMS | Amari et al. 1993, Hoppe et al. 1995, Travaglio et al. 1999 |
| diamond (bulk): C, N, | MS | Russell et al. 1996 |
| $Si_3N_4$ (single): C, N, Si | SIMS | Nittler et al. 1995 |
| oxides (single): O, Al-Mg (Ca, Ti) | SIMS | Huss et al. 1994, Hutcheon et al. 1994, Nittler et al. 1997, Choi et al. 1998, 1999 |
| silicates (single): O, Al-Mg | SIMS | Messenger et al. 2003, Nguyen and Zinner 2004, Nagashima et al. 2004 |
| *Isotopes of heavy elements* | | |
| SiC (bulk): Sr, Ba, Nd, Sm | TIMS, SIMS | Ott and Begemann 1990, Zinner et al. 1991, Prombo et al. 1993 |
| Sr, Zr, Mo, Ba, Ru | RIMS | Nicolussi et al. 1997, 1998a,c, Savina et al. 2003b, 2004 |
| graphite (single): Zr, Mo | RIMS | Nicolussi et al. 1998b |
| diamond (bulk): Te | TIMS | Richter et al. 1998 |
| *Noble gases* | | |
| SiC (bulk): He, Ne, Ar, Kr, Xe | NGMS | Lewis et al. 1990,1994 |
| graphite (bulk): Ne, Ar, Kr, Xe | NGMS | Amari et al. 1995a |
| SiC & graphite (single): He, Ne | NGMS | Nichols et al. 2005 |
| *Crystal structure* | | |
| SiC | TEM, Raman | Virag et al. 1992, Bernatowicz et al. 1992, Daulton et al. 2002, 2003 |
| graphite | TEM, Raman | Zinner et al. 1995, Bernatowicz et al. 1991,1996, Croat et al. 2003 |
| diamond | TEM, EELS | Bernatowicz et al. 1990, Daulton et al. 1996 |
| Multiple properties (grain size, morphology, elemental compositions) | SEM | Hoppe et al. 1994, 1995 |

Notes. [a] single: individual grain analysis. – bulk: analysis of aggregate grain samples
[b] EELS: Electron energy loss spectrometry. – INAA: Instrumental neutron activation analysis. - MS: Mass spectrometry. –NGMS: Noble gas mass spectrometry. - SEM: Scanning electron microscopy. - SIMS: Secondary ion mass spectrometry. - TEM: Transmission electron microscopy. - Raman: Raman spectroscopy. – RIMS: Resonant ionization mass spectrometry
[c] Only a few references are listed here to provide a starting point to analytical details

Heavy element isotopes in single grains were analyzed by resonant ionization mass spectrometry (RIMS) with the CHARISMA instrument at Argonne National Laboratory in Chicago (e.g., Nicolussi et al. 1998a). First, a plume of neutral atoms and molecules is generated by laser-induced thermal desorption. Then the atoms are resonantly ionized by tuned laser beams, and subsequently analyzed by a time-of-flight mass spectrometer. Although ionization schemes must be developed for each element, an advantage is that isotopes of interest can be analyzed without isobaric interference from other species.

# 5. The presolar grain "zoo"

The abundant presolar minerals in primitive meteorites are diamond, SiC, graphite, corundum, spinel, and silicates; less frequently found are $Si_3N_4$, hibonite, and $TiO_2$ (Table 3). Presolar graphite grains also often enclose small trace element carbide and Fe-Ni metal particles.

**Table 3**. Currently known presolar minerals

| Mineral | Characteristic size | Possible stellar source [a] | Discovery papers |
|---|---|---|---|
| Diamond | 2 nm | AGB?, SN? | Lewis et al. 1987 |
| SiC | 0.1 – 20 μm | AGB, SN, novae | Bernatowicz et al. 1987, Tang and Anders 1988 |
| graphite | 1 – 20 μm | AGB, SN | Amari et al. 1990 |
| carbides in graphite | 10 – 200 nm | AGB, SN | Bernatowicz et al. 1991, 1996, |
| metal grains in graphite | 10 – 20 nm | SN | Croat et al. 2003, 2004 |
| $Si_3N_4$ | 0.3 – 1 μm | AGB?, SN | Nittler et al. 1995 |
| corundum ($Al_2O_3$) | 0.2 – 3 μm | RGB, AGB, SN? | Hutcheon et al. 1994, Nittler et al. 1994 |
| spinel ($MgAl_2O_4$) | 0.2 – 3 μm | RGB, AGB, SN? | Nittler et al. 1997, Choi et al. 1998 |
| hibonite ($CaAl_{12}O_{19}$) | 0.2 – 3 μm | RGB, AGB, SN? | Choi et al. 1999 |
| $TiO_2$ | | | Nittler and Alexander 1999 |
| silicates (olivine, pyroxene) | 0.1 – 0.3 μm | RGB, AGB, SN? | Messenger et al. 2003 (in IDPs), Nguyen and Zinner 2004 (in chondrites) |

Notes. [a] AGB: asymptotic giant branch stars. RGB: red giant branch stars. SN: supernovae. A "?" indicates not known/uncertain.

By now, several thousand individual SiC and a few hundred graphite grains have been analyzed. Although diamond was discovered first and is the most abundant presolar mineral, its nm-size prevented individual grain analyses, and analyses are restricted to samples consisting of collections of grains (Lewis et al. 1987, 1991a, Richter et al. 1998). Most of the several hundred presolar oxide grains analyzed are corundum and spinel, and only some dozen are hibonite (Huss et al. 1994, Hutcheon et al. 1994, Nittler et al. 1994, 1997, 1998, Choi et al. 1998, 1999, Nittler and Alexander 1999, Krestina et al. 2002, Nguyen et al. 2003, Zinner et al. 2003). The first presolar silicates (olivine and pyroxene) in meteorites were detected by Nguyen and Zinner (2004). Nagashima et al. (2004) located presolar silicates in thin sections of two primitive meteorites.

Carbonaceous presolar grains are ubiquitous in essentially all types of primitive chondritic meteorites (Table 4). The observed abundance of presolar minerals depends not only on the resistance to chemical acid processing, which leaves abundances of diamond, graphite, and SiC essentially unaffected, depending on the specifics of the acid treatment (Amari et al. 1994), but also on grain survival during metamorphism on meteorite parent bodies.

Abundance estimates for presolar corundum, spinel, silicates, and $Si_3N_4$ are only available for a few meteorite groups. Corundum in unequilibrated H chondrites is ~0.03 ppm and 0.13 ppm in CM chondrites, which also contain 1-2 ppm spinel and 0.002 ppm $Si_3N_4$. A first lower abundance limit of ~25 ppm for presolar silicates in a very primitive carbonaceous chondrite was reported by Nguyen and Zinner (2004). In addition, presolar silicates were found in interplanetary dust particles (IDPs), where they appear to be more abundant (~5500 ppm in cluster IDPs) than in meteorites (Messenger et al. 2003).

**Table 4.** Abundance estimates of presolar minerals in different meteorites (ppm by mass)

| Chondrite group | Diamond | SiC | Graphite |
|---|---|---|---|
| CI | 940 –1400 | 14 | 10 |
| CM | 400 – 740 | 4–14 | 5-6 |
| CR | 400 | 0.6 | |
| CO | 300 –520 | 1-3 | <0.15? |
| CV-reduced | 545-620 | 0.17-0.39 | below detection limit |
| CV-oxidized | 240 –500 | 0.006-0.2 | ≤0.20 |
| CH | 87 | 0.41 | 0.13 |
| H 3.4 | ~36 | 0.063 | |
| L 3.4/3.7 | 54-64 | 0.008–0.08 | |
| LL3.0/3.1 | 100 – 130 | 0.39–1.52 | |
| EH3-4 | 50 –67 | 1.3–1.6 | |

Sources: Huss and Lewis 1995, Huss et al. 2003, Ott 1993, Zinner et al. 2003

Primitive meteorites of low petrographic type that never experienced high metamorphic temperatures or aqueous alteration (e.g., unequilibrated ordinary chondrites) provide the best time capsules for preserving presolar grains. Early studies showed that the C-isotopic compositions in the most primitive ordinary chondrites (petrologic type 3) varied more than those of higher petrologic types (4-6) (Swart et al. 1983). This could suggest selective destruction of presolar grain populations in ordinary chondrites of higher-metamorphic type, which is also indicated by noble gas measurements of diamond separates from different types of chondrites and provides a useful measure of meteorite parent body metamorphism (e.g., Huss and Lewis 1994a, 1995).

## 5.1. Presolar silicon carbide

Silicon carbide is the most extensively studied presolar mineral, both in the form of aggregated "bulk" samples and of single grains, for several reasons. First, the separation procedure of SiC is less complicated than that of graphite. Second, SiC is present in several classes of meteorites (Table 4). Third, although most SiC grains in meteorites are sub-micron in size, some grains larger than 1 μm are suitable for correlated isotopic analyses in single grains. Fourth, trace element concentrations are high enough to make elemental abundance and isotopic measurements with reasonable precision.

The **average size** of SiC grains is ~0.5 μm in the Murchison meteorite, with sizes ranging up to 20 μm, but grains >10 μm are extremely rare. The size distribution of SiC in different types of meteorites varies, e.g., the Murchison chondrite typically has larger-size SiC on average than other meteorites, for unknown reasons (Amari et al. 1994, Huss et al. 1997, Russell et al. 1997).

The **morphology** of most SiC grains extracted from meteorites by chemical procedures shows euhedral shapes with more or less pitted surfaces (Fig. 4a,b).

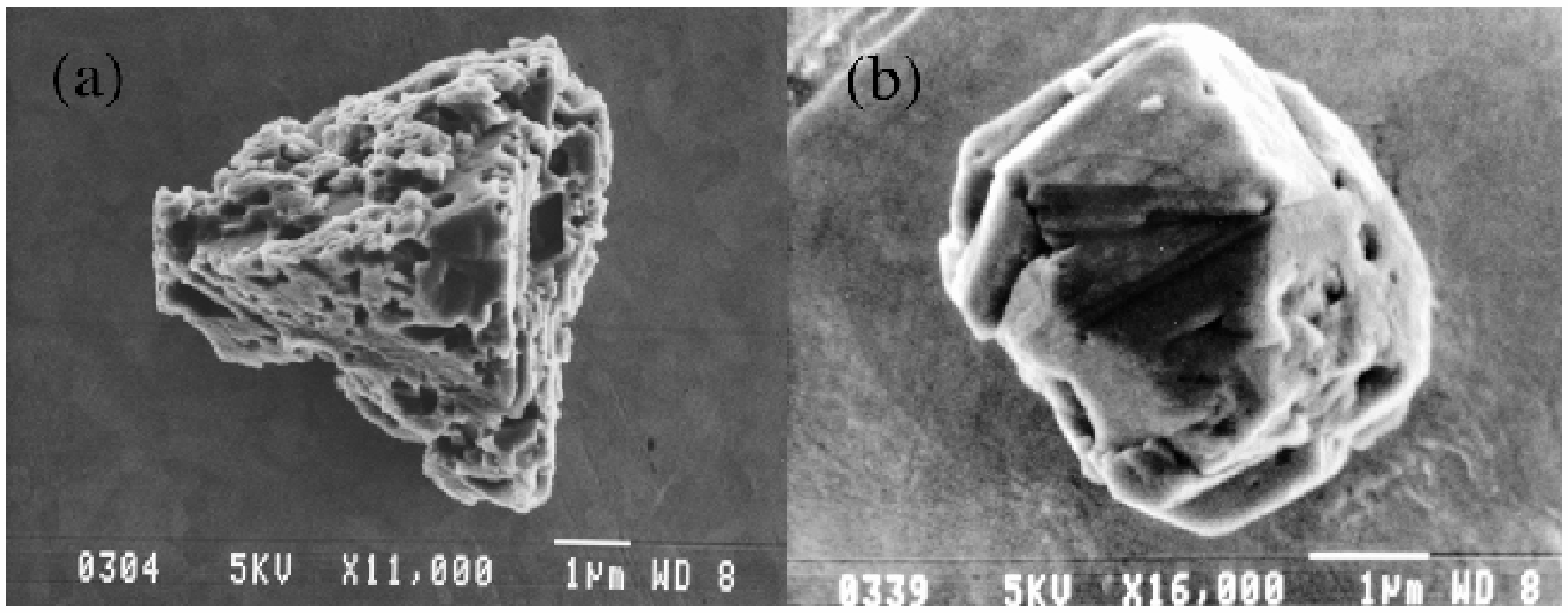

**Fig. 4.** Secondary electron images of SiC grains from the Murchison meteorite. Larger grains such as these shown are relatively rare. Scale bars are 1 μm. (a) The pitted surface structure is common for SiC grains, and most likely due to the harsh chemical treatments during the extraction from meteorites. The $^{12}C/^{13}C$ ratio of this grain is 55 (cf. solar = 89). (b) A SiC grain with a smooth surface. The $^{12}C/^{13}C$ ratio of this grain is 39.

In order to examine "pristine" SiC grains not subjected to chemicals, Bernatowicz et al. (2003) dispersed matrix material excavated from the interior of the Murchison meteorite onto polished graphite planchets and examined them by SEM. The pristine grains have less surface pits than grains isolated by chemical procedures, indicating that etching of surface defect structures occurs during chemical isolation. Of the 81 pristine grains studied by Bernatowicz et al. (2003), ~60% are coated with an amorphous, possibly organic phase. No differences in morphology other than those caused by sample extraction procedures have been observed among SiC grains.

Synthetic SiC has several hundred different crystallographic modifications but presolar SiC apparently only occurs in the cubic 3C and hexagonal 2H modification and intergrowths of these two (Daulton et al. 2002, 2003). This limited polytype distribution in presolar SiC suggests condensation of SiC at relatively low total pressures in circumstellar shells (Daulton et al. 2002, 2003).

### 5.1.1. Chemical and isotopic compositions of "bulk" SiC aggregates

The analyses of bulk samples have the advantage that data with high precision can be obtained. This provides well-defined *average* properties for whole suites of separates which allows one to examine systematic differences among the various fractions. On the other hand, there is the potential problem that aggregates may contain some contaminant, which can hamper the interpretation of bulk elemental abundances. However, any isotopic anomalies would only be diluted by contamination with normal material, and large overall isotopic anomalies can still be detected.

Silicon carbide is the carrier of Ne-E(H), Kr-S and Xe-S. Lewis et al. (1990, 1994) analyzed all noble gases in size-sorted SiC fractions (average size: 0.38-4.6 μm) extracted from the Murchison meteorite. They modeled the measured noble gas compositions by a mixture of two-components, which they designated

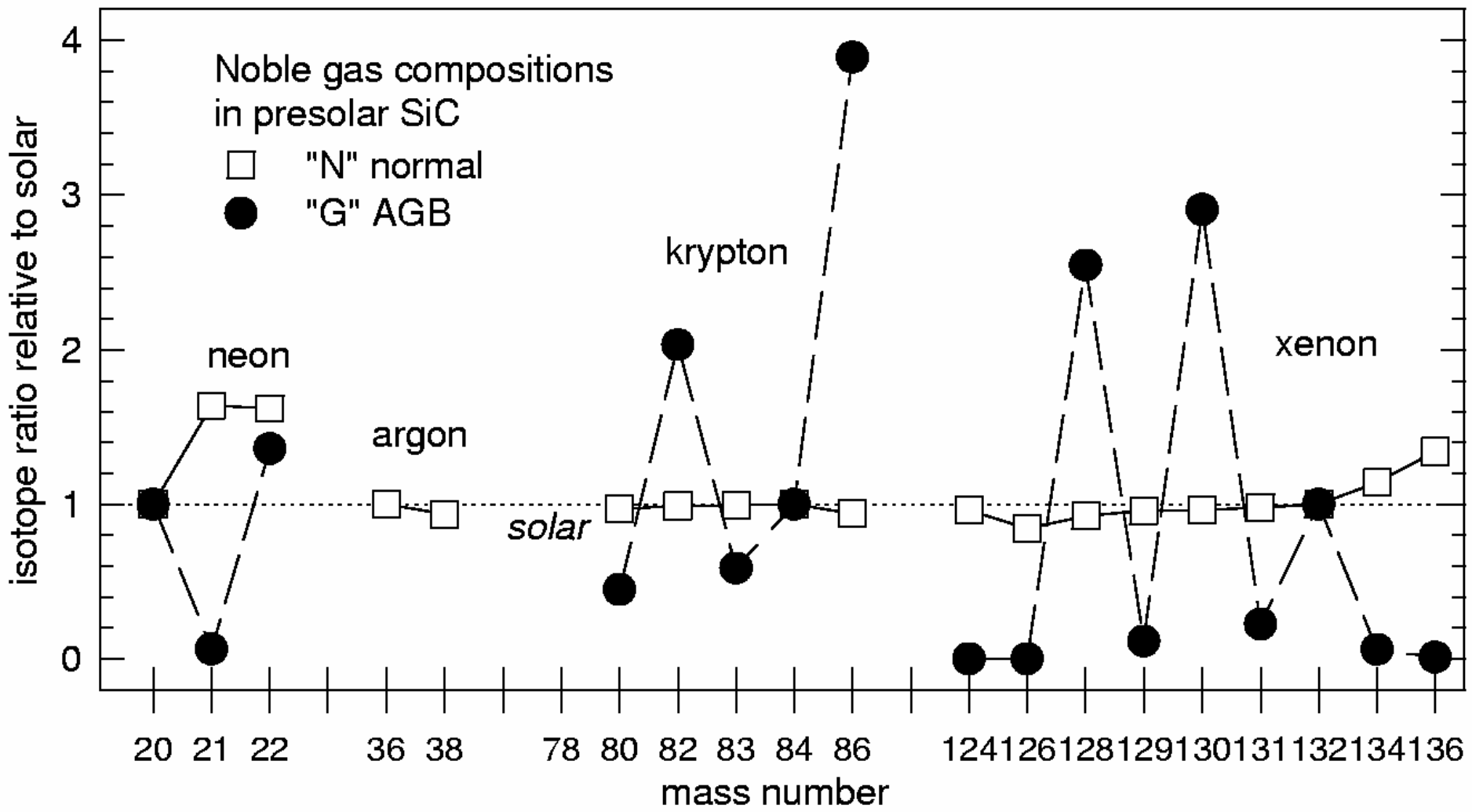

**Fig. 5.** Noble gas components in aggregates of SiC grains (Lewis et al. 1990, 1994, Ott 2002) normalized to solar isotopic composition (Wieler 2002). Isotope ratios are further normalized to $^{20}$Ne, $^{36}$Ar, $^{84}$Kr, and $^{132}$Xe, respectively. The dotted line shows solar composition.

"N" component for "*n*ormal", and "G" component for "A*G*B" (Fig. 5). The G component of Kr was found to be similar to the theoretical composition from nucleosynthesis by the *s*-process in AGB stars (Gallino et al. 1990, 1997), hence the name. Initially the terms "G" and "N" component were only used for noble gases but are also in use for other elements. The isotopic compositions of Sr (Podosek et al. 2004) and Ba (Zinner et al. 1991, Prombo et al. 1993) were analyzed for the same size-sorted SiC fractions (but for different suites) by TIMS and SIMS. In all cases, the abundant G component indicated that a large fraction of presolar SiC grains came from stars in which the *s*-process responsible for the G component operates.

### 5.1.2. Individual SiC grains: Clues to SiC sub-populations

Isotopic and trace element analyses of single grains by ion probe revealed the presence of different types of SiC. The isotopic compositions of C, N, and Si lead to five major SiC subtypes called mainstream, A+B, X, Y, and Z (Table 5, Figs. 6 and 7). The number of grains plotted in Figs. 6 and 7 does *not* represent their true distribution among the sub-populations because certain grains were preferentially searched for by ion imaging and then studied. The "true" distribution (by number) is noted in the legends of Figs. 6 and 7, and in Table 5.

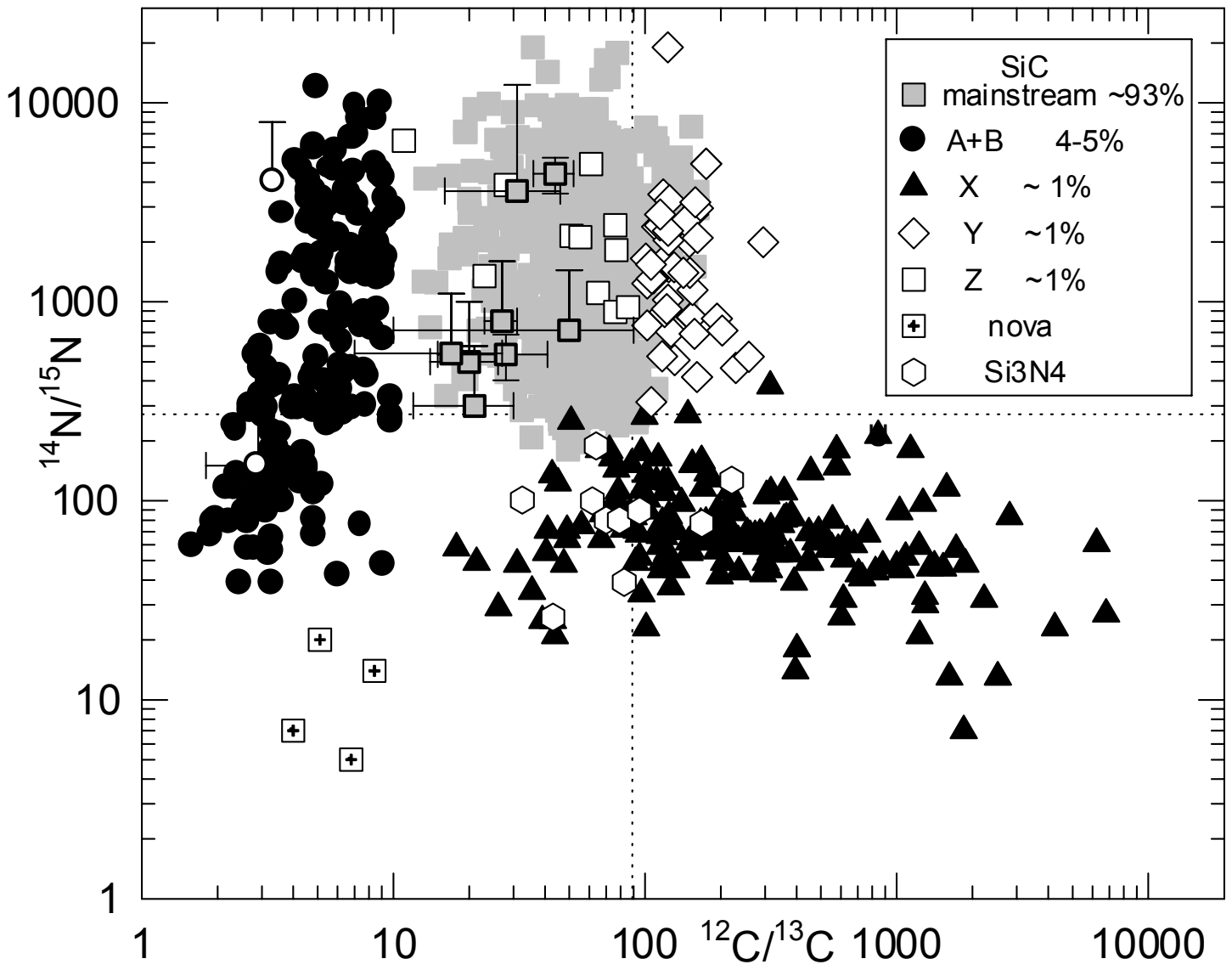

**Fig. 6.** SiC grains fall into different populations based on their C- and N-isotope ratios (Alexander 1993, Amari et al. 2001a-c, Hoppe et al. 1994, 1997,2000, Huss et al. 1997, Lin et al. 2002, Nittler et al. 1995). For comparison, stellar data are plotted with error bars and their N-isotopic ratios are typically lower limits (Wannier et al. 1991, Querci and Querci 1970, Olsen and Richter 1979). The dotted lines indicate solar isotope ratios.

About 93% (by number) of presolar SiC are **mainstream grains**, with lower $^{12}C/^{13}C$ and higher $^{14}N/^{15}N$ than the respective terrestrial reference ratios of $^{12}C/^{13}C$=89 and $^{14}N/^{15}N$=272. Their Si-isotopic composition is slightly $^{29}Si$- and $^{30}Si$-rich ($^{29}Si/^{28}Si$ and $^{30}Si/^{28}Si$ are up to 1.2×solar). In the three Si-isotope plot, mainstream grains define a line with $\delta^{29}Si = -15.9 + 1.31\ \delta^{30}Si$ (Lugaro et al. 1999). The fit parameters vary depending on the number of points included in the regressions, e.g., Hoppe et al. (1994) find $\delta^{29}Si = -15.7 + 1.34\ \delta^{30}Si$, and Nittler and Alexander (2003) obtained $\delta^{29}Si = -18.3(\pm 0.6) + 1.35(\pm 0.01)\ \delta^{30}Si$ .

Grains with $^{12}C/^{13}C < 10$ and $^{14}N/^{15}N = 40 - 12{,}000$ are called **A+B grains** (Amari et al. 2001a). Originally, it was thought that two distinct populations "A" and "B" existed, but these two belong to the same continuum spanned by C- and N-isotopes (Fig. 6). The Si-isotopes of A+B grains are similar to those of mainstream grains (Fig. 7). In the three Si-isotope plot A+B grains define a line with $\delta^{29}Si = -34.1(\pm 1.6) + 1.68(\pm 0.03)\ \delta^{30}Si$ (Amari et al. 2000b) with a small off-set in slope compared to the mainstream grains. The A+B grains are the second largest presolar SiC population and constitute 3-4% of all SiC grains.

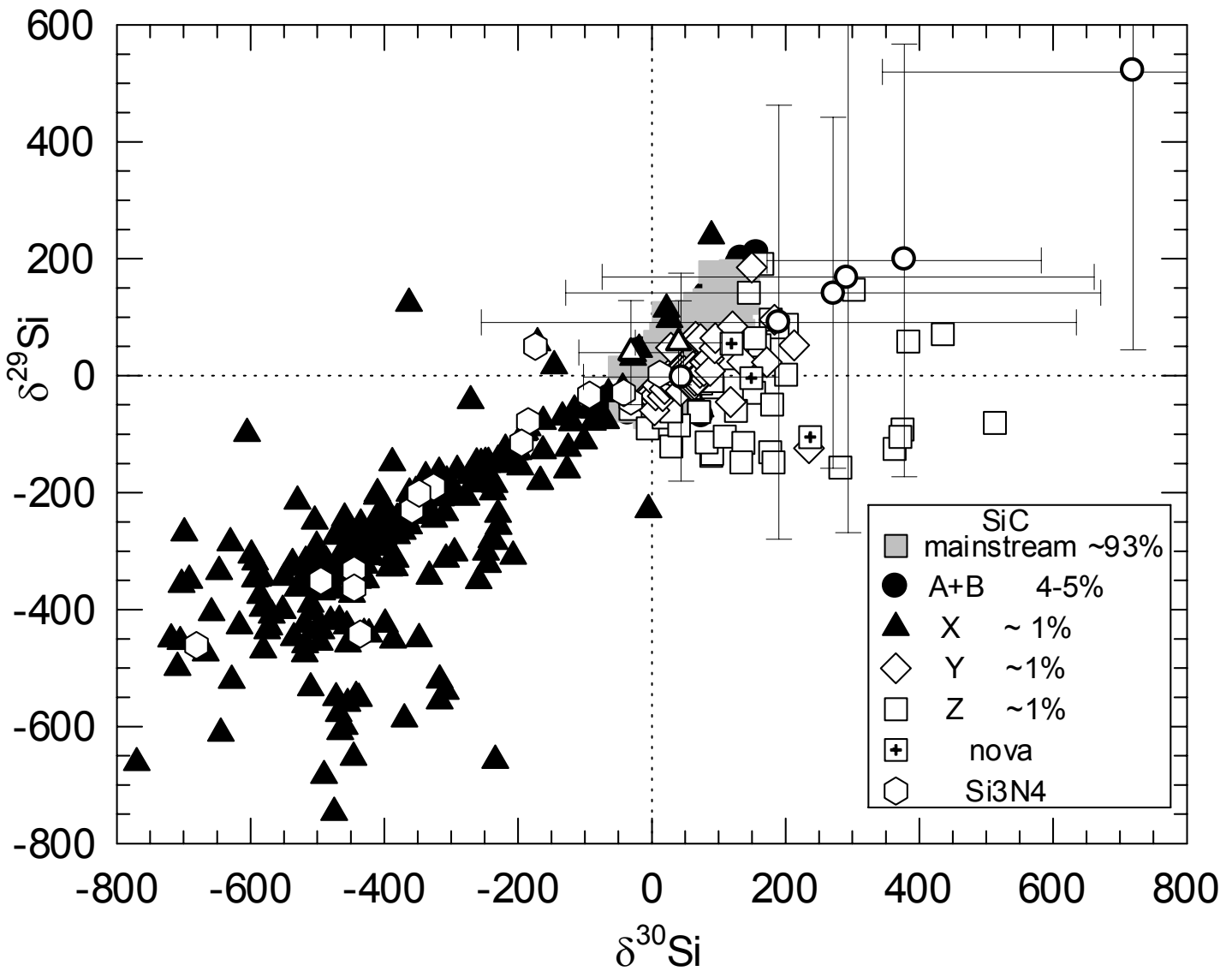

**Fig. 7.** The Si-isotopes for presolar SiC grains (references as in Fig. 6) and stars (symbols with error bars) are given in the δ–notation which describes the deviation of an isotope ratio ($^{i}N/^{j}N$) of a sample from the (terrestrial) standard ratio in per-mil: $\delta^{i}N$ (‰) = $[(^{i}N/^{j}N)_{sample} / (^{i}N/^{j}N)_{standard} - 1] \times 1000$. The grains fall into distinct populations. The triangles show two determinations for the C-star IRC+10°216 (Cernicharo et al. 1986, Kahane et al. 1988). The other stellar data (circles) are for O-rich M-giants (Tsuji et al. 1994).

The **SiC grains of type X**, ~1% of all SiC, have higher $^{12}C/^{13}C$ and lower $^{14}N/^{15}N$ than the respective solar ratios. The X grains have low $\delta^{29}Si$ and $\delta^{30}Si$ values, and $^{28}$Si excesses reach up to 5×solar (Amari et al. 1992, Hoppe et al. 2000, Amari and Zinner 1997). Another one percent of all SiC grains have $^{12}$C/$^{13}$C >100 and $^{14}$N/$^{15}$N above the solar ratio (Amari et al. 2001b, Hoppe et al. 1994). These **Y grains** appear to be $^{12}$C-rich mainstream grains but their $^{30}$Si/$^{28}$Si ratios are slightly larger than in mainstream grains, which merits placing them into a separate group. Up to 3% of all SiC grains, particularly among smaller size grain fractions, are **Z grains**. Their $^{12}$C/$^{13}$C and $^{14}$N/$^{15}$N ratios are similar to those of mainstream grains, but Z grains have large $^{30}$Si excesses (Alexander 1993, Hoppe et al. 1997). Only a few **nova SiC grains,** with $^{12}$C/$^{13}$C = 4-9, and $^{14}$N/$^{15}$N = 5-20, are known (Amari et al. 2001c, José et al. 2004, Nittler and Hoppe 2004a,b). Most nova grains have close-to-solar $^{29}$Si/$^{28}$Si but $^{30}$Si excesses.

**Table 5.** Some characteristics of presolar silicon carbide populations

| Designation | Mainstream | X | Y | Z | A+B [a] | Nova |
|---|---|---|---|---|---|---|
| Crystal type | 3C, 2H [b] | 3C, 2H [b] | 3C, 2H [b] | 3C, 2H [b] | 3C, 2H [b] | 3C, 2H? [b] |
| Heavy trace elements [c] | ~10-20× [c] | highly depleted | ~ 10× [c] | NA | solar or 10-20× [c] | NA |
| $^{12}$C/$^{13}$C | 10 – 100 | 20 – 7000 | 140 – 260 | 8 – 180 | < 3.5 (A) 3.5 – 10(B) | < 10 |
| $^{14}$N/$^{15}$N | 50 – 2×10$^{4}$ | 10 – 180 | 400 – 5000 | 1100 – 1.9×10$^{4}$ | 40–1.2×10$^{4}$ | < 20 |
| $^{29}$Si/$^{28}$Si [c] | 0.95-1.20× | $^{28}$Si-rich | 0.95-1.15× | ≈solar | 1.20× | ≈solar |
| $^{30}$Si/$^{28}$Si [c] | 0.95-1.14× | $^{28}$Si-rich | $^{30}$Si-rich | $^{30}$Si-rich | 1.13× | $^{30}$Si-rich |
| $^{26}$Al/$^{27}$Al | 10$^{-3}$ to 10$^{-4}$ | 0.02 to 0.6 | similar to MS | similar to MS | <0.06 | up to 0.4 |
| Other isotopic markers [c] | excess in $^{46}$Ti, $^{49}$Ti, $^{50}$Ti over $^{48}$Ti | $^{44}$Ca excess $^{41}$K excess | excess in $^{46}$Ti, $^{49}$Ti, $^{50}$Ti over $^{48}$Ti | excess in $^{46}$Ti, $^{47}$Ti, $^{49}$Ti over $^{48}$Ti | excess in $^{46}$Ti, $^{49}$Ti, $^{50}$Ti over $^{48}$Ti | $^{47}$Ti-rich |
| $^{22}$Ne [d] | yes | NA | NA | NA | NA | NA |
| Abundance | 87-94% | 1% | 1 – 2% | 0 – 3% | 2 – 5% | << 1% |

*Sources*: Amari et al. 2001a,b, Hoppe and Ott 1997, Hoppe and Zinner 2000, Nittler and Hoppe 2004a,b,Ott 2003, Zinner 1998

[a] Group A and B grains were initially separated but later found to form a continuum in composition.

[b] cubic 3C, hexagonal 2H; Daulton et al. (2002, 2003).

[c] Abundance compared to solar composition.

[d] $^{22}$Ne = Ne-E(H) = Ne(G); and NA: not analyzed.

Many SiC grains have $^{26}$Mg/$^{24}$Mg larger than the solar ratio but solar $^{25}$Mg/$^{24}$Mg within 10% (Amari et al. 1992, 2001a-c, Hoppe et al. 1994, 2000, Huss et al. 1997). Magnesium in some X grains is almost pure $^{26}$Mg, and $^{26}$Mg excesses are most likely from *in-situ* decay of $^{26}$Al ($t_{1/2}$ = 7.3×10$^{5}$ a) that was incorporated into grains at their stellar sources. The X grains have $^{26}$Al/$^{27}$Al of up to ~0.6 (Fig. 8), whereas ratios in A+B and mainstream grains typically do not exceed 0.01 (Hoppe et al. 1994, Amari et al. 2001a). Isotopic compositions were also measured for Ca and Ti (Ireland et al. 1991, Amari et al. 1992, 2001a,b, Hoppe et al. 1994, 1996, 2000, Nittler et al. 1996, Alexander and Nittler 1999, Hoppe and Besmehn 2002, Besmehn and Hoppe 2003) and Fe, Zr, Mo, Sr, Ba, and Ru (e.g., Nicolussi et al. 1997, 1998a, 1998c, Pellin et al. 2000a,b, Davis et al. 2002, Savina et al. 2003a,b, 2004).

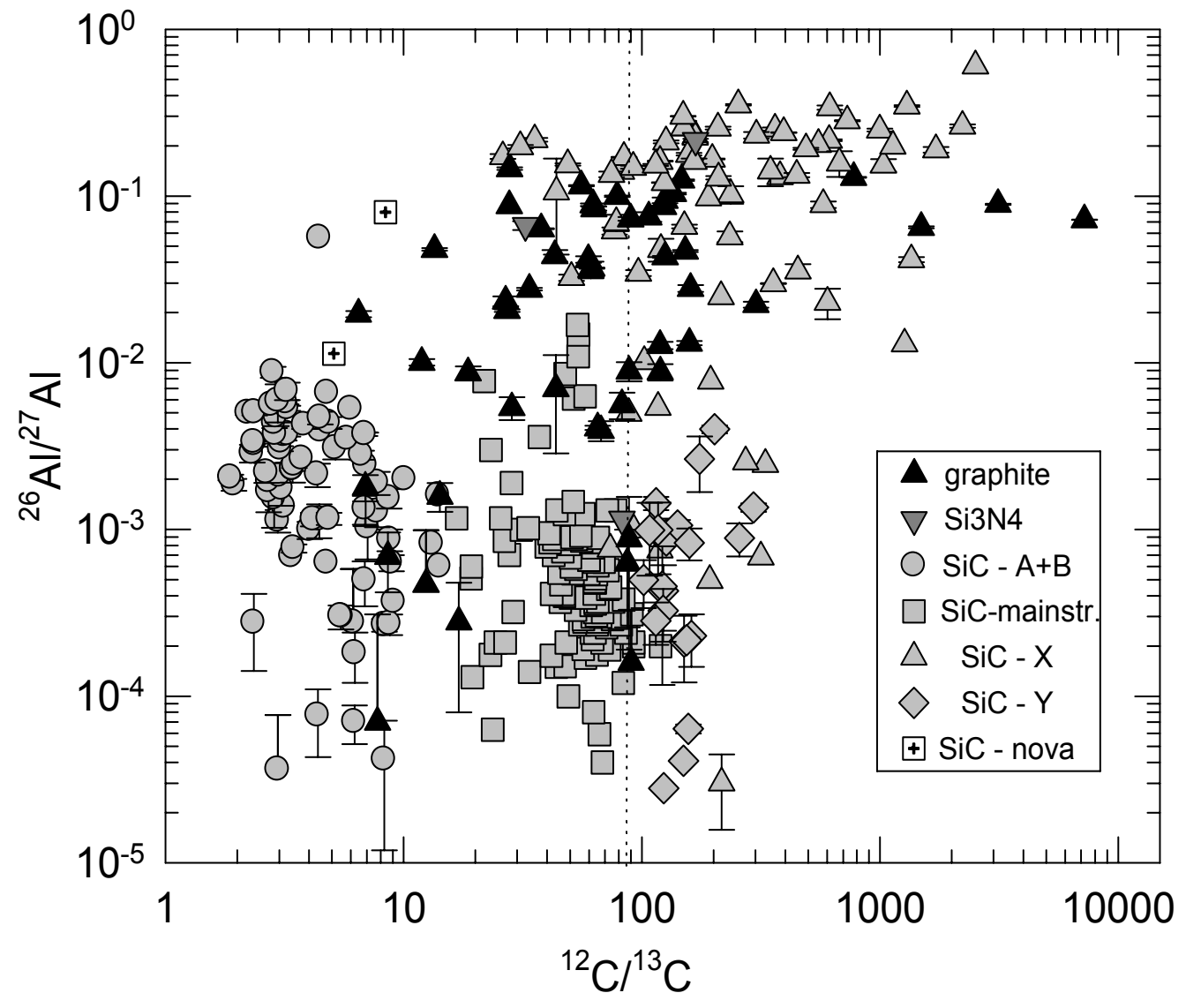

**Fig. 8.** Inferred $^{26}Al/^{27}Al$ ratios vs. $^{12}C/^{13}C$ ratios in SiC and low-density graphite grains. The SiC type X and graphite grains have the largest $^{26}Al/^{27}Al$. See text for data sources.

### 5.1.3. Trace elements in individual SiC grains

Silicon carbide grains contain several trace elements, some of them in considerable amounts. Magnesium concentrations are typically around 100 ppm and Al abundances can reach several mass-percent (e.g., Hoppe et al. 1994). Nitrogen, probably substituting for carbon in the SiC lattice, shows relatively high concentrations so the N-*isotopic* ratios can be analyzed with reasonable precision. However, determination of the absolute concentration of N is difficult as carbon must be present to produce $CN^-$ which is used to analyze N by the ion probe (Zinner et al. 1989). In addition to Al and Mg, concentrations of Ca, Ti, V, Fe, Sr, Y, Zr, Nb, Ba, Ce, and Nd were measured in 60 SiC grains (average size: 4.6μm) and in three size-sorted SiC aggregates of 0.49-0.81μm (Amari et al. 1995c). The general problem in determining multiple trace element abundances in presolar grains is that grains are partially consumed during the measurements. The first measurements are generally for C, N, and Si isotopic ratios to identify the grain's membership among the SiC populations. After this, only initially large grains have enough mass left for additional trace element analyses.

Trace element measurements of 34 mainstream grains define at least 8-9 different normalized abundance patterns (Fig. 9), and only 4 of these grains had unique abundances (Amari et al. 1995c). In mainstream grains, element/Si ratios of elements heavier than iron (Y, Zr, Nb, Ba, La, Ce, and Nd) show up to 35 times the solar ratio. These enrichments most likely reflect the composition at the stellar sources. However, the source composition can be fractionated according to volatility during trace element condensation into SiC (Lodders and Fegley 1995). In Fig. 9, the *s*-process elements Sr and Ba should be as abundant as their neighbors, but they are depleted in many grains because they are more volatile.

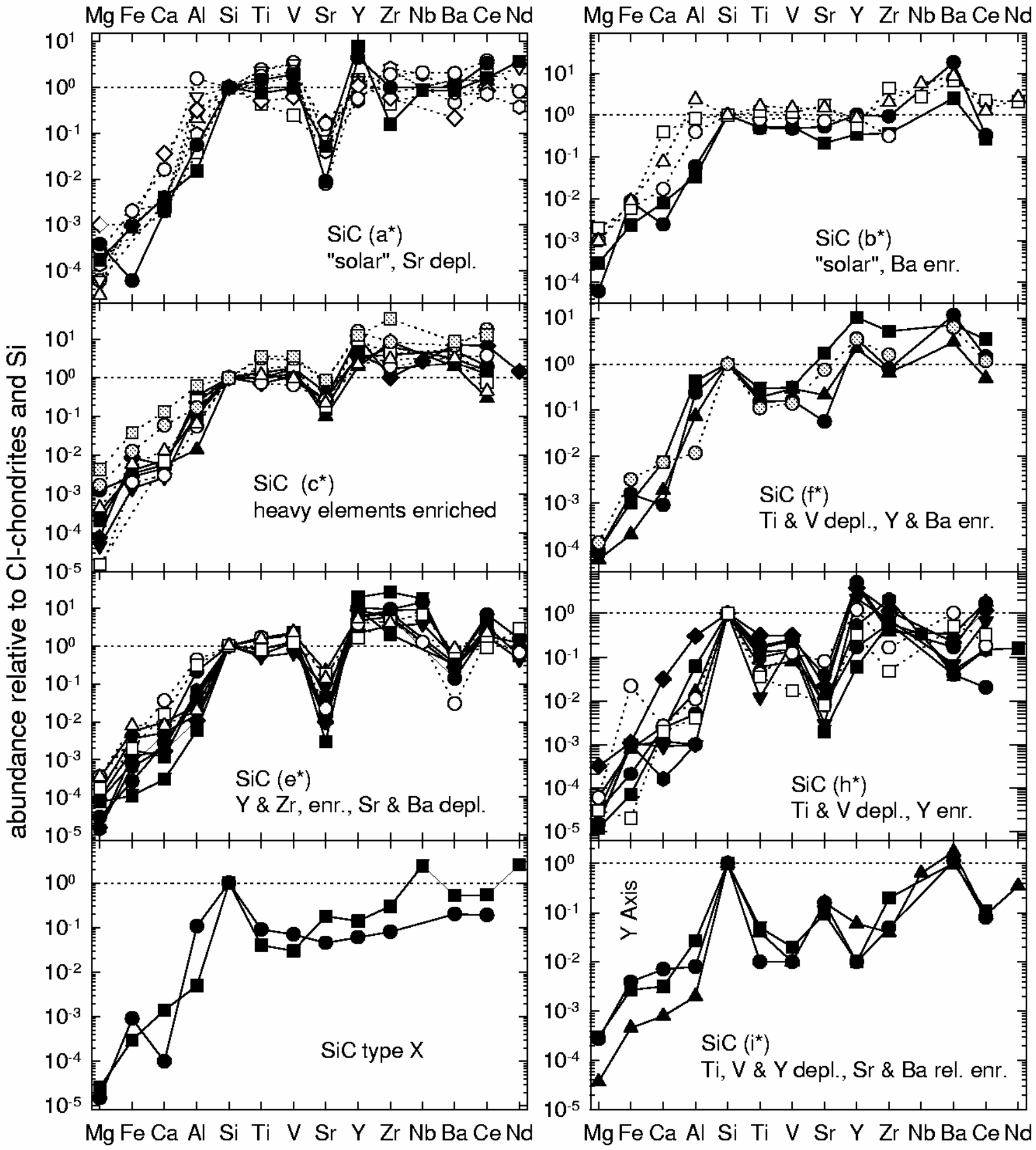

**Fig. 9.** Trace element abundances in individual SiC grains normalized to solar abundances and Si. Relative depletions or enrichments in Sr and Ba are most notable. Mainstream grains are shown by black symbols, open symbols are for A+B grains, and type Y grains are shown in grey. Abundances in two X type SiC grains are shown in a separate panel. The letters marked by a star refer to the original groupings in Fig. 1 by Amari et al. (1995c).

The enrichment of *s*-process elements in Y grains (Fig. 9) supports their close relationship to mainstream SiC. Trace elements have not yet been reported for Z type grains. Most of the trace element abundances measured in the two X grains are very low and data for Zr and Sr are upper limits. The trace elements of twenty A+B SiC grains (Fig. 9) give 4 different abundance patterns (Amari et al. 1995c, 2001a). This led to the surprising conclusion that A+B grains require at least two types of stellar sources with different characteristic heavy element abundances, which was not indicated by their isotopic compositions. In one of the patterns, all elements plot near the solar abundance ratio, and only Sr shows a relative depletion. The other 3 patterns show higher relative abundances of the heavy elements and various volatility-related fractionations, which makes these A+B grains more similar to mainstream grains.

### 5.2. Presolar silicon nitride

The few known $Si_3N_4$ grains share many properties of the SiC X grains (Figs. 6-8). Information about these rare grains is given by Amari et al. (1992), Besmehn and Hoppe (2003), Lin et al. (2002), Nittler and Alexander (1998), and Nittler et al. (1995).

### 5.3. Presolar graphite

Presolar graphite is present only in the least thermally altered primitive meteorites (Huss and Lewis 1995, Table 4). It is not as chemically resistant as SiC or the oxides corundum and spinel, and its isolation is complicated because other carbonaceous compounds with similar chemical and physical properties are present in primitive meteorites. The separation of graphite, which is achieved by a combination of mild oxidation and density separation, is far more elaborate than that of other presolar grains, and essentially all studies of presolar graphite have been performed on four graphite-rich fractions extracted from the Murchison meteorite (Amari et al. 1994). Some properties of these four fractions are given in Table 6.

Some structural, elemental, and isotopic features of presolar graphite grains vary with density. On average, the low-density separates contain more large grains (Hoppe et al. 1995). Amari et al. (1995a) and Hoppe et al. (1995) describe correlations between density and isotopic compositions of the noble gases and carbon. This is in marked contrast to SiC grains, where observed isotopic features in Kr, N, Sr, and Ba depend only on grain size. Graphite grains are much larger, typically >1μm, but ranging up to 20 μm, than most sub-micron SiC grains. Investigation by SEM and TEM show two **morphological** types (Fig. 10) - dubbed “cauliflower” and “onion”- (Bernatowicz et al. 1991, 1996, Hoppe et al. 1995, Bernatowicz and Cowsik 1997). Grains of the onion type have a concentric-layered structure (reminiscent to that of hailstone, although on a different absolute scale) of well-graphitized carbon and a core of randomly oriented, fine-grained crystalline aggregates (Bernatowicz et al. 1996).

**Table 6.** Some characteristics of presolar graphite density fractions [a]

| Designation | KE1, KE3 | KFA1 | KFB1 | KFC1 |
|---|---|---|---|---|
| Density ($g/cm^3$) | 1.6 – 2.05 | 2.05 – 2.10 | 2.10 –2.15 | 2.15 – 2.20 |
| Morphology | mainly cauliflower | | | mainly onion |
| $^{12}C/^{13}C$ | 3.6-7223 | 3.0-2146 | 3.8-3377 | 2.1-4064 |
| $^{14}N/^{15}N$ | mostly solar (28-306) | mostly solar (123-398) | mostly solar (153-315) | mostly solar [b] |
| $^{18}O/^{16}O$ | up to 184 × solar | up to 6.6 × solar | ~solar | ~solar |
| $^{29}Si/^{28}Si$ | 0.63-2.3 × solar | 0.54-1.57 × solar | ± solar [c] | 0.8-1.3 × solar [c] |
| $^{30}Si/^{28}Si$ | 0.46-1.94 × solar | 0.39-1.4 × solar | ± solar [c] | ± solar [c,d] |
| $^{26}Al/^{27}Al$ | up to 0.146 | up to 0.138 | up to 0.086 | |

Major sources: Hoppe et al. (1995), Travaglio et al. (1999)
[a] Amari et al. (1994).
[b] One grain in KFC1 with $^{14}N/^{15}N$ = 730±64.
[c] Very large uncertainties
[d] One nova grain with $^{30}Si/^{28}Si$ = 1.76 × solar.

The cauliflower type grains appear to be aggregates of small grains and consist entirely of turbostratic graphite (i.e., graphite with contorted layers not having long-range continuity, Bernatowicz et al. 1991, 1996). The onion type graphite is more abundant in high-density fractions, and the cauliflower type is more abundant in low-density fractions.

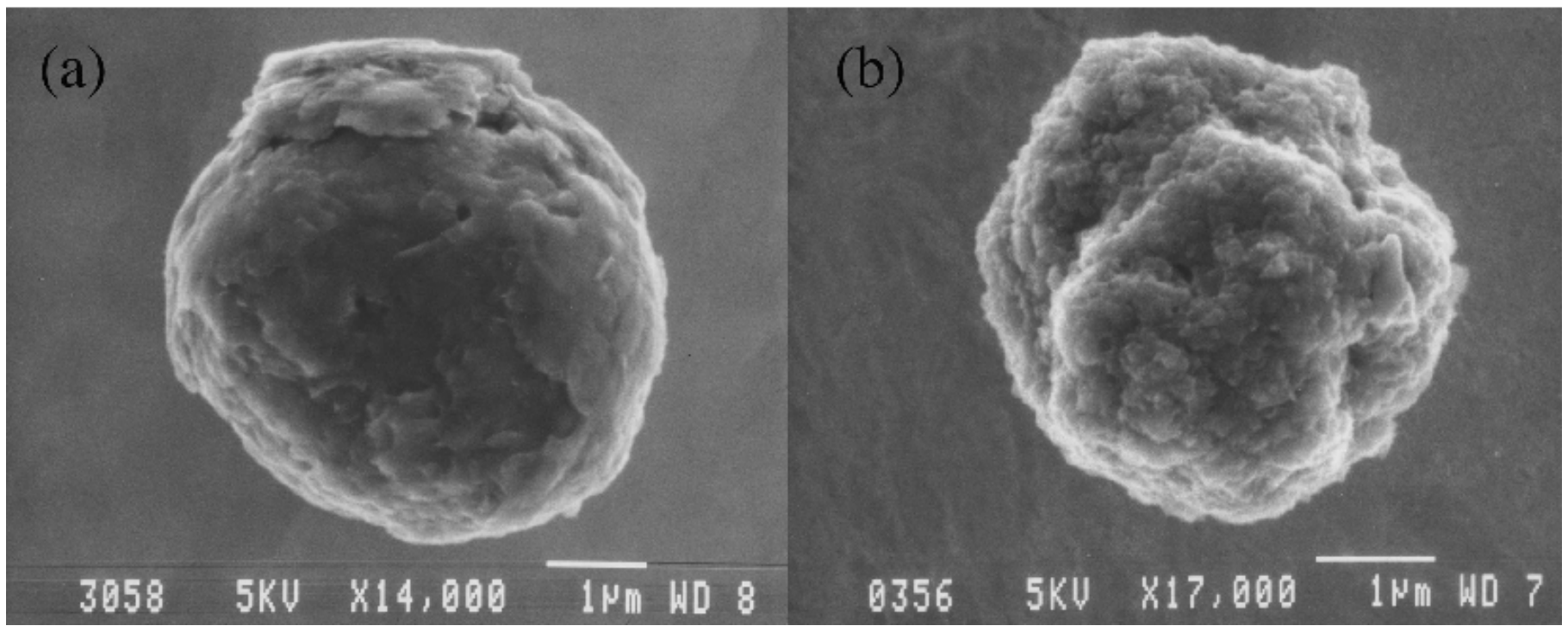

**Fig. 10.** Presolar graphite grains show two morphologies. (a) A graphite grain of the “onion” type, with a layered surface structure. (b) A graphite grain of the “cauliflower” type, which appears as aggregates of small grains.

Fig. 11 shows the C- and N-isotopic composition of individual graphite grains from the four density fractions. Overall, the $^{12}C/^{13}C$ ratios range from 2.1 to 7223, demonstrating their presolar origin. Grains with $^{12}C/^{13}C$ higher than the solar ratio are most abundant (~75%) in the highest density separate KFC1. In contrast, the isotopic composition of N, one of the most abundant trace elements in graphite, is bewilderingly solar in many grains. This has been interpreted that at least part of the indigenous N in graphite equilibrated with air (Hoppe et al 1995) The graphite from the lowest density separate (KE3) is a possible exception because there is a slight decrease in $^{14}N/^{15}N$ with increasing $^{12}C/^{13}C$ (Fig. 11). An outstanding feature of graphite grains is the presence of almost pure $^{22}Ne$ (=Ne-E(L)). If N isotopically equilibrated with air the noble gases should also equilibrate, thus erasing any isotopic anomalies. In particular, Ne should be affected because it has adsorption properties similar to N, but Ne is anomalous. This compounds the mystery of the 'normal' N-isotopic composition and requires more investigation. Abundances of H, O, N, Al, and Si in individual graphites are typically about one-mass percent (Hoppe et al. 1995). With increasing density of the grain fractions, the concentrations of H, N, O, and Si decrease. Isotopic compositions of O, Mg, Al, Si, Ca, Ti, Zr and Mo in individual graphite grains from the KE3, KFA1 and KFC1 density fractions were measured by e.g., Hoppe et al. (1995), Amari et al. (1995b, 1996), Nicolussi et al. (1998b), and Travaglio et al. (1999).

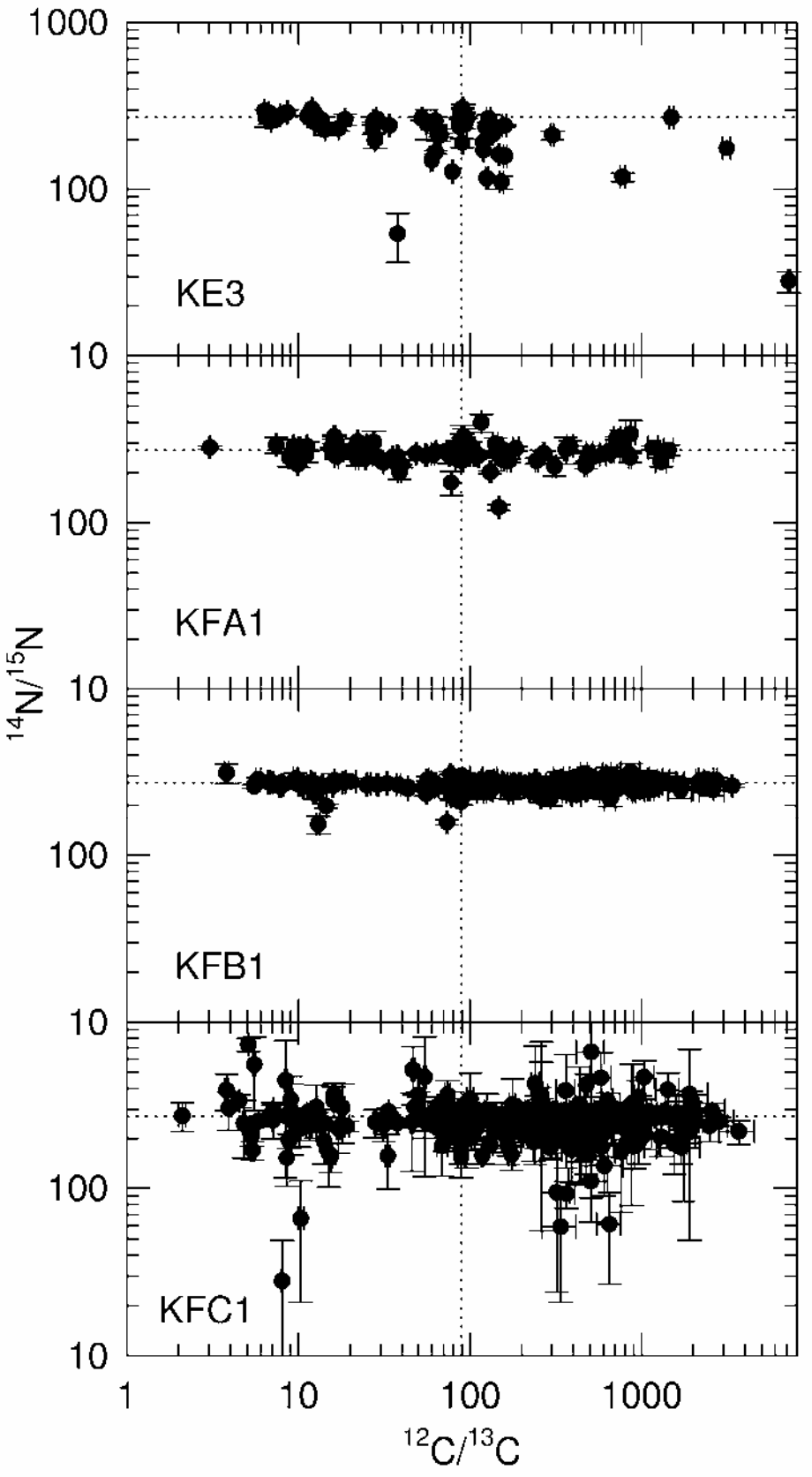

**Fig. 11.** Carbon and N-isotopes in individual graphite grains from 4 density fractions. Density increases in alphabetical order (E<FA<FB<FC).

The Si-isotopes in graphite grains (Fig. 12) are similar to those observed for the different SiC populations. Many grains from the KE3 separate are rich in $^{28}Si$, like the SiC X grains (Fig. 7). This kinship also follows from the $^{26}Al/^{27}Al$ ratios (Fig. 8). A significant fraction of grains from the low-density fractions KE3 (Travaglio et al. 1999) and KFA1 (Amari, unpublished) appear normal in $^{17}O/^{16}O$ but show large enrichments in $^{18}O/^{16}O$ relative to solar (Fig. 14). This composition is remarkably different from that of presolar oxides.

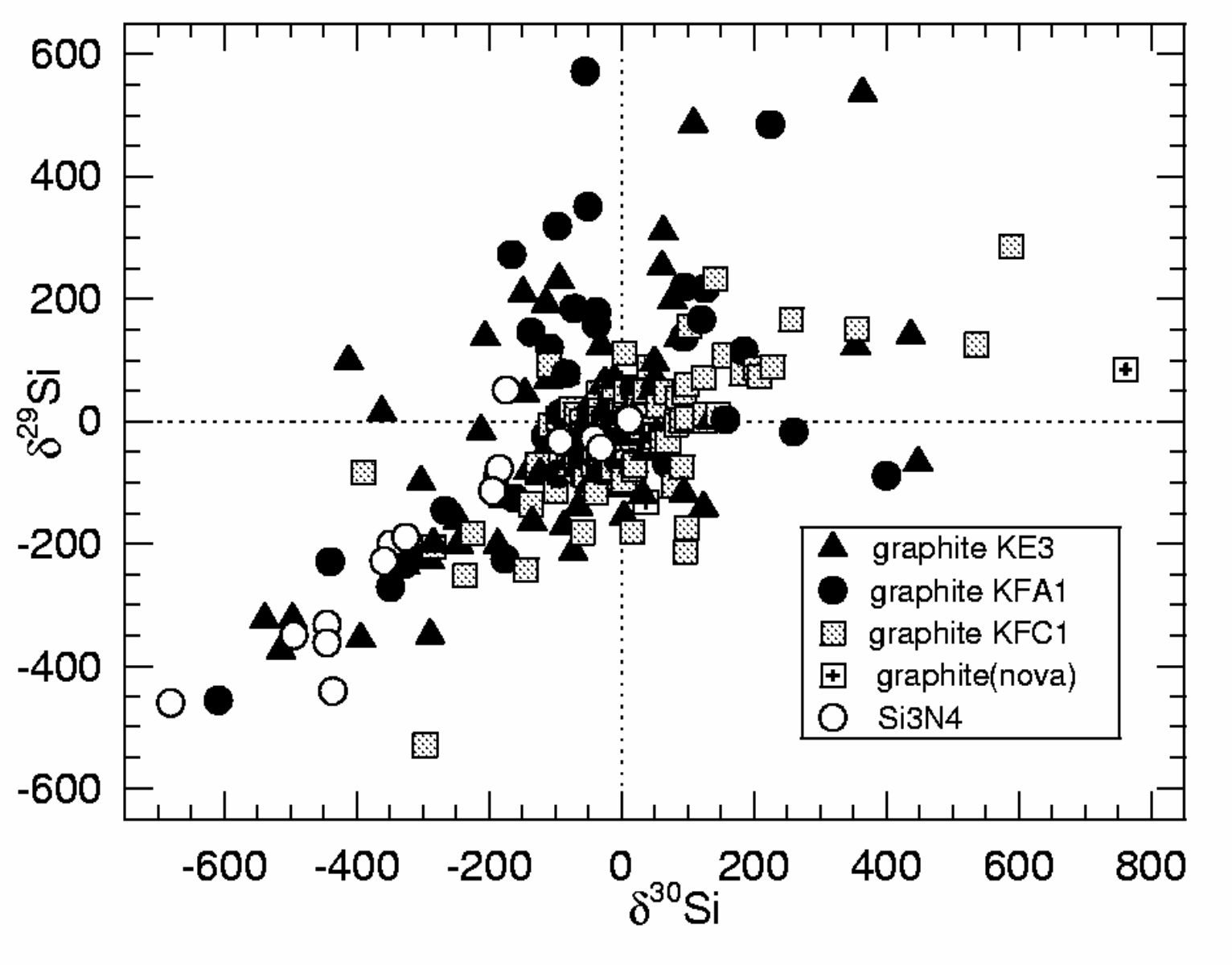

**Fig.12.** Silicon isotopes in presolar graphite grains from the low-density fractions define arrays similar to those given by presolar SiC (Fig. 7). Note that the measurements may have relatively large uncertainties, see e.g., Hoppe et al. (1995).

The KE3 grains, and a few of the KFA1 grains are believed to have formed in supernovae, whereas the high-density KFC1 grains most likely formed in low-metallicity (i.e., less metal-rich than the sun) AGB stars. In addition, novae can account for isotopic features of a few grains.

There are no equivalent designations for the different graphite populations as for the mainstream, A+B, X, Y, and Z populations of SiC. It is still unknown how many presolar graphite populations exist and how many types of stars are required to account for the presolar graphite. The C-isotope ratios alone are not diagnostic enough to distinguish the possible stellar sources, and the isotopic compositions of other elements are needed to identify the graphite grains' parent stars. However, with the exception of the KE3 grains, the low trace element contents make isotopic analyses challenging.

### 5.4. Sub-grains in graphite and SiC

Nanometer size sub-grains hidden in presolar graphite and SiC grains made their debut during TEM studies by Bernatowicz et al. (1991, 1992). A SiC grain about 5.3 µm in size of clear presolar origin ($^{12}C/^{13}C = 51.6\pm0.4$) contained several sub-grains ranging in size from 10-70 nm. Diffraction patterns and energy-dispersive

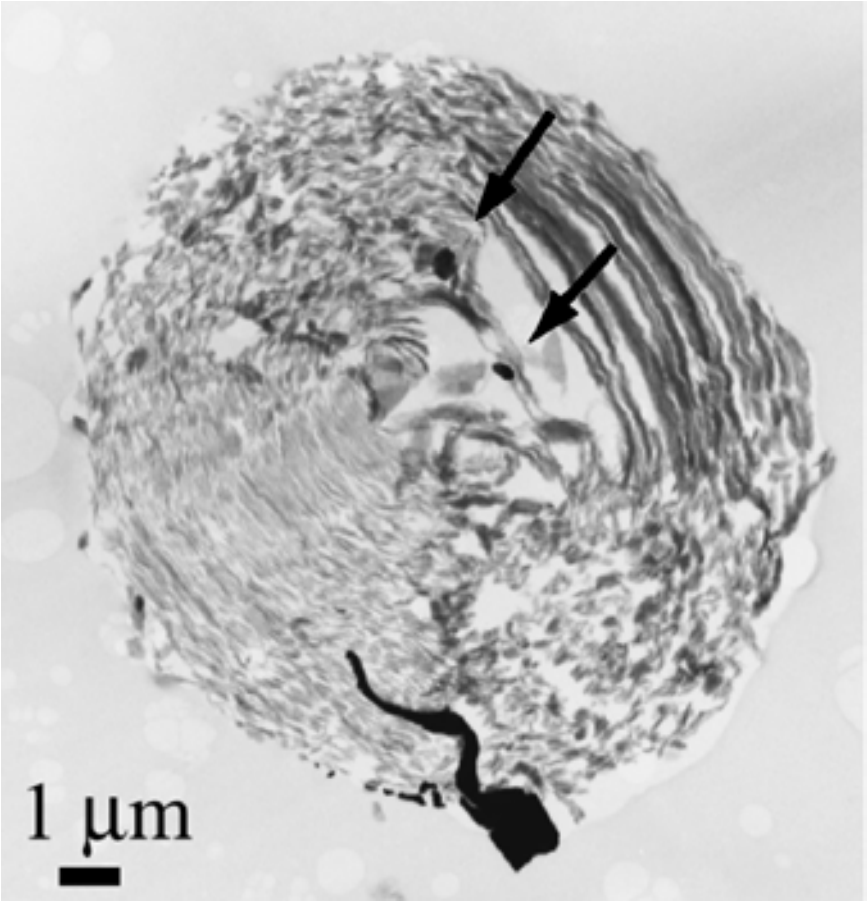

**Fig. 13**. A TEM image of two TiC sub-grains in a slice of a presolar KE3 graphite grain. Photo courtesy of K. Croat and T. Bernatowicz.

X-ray spectrometry confirmed that they are TiC. An epitaxial relationship between the TiC and the SiC host suggests that SiC and TiC either grew simultaneously, or that TiC exsolved from SiC (Bernatowicz et al. 1992).

More extensive studies have been performed on inclusions in graphite grains from the lowest density separate KE3 (Croat et al. 2003) and the highest density separate KFC1 (Bernatowicz et al. 1996). The tiny sub-grains in graphite are shown in Fig. 13. The sub-grains in KFC1 graphites are smaller (5-200 nm) than those in KE3 graphites, which contain ~10 to ~400 nm-size TiC grains.

The variable V/Ti ratios of 0.07–0.2 in sub-grains indicate an independent origin from the host graphites. In contrast to sub-grains in SiC, those in graphite show no crystallographic (epitaxial) relationship with the host graphite, suggesting that they formed prior to graphite and were randomly incorporated into graphite grains later. About 30% of the TiC grains have partially amorphous rims (3 to 15 nm thick) that could be the result of atom bombardment when the grains were adrift before they were embedded into the growing graphite grains (Croat et al. 2003).

Many graphite grains from the high-density KFC1 fraction contain sub-grains consisting of refractory carbides with compositions ranging from nearly pure TiC to nearly pure Zr-Mo carbides (Bernatowicz et al. 1996; Croat et al. 2004). In several graphite grains, a carbide particle is located in the center of the grains, indicating that the carbide served as a nucleation site (see Figure 13 and the Figure 7(a) in Bernatowicz et al. 1996). However, there are also graphite grains with sub-grains of TiC, iron-nickel metal, and cohenite, which contain no Mo-Zr carbides. This points to different stellar sources, possibly supernovae, for these graphite grains (Croat et al. 2003).

### 5.5. Presolar oxide grains: Corundum, spinel, hibonite

Oxide grains are resistant to chemicals used to isolate carbonaceous presolar grains, and they are also concentrated in SiC-rich residues. The problem of identifying presolar oxides is that the vast majority of oxides in meteoritic residues is isotopically normal because of its solar system origin. For example, in the Tieschitz (H3.6) chondrite presolar $Al_2O_3$ is estimated to be 0.03 ppm (Nittler et al. 1997), and in CM2 chondrites less than 1% of all corundum grains are expected to be presolar (Zinner et al. 2003), so the ion imaging technique is valuable in locating presolar oxides among the normal ones.

The first presolar oxides were found during ion probe studies on individual grains (Huss et al. 1994, Hutcheon et al. 1994). Most presolar oxides were found by ion imaging and subsequently analyzed by conventional ion probe in high-mass-resolution mode. The oxides are mainly corundum and spinel (Huss et al. 1994, Hutcheon et al. 1994, Nittler et al. 1994, 1997, 1998, Nittler and Alexander 1999, Choi et al. 1998, 1999, Krestina et al. 2002), a few hibonite grains (Choi et al. 1999, Krestina et al. 2002), and one $TiO_2$ grain (Nittler and Alexander 1999).

Corundum was the dominant presolar oxide known until Zinner et al. (2003) analyzed individual grains from the spinel-rich separates CF (mean diameter 0.15 μm) and CG (mean diameter 0.45 μm) of the Murray meteorite with the NanoSIMS. Presolar spinel is more abundant among oxide grains smaller than 0.5μm (see Table 2 by Zinner et al. 2003). Nguyen et al. (2003) obtained O- isotopic images of tightly packed CF and CG grains and found about 284 presolar oxide grains by ion imaging, capitalizing on the high spatial resolution from the small beam diameter of the NanoSIMS. While isotopic anomalies in very small grains tend to be diluted by normal, surrounding grains, ion imaging still remains an efficient method to quickly locate anomalous grains.

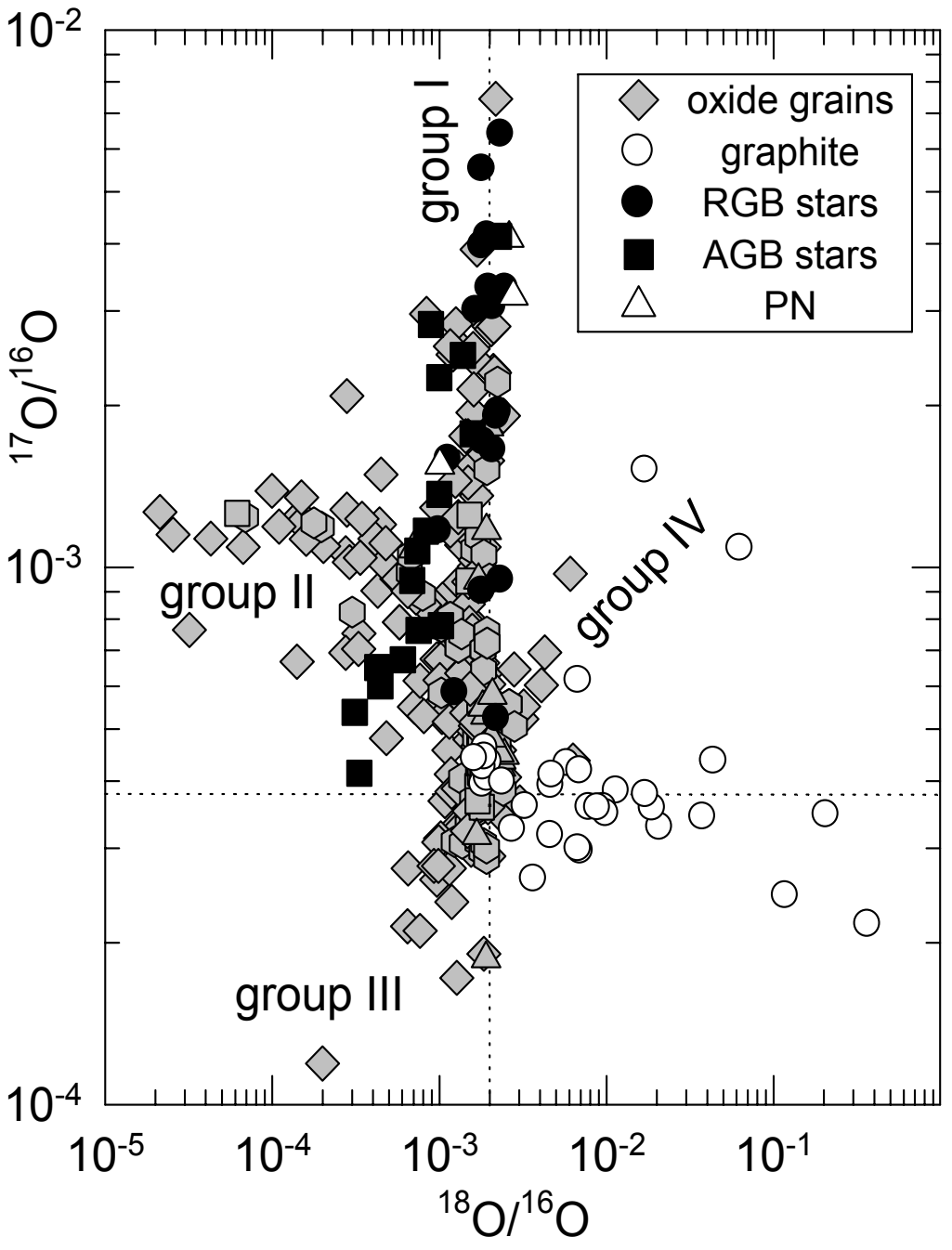

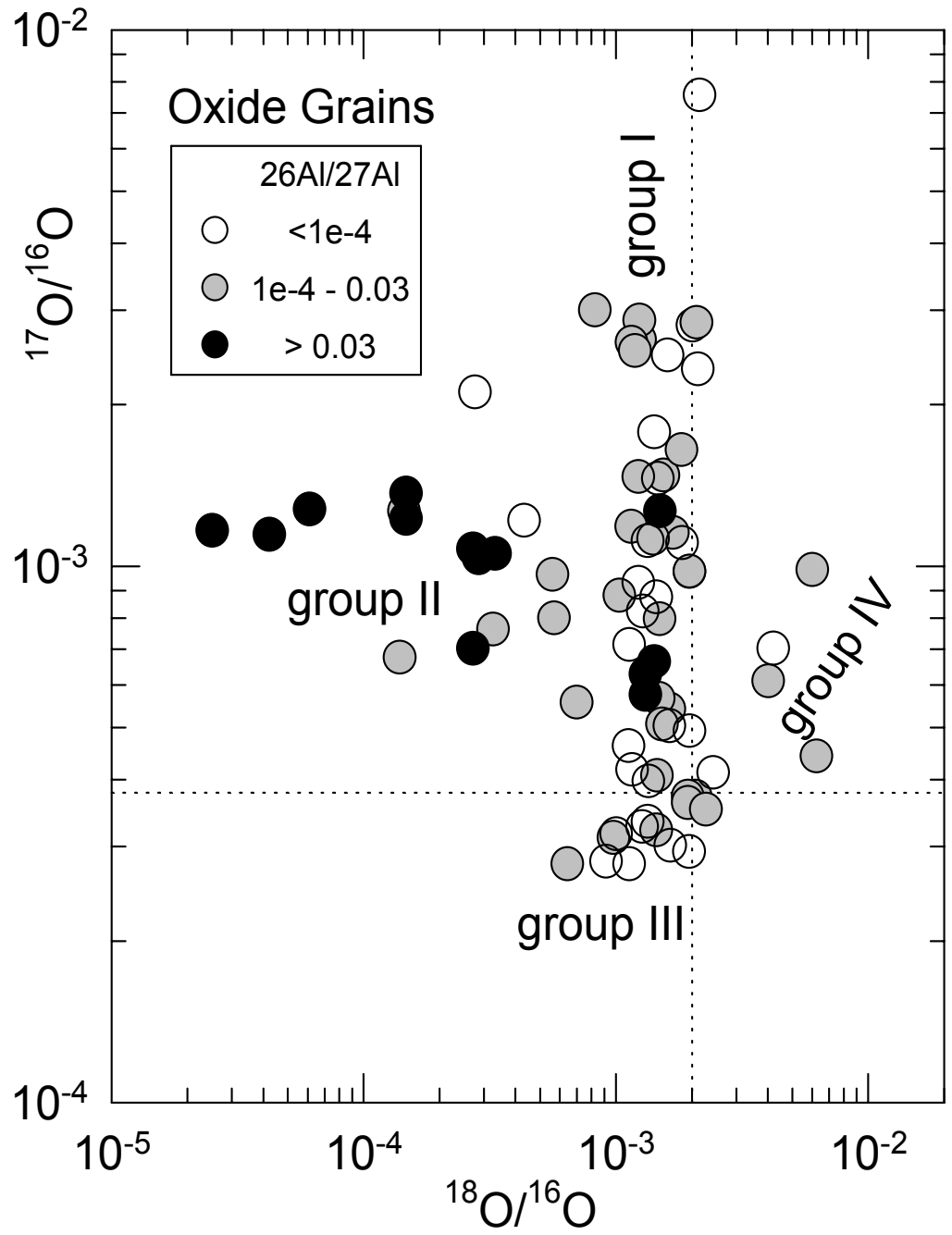

**Fig. 14.** The O-isotopes in presolar oxide and silicate grains define 4 groups (see text for references). Also shown are data for graphite grains from the low-density fractions KE3 (Travaglio et al. 1999) and KFA1 (Amari, unpublished); and planetary nebulae (PN), RGB, and AGB stars (Harris & Lambert 1984, Harris et al. 1985, 1987, 1988, Kahane et al. 1992, Lambert et al. 1986, Smith & Lambert 1990, Wannier and Sahai 1987). Dotted lines indicate solar ratios.

**Fig. 15.** Same as Fig. 14, but symbol colors indicate the $^{26}Al/^{27}Al$ content of oxide grains. Black symbols show grains with $^{26}Al/^{27}Al > 0.03$, grey with $0.03 < ^{26}Al/^{27}Al < 10^{-4}$, and white with $^{26}Al/^{27}Al < 10^{-4}$ and grains for which only upper limits have been determined.

Oxygen isotopic compositions are used to classify the oxide grains into four groups (Table 7, Fig. 14). Grains in group I exhibit $^{17}O$ excesses and/or modest $^{18}O$ deficiencies relative to SMOW (Standard Mean Ocean Water, generally taken to be representative of the solar O-isotopic composition). Grains of group II are characterized by large $^{18}O$ deficiencies ($^{18}O/^{16}O$ < 0.5×solar) and $^{17}O$ excesses (up to 3.5×solar). Grains with moderate depletions in $^{17}O$ and $^{18}O$ (equivalent to small $^{16}O$ excesses) are in group III, and those with $^{17}O$ and $^{18}O$ excesses are in group IV. Corundum and spinel are present in all four groups. The few hibonite grains known fall into groups I to III. The $^{26}Al/^{27}Al$ ratios (Fig. 15) inferred from $^{26}Mg$ excesses measured in many oxide grains are useful to further classify the oxide grains (Hutcheon et al. 1994, Nittler et al. 1994, 1997). Many grains in group II of the O-isotopic classification also show the highest $^{26}Al/^{27}Al$ ratios.

**Table 7**. Isotopic characteristics of presolar oxides

| Designation | Group I | Group II | Group III | Group IV |
|---|---|---|---|---|
| Range in $^{17}O/^{16}O$ | (0.45–2.9)×$10^{-3}$ | (0.55–1.4)×$10^{-3}$ | (1.9–4.15)×$10^{-4}$ | (5.2–98)×$10^{-4}$ |
| Range in $^{18}O/^{16}O$ | (0.89–2.2)×$10^{-3}$ | ≤ 7.1×$10^{-4}$ | (0.65–1.9)×$10^{-3}$ | (3.1–6.1)×$10^{-3}$ |
| Mean initial $^{26}Al/^{27}Al$ | 0.0023 | 0.0060 | 0.0004 | 0.0021 |

*Sources*: Nittler et al. (1997), Nguyen et al. (2003), Zinner et al. (2003)

### 5.6. Presolar silicates

Not many presolar silicate grains are known because of two major difficulties. Silicates are more susceptible to metamorphism and chemical processing than presolar carbonaceous and oxide grains. For example, CI chondrites, with the highest level of carbonaceous presolar grains, have experienced aqueous alteration on their parent bodies which transformed silicates into hydrous silicates. This process erases any presolar isotopic signatures in the silicates so meteorites from parent bodies that never experienced (much) aqueous alteration are best to search for presolar silicates. The second difficulty is to locate presolar silicates among boundlessly abundant solar system silicates that are major constituents of meteorites. The ion imaging technique using the CAMECA-3f was applied to search for presolar silicates in the Murchison CM-chondrite (Nittler, private communication) but the spatial resolution and sensitivity of the instrument were not high enough to identify sub-micron presolar silicate grains.

The first identification of six presolar silicates of 0.3–0.9 µm in diameter was in IDPs by O-isotope ion imaging with the NanoSIMS (Messenger et al. 2003). One of them is forsterite and two are amorphous silicates called GEMS (glass with embedded metal and sulfides). Using the same technique, Nguyen and Zinner (2004) identified 9 presolar silicate grains ranging between 0.2 and 0.6µm in the size-separated disaggregated matrix from the Acfer 094 chondrite, which has experienced minimal thermal or aqueous processing. The grains are tentatively identified (because of interference from underlying or surrounding grains) as pyroxene, olivine, and Al-rich silicate. Also employing advanced techniques for high spatial resolution, Nagashima et al. (2004) identified one olivine and five mineralogically-uncharacterized presolar silicate grains in thin sections of the Acfer 094 and NWA530 meteorites. The available O-isotope data on presolar silicates from IDPs and meteorites are similar to those of presolar oxides (Fig. 14).

## 5.7. Presolar diamond

Diamond was the first presolar mineral identified in meteorites (Lewis et al. 1987), and it has the highest relative abundance among carbonaceous presolar grains. Still, it remains the least understood, mainly because the diamond grains are only 1-3 Ångstroms in size (Lewis et al. 1987, Daulton et al. 1996), which makes them too small for individual analysis. The presolar diamond separates obtained by chemical acid treatment are not 100% pure carbon and are of lower density than normal crystalline diamond (3.51 g/cm$^3$). The diamonds contain N and O, probably in chemical functional groups on their surfaces as inferred from infrared spectra (Lewis et al. 1989, Mutschke et al. 1995, Andersen et al. 1998, Jones et al. 2004). Diamonds from ordinary-, enstatite-, and CV-chondrites contain N ranging from 2200 to 42100 ppm (by mass) and those from CI- and CM-chondrites have 7500 to 9200 ppm (Russell et al.1991). In addition, diamond separates contain noble gases, H enriched in deuterium up to 250-340‰ (Carey et al. 1987, Lewis et al. 1989), and some trace elements (Lewis et al. 1991b).

The $^{12}C/^{13}C$ ratio of 92-93 of bulk diamond in primitive meteorites (Lewis et al. 1987, 1989, Russell et al. 1991, 1996) is surprisingly close to the solar ratio of 89, considering the wide range in C-isotopic compositions of presolar SiC and graphite grains. The $\delta^{15}N$ in diamonds ranges from –330‰ to –350‰ (Lewis et al. 1987, Russell et al. 1996). This range is below the values observed in various types of meteorites (~—100 to +200‰) but is close to the upper limit of <–240‰ for solar wind implanted into the lunar regolith, and to the N-isotopic composition of Jupiter ($\delta^{15}N$ = -374‰, Owen et al. 2001), which is probably more representative for the solar system than that of the terrestrial atmosphere, which is influenced by atmospheric escape processes.

The similarity of C- and N-isotopes of presolar diamonds and the solar system supports the idea that a large diamond fraction originated within the solar system and is not presolar. However, the complexity of the carbon release patterns and C/N ratios observed during stepwise heating by Russell et al. (1996) suggests the presence of more than one type of diamond. It is also possible that another phase remained in the diamond separate after chemical and physical isolation.

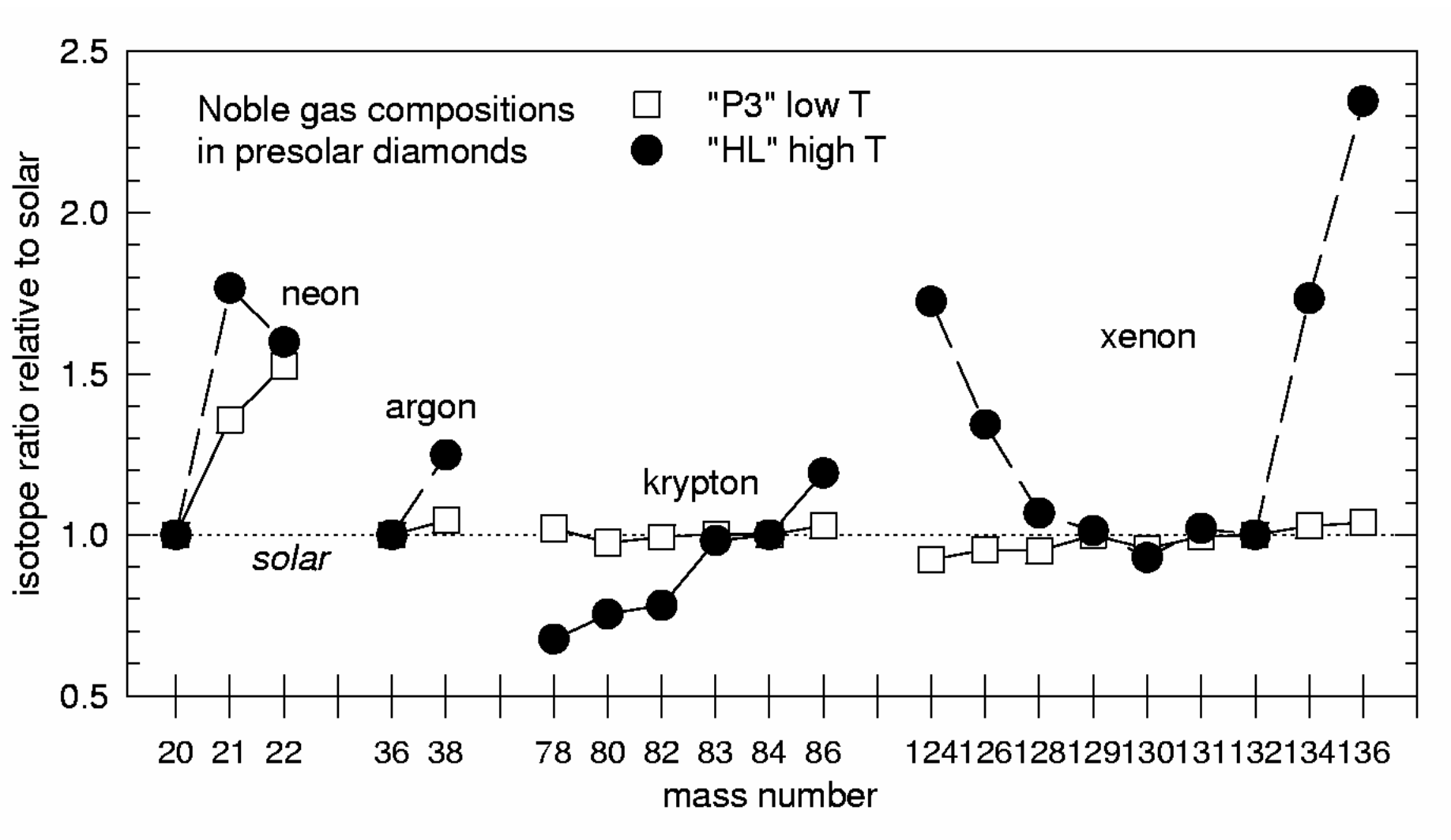

**Fig. 16.** Noble gas isotopic composition in presolar diamonds. Data are from Huss and Lewis 1994a (see also Ott 2002) normalized to the solar isotopic composition from Wieler (2002). The isotope ratios are further normalized to $^{20}$Ne, $^{36}$Ar, $^{84}$Kr, and $^{132}$Xe, respectively. The dotted line indicates solar composition.

Detailed studies show that at least two clearly resolved noble gas components, called “P3” and “HL”, reside in meteoritic nanodiamonds (Fig. 16). In addition, a component called “P6”, less clearly resolved, is released at the highest temperatures from diamond separates (Tang et al. 1988, Tang and Anders 1988, Huss and Lewis 1994b, Ott 2002). The first component, “P3” is released around 500°C. Except for Ne, the isotopic ratios in P3 are close to solar, without any major fractionations among individual isotopes of a given element (Fig. 16). The second component, HL, released around 1300°C, shows stronger fractionations relative to solar. Its most prominent feature is the relative enrichment in the light and heavy Xe-isotopes, the Xe-HL. The Kr isotopic ratios increase with increasing mass number. The $^{38}$Ar/$^{36}$Ar is higher than solar, and the Ne isotopic composition is probably a mixture of “true” Ne-HL and Ne from the third component “P6”, which is not well characterized (Huss and Lewis 1994b, Ott 2002). In contrast to Ne in the P3 component, the solar-normalized $^{21}Ne/^{20}Ne$ ratio in the high temperature fraction is larger than the $^{22}Ne/^{20}Ne$ ratio (Fig. 16).

The low temperature component P3 is mainly seen in diamond separates from CI- and CM-chondrites and is only marginally present in CV-, ordinary-, and enstatite chondrites (Huss and Lewis 1994b, Ott 2002). This suggests preferred removal of the P3 component during thermal metamorphism, and its presence or absence is used to estimate the degree of parent body metamorphism (Huss and Lewis 1994a, 1995).

The two major noble gas components suggest that at least two different populations of diamond exist, but the reason why one diamond population is less thermally stable than the other is not entirely clear. However, the N concentrations in the diamonds of CI- and CM-chondrites are also higher than in diamond from CV-, ordinary-, and enstatite chondrites. If the diamonds carrying the low temperature P3 component have more N as impurity in solid solution, their crystal structure is plausibly less stable against thermal destruction.

The tracer of presolar diamond, Xe-HL, indicates a supernova origin of diamonds, and isotope anomalies in other heavy elements, i.e., Te and Pd (Richter et al. 1998, Maas et al. 2001) also point to supernova nucleosynthesis. On the other hand, the C- and N-isotopic signatures indicate that not all presolar diamonds originate from supernovae and that the supernova contribution to the diamonds is probably not very large. A huge isotopic anomaly in Xe-HL within a small grain fraction can easily dominate the overall observed Xe-isotopic composition but not the C- and N-isotopic composition. Carbon (and likely N) is present in all diamonds grains, and the C and N signatures from a small fraction of supernova grains could easily be masked by those from more abundant grains. In contrast, the noble gases in the presolar diamond fraction could be dominated by a small population of gas-rich diamond.

There is no shortage of suggested origins of presolar diamonds, which remains enigmatic. These include an origin in supernovae (Clayton 1989, Clayton et al. 1995), within the solar system itself (Dai et al. 2002), novae (Clayton et al. 1995), C-rich giant stars (Clayton 1975, Lewis et al. 1987, Andersen et al. 1998), Wolf-Rayet stars (Tielens 1990, Arnould et al. 1997), and binary star scenarios, where mass from a carbon-rich giant star is transferred onto a white dwarf, which then explodes as a type Ia supernova (Jørgensen 1988).

# 6. Giant stars and their grains

The majority of known presolar grains apparently came from giant stars, which are steady dust contributors to the ISM. We describe some of the evolution of giant stars because their nucleosynthetic products condense into grains in the stars' circumstellar shells. One should keep in mind that the *broad* theoretical picture of nucleosynthesis and evolution of giant stars is consistent with astronomical observations and measurements of presolar grains, but that many details are not yet completely understood.

### 6.1. Chemical and isotopic markers of giant star evolution

Observationally the different stages of stellar evolution are tracked in the Hertzsprung-Russell (HR) diagram (Fig. 17). It shows the difference in the flux of star light between certain optical wavelengths (the "color difference"), versus absolute magnitude. The color difference is a measure of stellar effective (photospheric) temperatures. The absolute magnitude is the energy flux at a given wavelength normalized to a standard distance to take distance-related dimming of starlight into account, and it provides a measure for stellar brightness or luminosity. In the HR diagram, cool stars plot on the right side, and the brightest stars plot at the top. The diagonal band of stars is the "main sequence" and consists of dwarf stars (including the Sun) with increasing mass from bottom to top. The branches off the main sequence are occupied by giant stars.

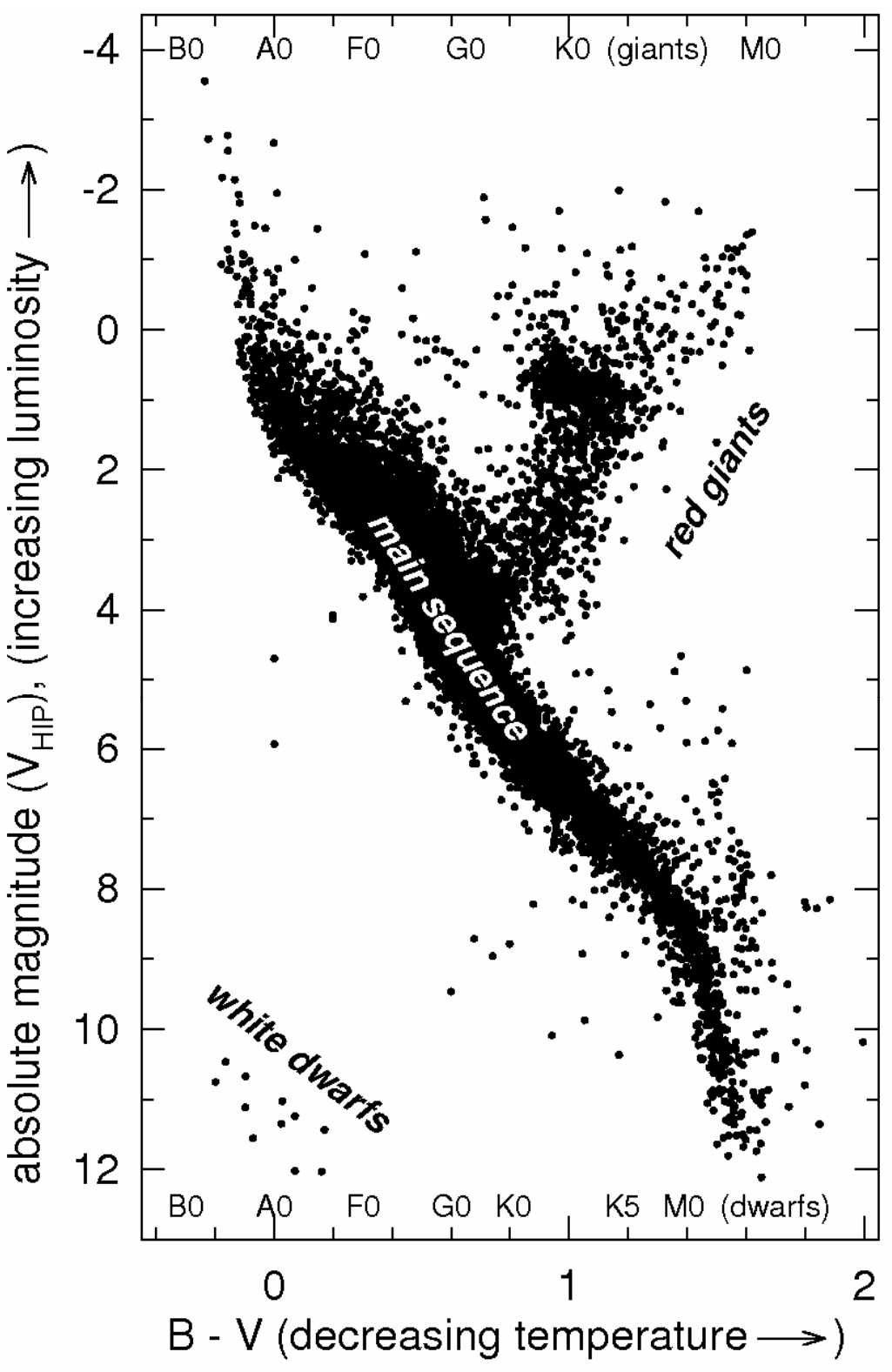

**Fig. 17.** The Hertzsprung-Russell diagram separates dwarf stars on the main-sequence from stars on the giant star branches. Data are for 11760 stars within 100 parsec with distance measurement uncertainties <5% (HIPPARCHOS catalogue).

Red giants are in late stages of stellar evolution, and have evolved from dwarfs with main sequence masses of ~1 to ~8 $M_\odot$ (the "$\odot$" refers to solar units, e.g., 1 $M_\odot$ = 1 solar mass). The evolution of low- to intermediate mass stars is sketched in Fig. 18, and more details are described by Iben and Renzini (1983), Iben (1991), Busso et al. (1995), Arnett (1996), Gallino et al. (1997), Wallerstein et al. (1997), Lattanzio and Forestini (1999), and Pagel (1997).

Stars spend most of their life on the main-sequence where they burn H to He in their cores. This hydrostatic burning lasts ~10 Ga in a 1 $M_\odot$ star but is shorter in more massive stars. After exhaustion of H in the core, the remaining He-rich core (the "He-core") contracts, and rising temperatures ignite H-burning in a thin shell around the He-core. The star's radius expands to sizes of 10-50 $R_\odot$ (e.g., Jura 1999) and the star becomes more luminous as it moves onto the "red giant branch" (RGB).

Hydrogen-burning proceeds through the proton-proton (pp)-chain, and, more importantly in higher mass stars ($\gtrsim$1.1 $M_\odot$) with higher core temperatures, through the carbon-nitrogen-(CN) or carbon-nitrogen-oxygen-(CNO) cycle, where isotopes of C,N, and O act as catalysts to convert four protons into He. In this process, the sum of C, N, and O nuclei remains about constant but it increases the N abundance and decreases the C and O abundances relative to the main-sequence composition (usually taken as solar if not stated otherwise). Steady-state (or "equilibrium") values of the CNO-cycle are $^{12}C/^{13}C$ ~3.5, and, depending on temperature, $^{14}N/^{15}N$ ~30,000 (at low T) to $^{14}N/^{15}N$ <0.1 (at high T).

**Table 8.** Some properties of giant stars of < 9 $M_\odot$ and the Sun

| | Sun | M | M | S | C | J |
|---|---|---|---|---|---|---|
| Status | main-sequence | red giant branch, RGB | asymptotic giant branch, AGB | | | possibly RGB? |
| C/O | 0.5 | $\leq$ 0.5 | ~0.5–0.7 | 0.6–1 | 1.2 | 1.1 |
| *s*-process elements [a] | solar | solar | > solar | > solar | > solar | ~solar |
| $^{12}C/^{13}C$ | 89 | 6 - 20 | 10-30 | 50-70 | 30-80 | 3 – < 10 |
| $^{14}N/^{15}N$ | 272 | ? | ? | ? | >500 (4–12)×$10^3$ | ~70, ~150 |
| $^{17}O/^{16}O$ | 3.78×$10^{-4}$ [b] | (0.91–6.25) ×$10^{-3}$ | (0.9–6.3) ×$10^{-3}$ | (0.33–1) ×$10^{-3}$ | (0.18–2.4) ×$10^{-3}$ | (0.24–1.4) ×$10^{-3}$ |
| $^{18}O/^{16}O$ | 2.01×$10^{-3}$ [b] | 2.0×$10^{-3}$ | (0.21–1) ×$10^{-3}$ | (0.2–1) ×$10^{-3}$ | (0.42–1.4) ×$10^{-3}$ | 6.35×$10^{-4}$ |
| Isotopic changes [c] | - | +$^{13}$C, +$^{14}$N, +$^{17}$O | | +$^{12}$C, + $^{22}$Ne | | +$^{13}$C, +$^{14}$N |

*Major sources*: Dominy et al. (1986), Harris and Lambert (1984), Harris et al. (1985, 1987, 1988), Lambert et al. (1986), Olson and Richter (1979), Querci and Querci (1970), Smith and Lambert (1990), Wannier et al. (1991)

[a] Relative to solar composition.

[b] Assuming values of standard mean ocean water are representative of the solar isotopic composition.

[c] Isotopic changes from solar that may be observable in presolar grains.

During the RGB stage, lasting about 500Ma for a solar mass star, the outer envelope becomes convective and penetrates into the region where partial H-burning took place. Thereby, the unprocessed envelope material becomes polluted with the byproducts of the CNO-cycle (the "**first dredge-up**"). Observed $^{12}C/^{13}C$ ratios in RGB stars range from 6 to 20 (Table 8) - significantly reduced from the (solar) main-sequence ratio of 89 - and indicate dredge-up of CNO-processed material. The N-isotopes are difficult to measure in stars.

Measured $^{17}O/^{16}O$ ratios in RGB stars are larger than solar, and $^{18}O/^{16}O$ ratios are below or near the solar ratio (Table 8).

Compared to standard evolution models, stars less massive than ~2.3 $M_\odot$ show lower $^{12}C/^{13}C$ and Na and Al abundances than expected from the first dredge-up of CNO-processed material (e.g., Lambert 1981, Gilroy and Brown 1991, Sneden 1991). In order to bring nucleosynthesis models into accord with stellar observations, an extra mixing process was postulated (Charbonnel 1994, 1995). This is known as **cool bottom processing**, "CBP" (Wasserburg et al. 1995, Boothroyd and Sackmann 1999). In CBP, the base of the inert convective envelope cycles through the regions that are on top of the H-burning shell so that CN-processing of envelope material can occur. This leads to $^{12}C/^{13}C$ = 3.5 and high $^{14}N/^{15}N$ ratios in the observable stellar envelope. CBP is also assumed to operate at later stages of stellar evolution (Nollett et al. 2003).

The RGB phase terminates when the He-rich core, left behind after main-sequence burning, ignites. This violent explosion is known as "He-core flash" for stars <2.5 $M_\odot$. The He-core burning may occur concurrently with H-shell burning in stars of >2.5 $M_\odot$. The interior structure turns to a He-burning core, an overlying H-burning shell, and a convective envelope. At this stage, lasting ~50 Ma in a 1 $M_\odot$ star, the luminosity drops and the effective temperature slightly increases so that the star "moves" from the RGB back to near the main sequence in the HR diagram. Helium-burning in the core creates $^{12}C$ by fusion of three $^{4}He$ nuclei, known as the 'triple-α' reaction: $^{4}He(^{4}He,\gamma)^{8}Be(^{4}He,\gamma)^{12}C$. In the most massive giant stars (~8-9 $M_\odot$) higher core temperatures may lead to α-capture on $^{12}C$ to produce $^{16}O$ (i.e., $^{12}C(^{4}He,\gamma)^{16}O$). This creates a C- and O-rich 'CO core'.

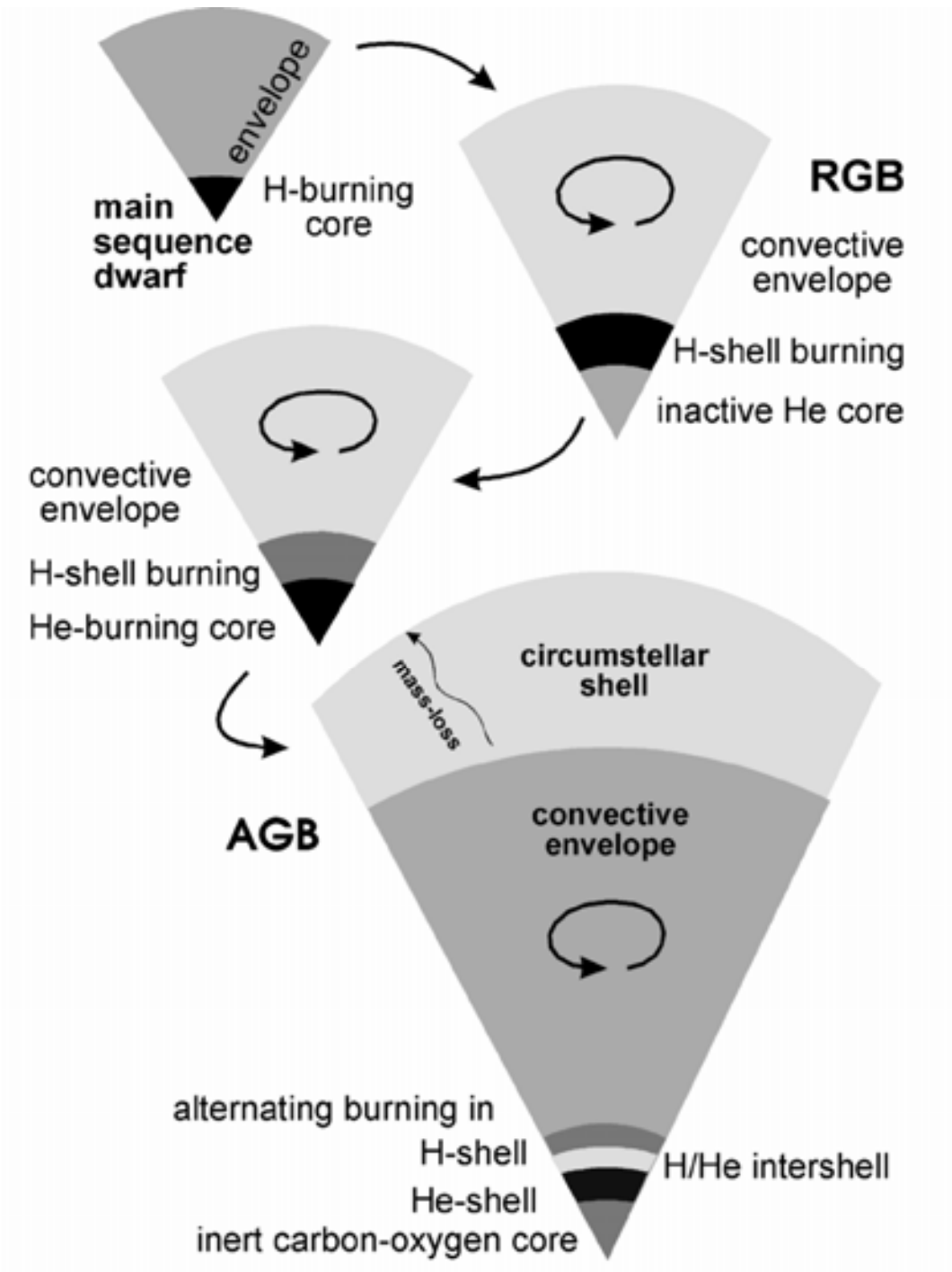

**Fig. 18.** A cartoon of the different interior zones in giant stars. The diagrams *are not* to scale and the zones where nucleosynthesis takes place are much smaller.

After He is exhausted in the core, the star approaches the asymptotic giant branch (AGB).

This name comes from the location of these stars in the HR diagram, where AGB stars plot asymptotically to the red giant branch. Hydrogen-shell burning stops, and He-shell burning sets in. In larger mass stars ( >4-5 $M_{\odot}$) the convective envelope can penetrate deep enough to transport more CNO-processed material (rich in $^{4}$He, $^{13}$C, $^{14}$N, $^{17}$O) previously made in the H-burning shell to the stellar surface (the '**second dredge-up'**).

On the AGB, burning alternates between a thin H- and a He-shell surrounding the inert core. The two shells are separated by an intershell region. If the He-burning shell is ignited, temperatures raise and overlying zones expand. Expansion of the He-shell moves the intershell and the H-burning shell into cooler regions. Then H-burning is extinguished, the temperature-dependent energy production in the expanding He-shell drops. At some point, He-burning ceases, and subsequent contraction leads to ignition of a new H-burning shell. The successive re-ignitions of He-shell burning occur in intervals of ~$10^4$ a and are known as thermal pulses (TP).

Thermally pulsing 'TP-AGB' stars consist of an inert C-O-rich core, surrounded by a He-burning shell, a He-intershell layer, a H-burning shell, and the convective envelope (Fig. 18). When H-shell burning is inactive, convection transports the carbon made by partial triple α-burning in the He-shell to the stellar surface (the '**third dredge-up'**). This increases the observable C/O and $^{12}$C/$^{13}$C ratios. If a star experienced several thermal pulses and third-dredge-up episodes it will show relatively larger amounts of He-burning products at its observable atmosphere.

The third dredge-up also brings up products made during H-shell burning, including $^{23}$Na (the only stable Na isotope) and radioactive $^{26}$Al. These are made in the NeNa- and MgAl-cycles that are linked to the CNO-cycle (Forestini et al. 1991, Gallino et al. 1994, Wasserburg et al. 1994, Guélin et al. 1995). The observation of $^{26}$Al in giant stars would provide another strong link to presolar grains because many different grains show evidence of incorporated $^{26}$Al. However, searches for $^{26}$Al in giant stars (using the isotopic shifts from $^{26}$Al and $^{27}$Al in spectral bands of AlH, AlCl, or AlF, e.g., Branch and Perry 1970) give no firm results. Guelin et al. (1995) obtained $^{26}$Al/$^{27}$Al $\leq$0.04 for the closest, well-studied C-star CW Leo (IRC+10°216), but more measurements are required to check for $^{26}$Al in other giant stars.

Cool bottom processing may operate in low mass stars on the AGB during H-shell burning. This process is invoked to explain depletions in $^{18}$O/$^{16}$O (relative to solar) in some presolar grains (e.g., Wasserburg et al. 1995, Nittler et al. 1997), and it may also account for the production of $^{26}$Al (e.g., Nollett et al. 2003).

The TP-AGB stars show nucleosynthesis products made by neutron capture reactions, which are of particular interest here. The neutrons necessary for this are produced by α- capture on $^{13}$C by the reaction $^{13}$C(α,n)$^{16}$O. During a third dredge-up episode, the H-rich envelope is in contact with the $^{12}$C- and He-rich intershell. There $^{12}C$ becomes abundant from the previous convective thermal pulse as fresh $^{12}C$ is produced by partial triple alpha burning at the bottom of the thermal pulse and then spread into the He intershell. A small amount of protons likely diffuses from the H-shell into the intershell, and with protons and $^{12}C$, the $^{13}C$ is made via by the reaction $^{12}C(p,\gamma)^{13}N(\beta^+\nu)^{13}C$ forming a "$^{13}C$-pocket" at the top of the intershell.

In intermediate-mass stars (>5 $M_\odot$) the reaction $^{22}Ne(\alpha,n)^{25}Mg$ becomes a more important source of neutrons during later thermal pulses, and the neutron density strongly depends on the maximum bottom temperature in the convective shell. The $^{22}Ne$ for this neutron source comes from the reaction $^{14}N(\alpha,\gamma)^{18}F(\beta^+\nu)^{18}O(\alpha,\gamma)^{22}Ne$, which can occur in both low- and intermediate-mass stars. This reaction is the source of the Ne-E(H), which was one of the beacons during the search for presolar grains. Lewis et al. (1990, 1994) argue that Ne-E(H) in SiC is unlikely to originate from decay of $^{22}Na$ (which only comes from novae or supernovae), because $^{22}Ne$ correlates with $^{4}He$, which is abundant in the He-rich shell.

Neutron capture by elements around the iron peak builds up the abundances of heavy elements (e.g., Iben and Renzini 1983, Gallino et al. 1990, 1997, 1998, Busso et al. 1999). In AGB stars, neutron capture proceeds on a slow time scale (relative to the time scales of beta-decay of the radioactive nuclides produced) and is called the "*s*-process". Elements with relatively abundant stable isotopes made by the *s*-process include Sr, Ba, Zr, Y, and the light rare earth elements, as well as several isotopes of Kr and Xe. AGB stars that experienced many third dredge-up episodes should show higher C/O and higher *s*-process element abundances than stars like the sun or RGB stars of solar metallicity.

Another mechanism proposed to operate in stars of 4–7 $M_\odot$ on the AGB is **hot bottom burning (HBB)** or envelope burning (Becker and Iben 1980, Renzini and Voli 1981, Boothroyd et al. 1995, Lattanzio and Forestini 1999). In such more massive stars, the base of the convective envelope is hotter and CNO-processing can occur in the portion of the envelope located above the H-burning shell. This model predicts a decrease of $^{12}C/^{13}C$ to ~3.5 and increases of $^{14}N/^{15}N$ up to 30000. The freshly produced $^{12}C$ from triple α-burning in the He-shell is thus converted to $^{13}C$ and $^{14}N$. However, the C to N conversion may prevent an increase in the C/O ratio and the AGB star cannot become a C-star, or depending on timing, HBB could change a C-star back to an O-rich AGB star.

### 6.2. Types of cool giant stars

The chemistry of cool giant stars divides them into at least three major categories, with transitional types in-between. Starting with solar-like elemental composition and moving to more carbon and *s*-process element-rich objects, the giant star spectral sequence is M-MS-S-SC-C. The **M-stars,** which comprise two-thirds of giants, are either RGB stars (showing only products from the CNO-cycle) or AGB stars whose envelopes have just begun to become polluted by the third dredge-up. The spectra of M-giants are dominated by metal oxide bands such as VO and TiO, which shows that these stars are relatively cool (~3000 K) near their surfaces. The **S-stars** are clearly on the AGB because they have C/O ratios near unity and over-abundances of *s*-process elements relative to solar. In addition to TiO and VO bands, S-star spectra show strong ZrO bands because Zr is increased by the *s*-process. The discovery of a pure *s*-process product, radioactive $^{99}$Tc ($t_{1/2}$= 2.1×10$^5$ a) by Merrill (1952) proves that the *s*-process indeed operates in these relatively low mass stars. Note that some S-stars are enriched in s-process elements but not in Tc (Little et al. 1987). Jorissen et al. (1993) call them "extrinsic S-stars" and use "intrinsic S-stars" for objects with Tc. The extrinsic S-stars probably accreted the s-process elements from a more massive companion star that already evolved through the AGB stage and is now a white dwarf. Extrinsic S-stars may not be on the AGB.

Carbon stars atmospheres have C/O> 1 from multiple third dredge-up episodes and they show bands from abundant carbon-bearing molecules such as $C_2$, CH, and CN as well as atomic metal lines. Of the several sub-types of C-stars (N, J, R, and CH stars) only the **N-type carbon stars** are thought to be true TP-AGB stars. The evolutionary status of the other types of C-stars is not well known (Wallerstein and Knapp 1998).

Convective mixing brings the products of nucleosynthesis from the stellar interior to the stellar surface, but these products need to get into dust grains and become part of the interstellar matter. This is accomplished by stellar winds that drive mass-loss from AGB star envelopes. The accumulation of gas and condensates from the cooling gas creates **circumstellar shells** (Lafon and Berruyer 1991, Lamers and Cassinelli 1999). Sometimes "circumstellar shells" are referred to as "circumstellar envelopes" but here we use "circumstellar shells" to avoid confusion with the stellar envelope that directly surrounds the interior nuclear burning zones. The shells can extend to several hundred stellar radii, with stellar radii themselves ranging from 100 to 500 $R_\odot$. They are divided into an inner shell in which thermochemical (e.g., condensation) processes are more important, and an outer shell where UV-driven photochemical processes dominate (Glassgold 1996).

Circumstellar shells are detected by excesses in infrared radiation as light from the central star is absorbed and re-emitted by dust at longer red and infrared wavelengths (e.g., Olnon et al. 1986, Gezari et al. 1993, Kwok et al. 1997). In some instances, enshrouded stars are only detected by the infrared radiation from their thick surrounding shells (e.g., Guglielmo et al. 1998, Volk et al. 2000). Typical mass-loss rates are ~10$^{-4}$ to 10$^{-7}$ $M_\odot$ per year for M- and C-type AGB stars, and about a factor of ten less for S-stars (e.g., Loup et al. 1995, Olofsson et al. 1993). In contrast to the extensive mass-loss experienced by TP-AGB stars, giants on the RGB do not show large infrared excesses, and their mass-loss is not very productive (e.g., Jura 1999). Typical mass-loss rates of RGB stars are ~$10^{-9}$ to$10^{-8}$ $M_{\odot}$ per year (e.g., Mauron and Guilain 1995). At some point, mass-loss has removed an AGB star's envelope, and after several one to ten million years of alternating H- and He-shell burning in the inside and continuous mass-loss on the outside, the AGB stage of stellar evolution comes to an end.

During the post-AGB stage (van Winckel 2003), mass-loss rates drop because most of the envelope is already gone. The circumstellar shell continues to expand with terminal velocities of 10-30 km/s. As it separates more from the central star, the veil of dust and molecular absorptions around the hot stellar remnant is lifted. The last envelope matter leaves the star in a fast wind of 1000-4000 km/s and once this much faster wind collides with the older, slower expanding material a strong shock front is created. The shock interaction of the ejecta and the radiation from the hot central star creates a **planetary nebula,** which continues to expand with a final velocity of ~40 km/s (Kwok et al. 1978, Kwok 1994). One may speculate if the dust in the older shell is destroyed when it is hit by the shock wave. However, dust is still observed in planetary nebulae.

The remains of a giant star are a white dwarf consisting of the C-O-rich core, and a planetary nebula created from material of the former stellar envelope. Of an initially 1 $M_{\odot}$ star, about half remains, whereas stars of initially 4-8 $M_{\odot}$ loose around 80% of their main sequence mass to the ISM (e.g., Weidemann 2000). The local mass return from M-giants is estimated 1-2×$10^{-4}$ $M_{\odot}$/year/kpc$^2$ and C-giants provide about the same amount. The total mass returned to the galactic ISM is ~0.3-0.6 $M_{\odot}$/year (e.g., Wallerstein and Knapp 1998).

### 6.3. Dust condensation in circumstellar shells

Dust grains condense in circumstellar shells around giant stars and then enter the ISM. Temperatures in the expanding circumstellar shell drop by adiabatic cooling, and, at a far enough distance from the star, become low enough (<2000 K) for grain condensation. An important property of giant stars related to grain formation is **stellar variability**. Many AGB stars are "pulsating variable stars" and show more or less regular periodic variations in brightness (e.g., Hoffmeister et al. 1985). The variations are caused by thermal ionization and de-ionization processes in the stellar envelope relatively close to the photosphere (e.g., Bowen 1988). These processes regulate the opacity and energy transfer through the envelope, and lead to radius expansions (dimmer star) and contractions (brighter star). The variability cycles last 100 to 600 d, with Mira-type variables showing the most regular periods of 400-500 d (e.g., Hoffmeister et al. 1985). (Note that stellar variability is not related to the thermal pulses at the TP-AGB stage, where cyclic ignitions of He-shell burning occur at a frequency of about $10^4$ a.)

Hence the dynamic nature of the envelope responsible for giant star variability also affects the radial temperature and pressure structure within the circumstellar shell. Because condensation is a function of temperature and total pressure, stellar variability influences the absolute distance from the central star at which grain formation occurs. Fig. 19 shows how condensation may proceed during the variability cycle (Lodders and Fegley 1997).

During maximum light, the star has the smallest radius and the highest photospheric temperature, and high temperatures within a few stellar radii do not favor condensation. When the star changes phase, temperatures drop in the expanding circumstellar shell. Dust formation should be most efficient at the low temperatures during minimum brightness. Because of the large radius expansion, the absolute grain-forming location is also further away from the center of the star (but still within a few radii of the expanded star). After passing through the minimum phase, temperatures increase, the stellar radius decreases, and gas and dust from the circumstellar shell can "fall back" onto the star, which may lead to dust evaporation. However, this can be prevented if dust grains are accelerated in the circumstellar shell by radiation pressure from the star and leave into the ISM.

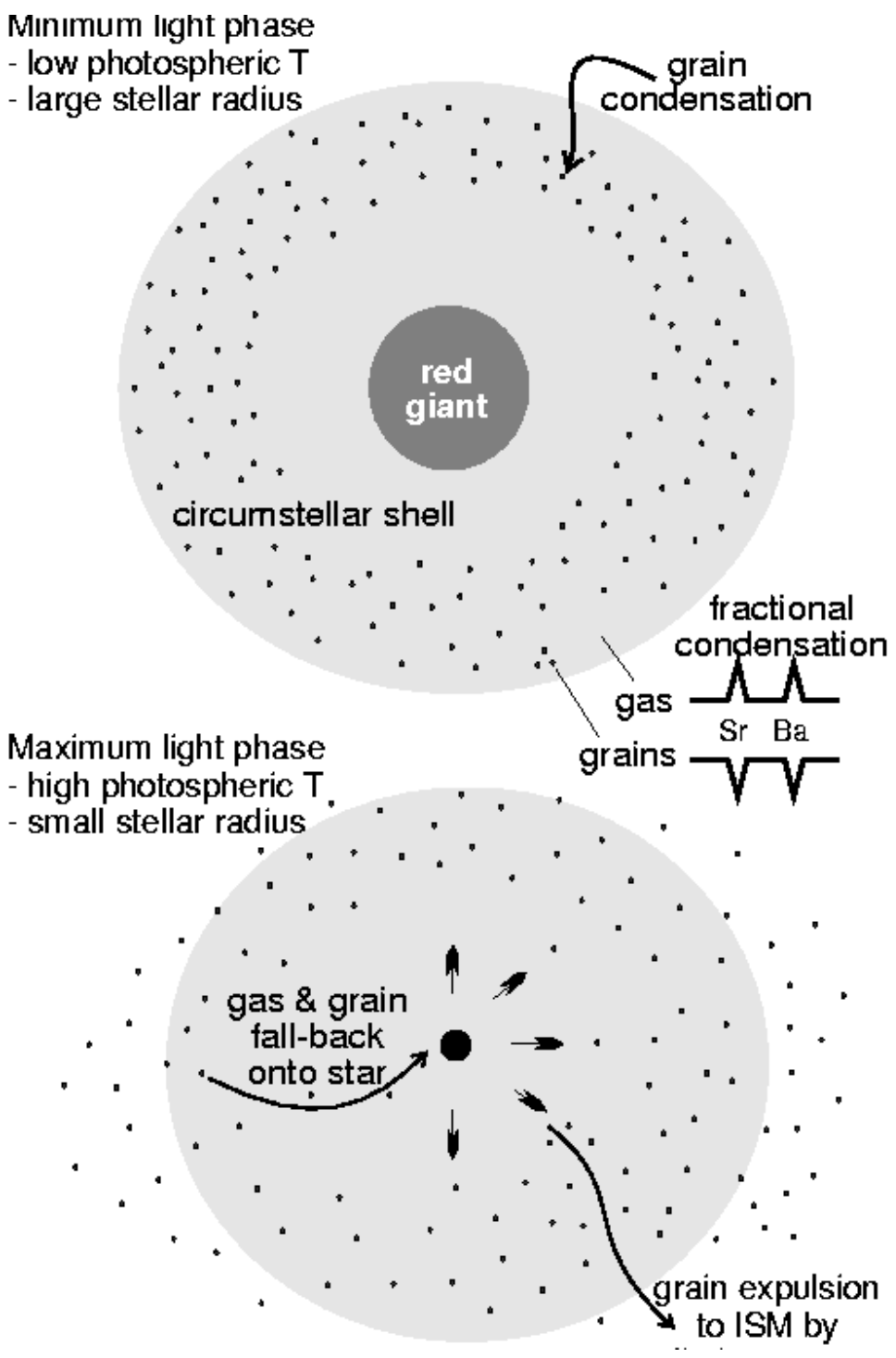

**Fig. 19.** A sketch of grain condensation in circumstellar shells of giant stars during the variability cycle.

Considering the observed variability periods (e.g., <600 days for minimum to minimum), grain formation must occur on time scales of at least half the variability period (between minimum and maximum).The nature of the dust condensing from ejected envelope material determined by the C/O ratio in the gas, which ranges from ~ 0.5 (solar) to ~1.2 in giant stars. Changes in other element abundances introduced by nucleosynthesis (e.g., higher N abundances in RGB stars, and higher *s*-process element abundances in AGB stars) should not affect the oxidation state. Under "oxidizing" conditions (C/O < 1) major element oxides and silicates condense (e.g., Lord 1965, Larimer 1969, Grossman 1972, Ebel and Grossman 2000, Lodders 2003), and carbonaceous dust forms under "reducing" conditions (C/O> 1) from a gas of otherwise solar elemental composition (Friedemann and Schmidt 1967, Friedemann 1969, Gilman 1969, Larimer 1975, Lattimer et al. 1978, Larimer and Bartholomay 1979, Lodders and Fegley 1995, 1997).

Table 9 lists condensates of the more abundant elements in circumstellar shells of M-and C-stars expected from thermochemistry, the minerals seen in circumstellar shells and planetary nebulae, and the known presolar minerals. Several minerals are both known in circumstellar environments and as presolar grains and are the first condensates to appear under oxidizing and reducing conditions, respectively. The **condensation temperatures** as a function of total pressure are shown in Figs. 20 and 21. The range in total pressures shown covers that expected in circumstellar shells, e.g., $10^{-5}$ to $10^{-8}$ bar at $T<2000$ K. With the notable exception of carbon, the condensation temperatures generally increase with increasing total pressure.

**Condensates for M-stars** are similar as for a solar composition gas because their C/O ratios result in comparable oxidation states (e.g. Dominy et al. 1986, Smith and Lambert 1985, 1986, 1990, Smith et al. 1987). Condensates of Ca and Al include corundum, spinel, hibonite, gehlenite, and anorthite. Silicon and Mg condense as forsterite and enstatite, and Fe forms an FeNi-alloy and troilite. Several of these condensates are present in circumstellar shells and in planetary nebulae by infrared and mid-infrared observations. Abundant silicate grains are expected in O-rich shells, and the 9.7 μm emission in M-giant spectra observed by Gillett et al. (1968) was ascribed to silicates by Woolf and Ney (1969). This silicate feature is observed in shells of many O-rich stars (e.g., Little-Marenin and Little 1988, 1990, Sloan et al. 1996). The 20 and 28 μm emissions identify Mg-rich pyroxene and forsteritic olivine (e.g., Molster et al. 2002a-c, Suh 2002). Circumstellar silicates seem to be both crystalline and amorphous, with thicker and cooler shells favoring crystalline silicates.

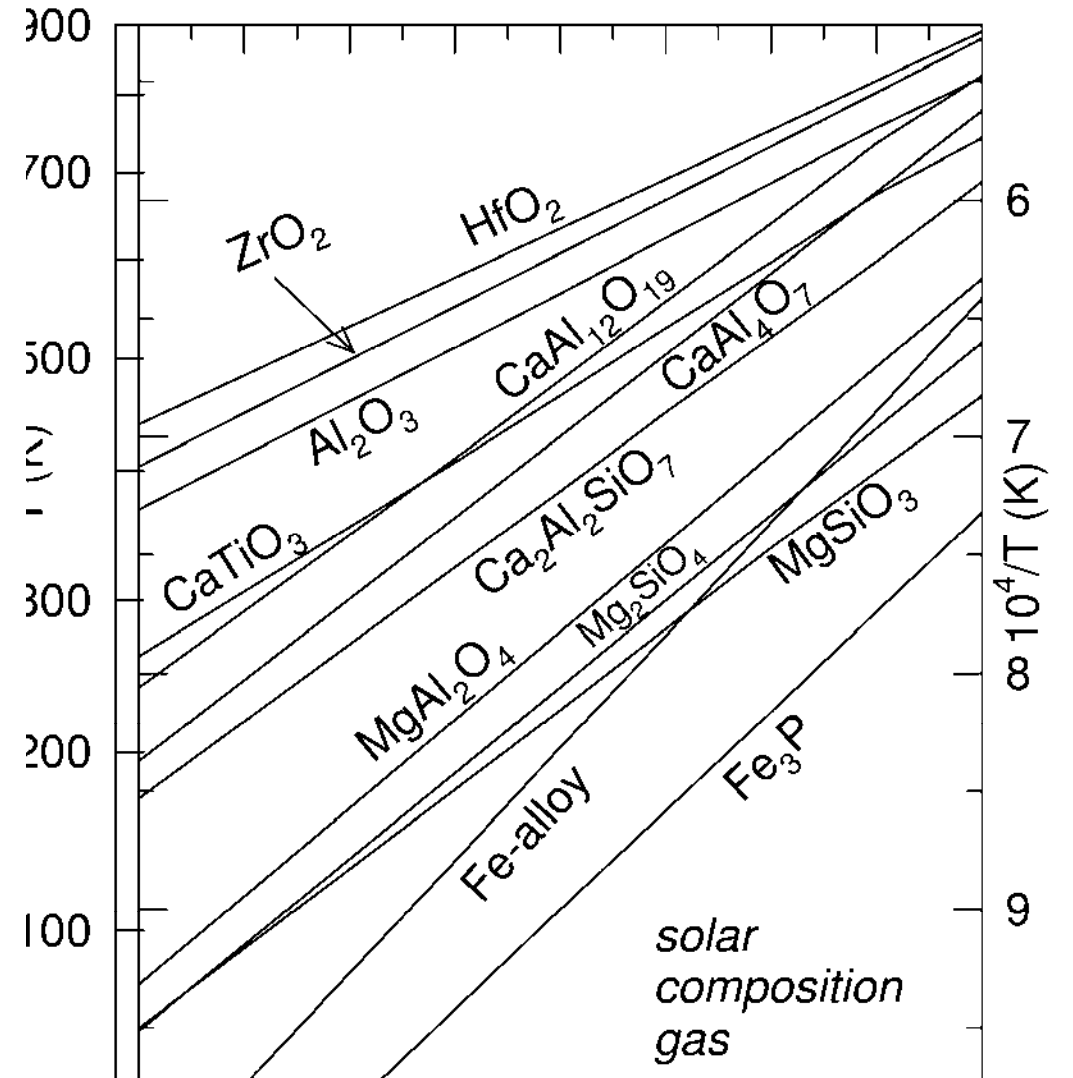

**Fig. 20.** Condensation temperatures as a function of total pressure for a gas with C/O = 0.5.

Schmid-Burgk and Scholz (1981) already discussed $Al_2O_3$ formation in M-giants and an emission feature at 13 μm, present in 40-50% of all O-rich AGB stars (Sloan et al. 1996), was identified as corundum (Onaka et al. 1989, Begemann et al. 1997, Kozasa and Sogawa 1997, Fabian et al. 2001). Corundum may have an additional emission at 21 μm (Begemann et al. (1997), which is seen in several post-AGB objects. However, in these stars the 21 μm feature is not accompanied by one at 13 μm, and the origin of the 21 μm emission remains mysterious. Another possible source for the 13 μm feature is magnesian spinel, especially if it is accompanied by 16.8 and 32 μm emissions (e.g., Posch et al. 1999, Fabian et al. 2001). However, the identification of spinel is uncertain because silicate emissions may interfere and other interpretations were suggested (e.g., Sloan et al. 2003).

Metallic iron, an expected major dust component, is infrared inactive. Kemper et al. (2002) infer its presence in the dusty circumstellar shell of an evolved O-rich giant because an abundant infrared opacity source is needed to fit the spectrum. Troilite, a low temperature condensate, is not yet identified in M-giants. Perovskite is also not yet detected circumstellar shells. Posch et al. (2003) found that the infrared signatures of perovskite, other Ca-titanates, and Ti-oxides coincide with those from silicates and aluminum oxides. More abundant Si and Al condensates likely reduce the spectral band strengths of Ti-bearing condensates so that Ti-bearing dust may not be resolvable.

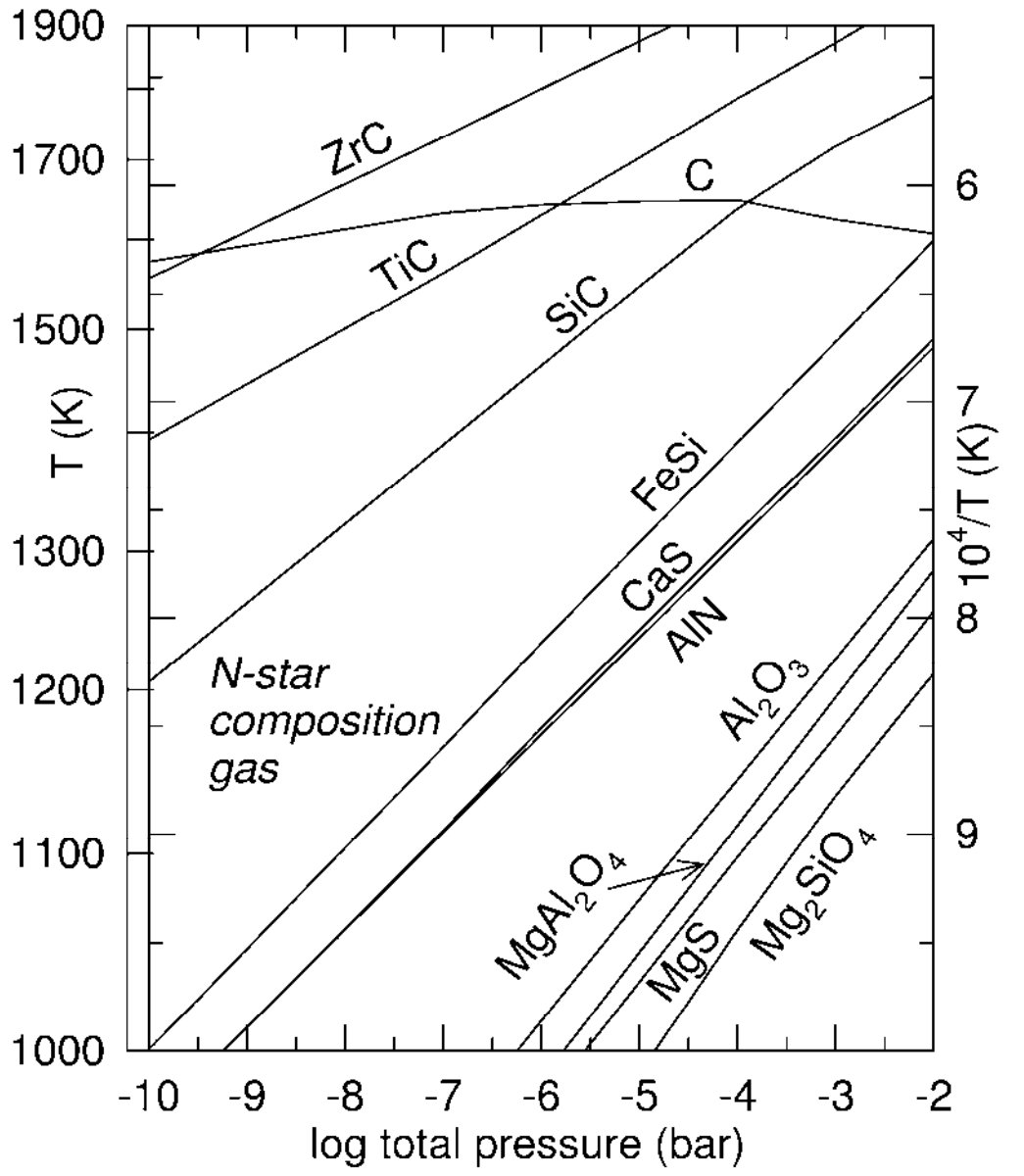

**Fig. 21.** Condensation temperatures as a function of total pressure at C/O = 1.1.

**Condensates for reducing conditions** (Fig. 21) are calculated for the average composition of N-stars by Lambert et al. (1986). Condensates are carbon and carbides (SiC, TiC), sulfides (CaS, MgS), and nitrides (AlN, TiN). Iron condenses as silicide (FeSi) and troilite. Silicates also condense at C/O >1, despite a persistent misconception among some researchers that silicates do not form under reducing conditions. (The reduced enstatite chondrites with mainly pure enstatite show that Mg-silicates formed at C/O >1, e.g., Krot et al. 2000).

The 11.3 μm emission of SiC is prominent in circumstellar shells of C-stars (e.g., Gilra 1971, Treffers and Cohen 1974, Forrest et al. 1975, Little-Marenin 1986, Little-Marenin et al. 1987). Carbon, either amorphous and/or in graphitic form, is believed to be present around many C-stars (Rowan-Robinson and Harris 1983, Jura 1986, Martin and Rogers 1987, Orofino et al. 1987, Blanco et al. 1994). Diamond is not an expected condensate under equilibrium thermodynamic conditions. However, emissions of surface-hydrogenated diamond at 3.43 and 3.53 μm (Guillois et al. 1999) are seen in the C-rich shell of HR4049, a very metal-poor post-AGB star (Geballe et al. 1989, van Kerckhoven et al. 2002), and in the proto-planetary nebula CRL 2688 (Geballe et al. 1992).

The 30 μm emission observed in C-rich AGB stars and planetary nebulae is attributed to MgS (Goebel and Moseley 1985, Nuth et al. 1985, Hony et al. 2002a, Hony and Bouwman 2004). Another expected abundant condensate, FeS, was identified in planetary nebulae (Hony et al. 2002b, Begemann et al. 1994).

The phase responsible for the 21 μm emission from several post-AGB stars and planetary nebulae mentioned before has been a puzzle ever since it was first described (Kwok et al. 1989). Among other suggested phases (e.g., spinel, nanodiamonds, $SiS_2$, polyaromatic hydrocarbons), this feature was ascribed to TiC nano-particles (von Helden et al. 2000), but this assignment is debated (e.g., Li 2003). Although TiC is one of the first condensates from a reducing gas, the low Ti abundance (relative to Si or C) may prevent detection of TiC in circumstellar shells. In addition, presolar graphite grains contain TiC as inclusions, which implies that detection of TiC in circumstellar shells is limited because TiC is hidden within other grains.

The total pressure and temperature conditions under which presolar grains formed can be estimated from thermochemical calculations. Many transition element carbides (e.g., TiC, ZrC, MoC) are quite refractory (e.g., Larimer 1975, Lattimer et al. 1978, Lodders and Fegley 1995, 1997) and tiny inclusions of such carbides within graphite grains apparently acted as nucleation seeds for the surrounding carbon (Bernatowicz et al. 1991, 1996, Croat et al. 2003). The presence of refractory carbide inclusions in graphite and the absence of SiC inclusion corresponds to the condensation sequence transition metal carbides-C-SiC. A detailed study using equilibrium and non-equilibrium chemistry by Bernatowicz et al. (1996) shows that favorable conditions for the observed condensation sequence of metal carbides and carbon in the stellar outflows require total pressures $>10^{-7}$ bar and C/O > 1.05. Chigai et al. (1999) considered non-equilibrium conditions and inferred C/O ratios less than 1.26 to 1.46 and total pressures between $10^{-7}$ to $2\times10^{-9}$ bars.

**Table 9.** Expected and observed major element condensates for giant stars and presolar grains [a]

| | | M-stars | | | C-stars | | |
|---|---|---|---|---|---|---|---|
| | Abundance [b] | Mineral and ideal formula | Stellar shell | Presolar Grains | Mineral and ideal formula | Stellar shell | Presola rGrains |
| O | $1.41\times10^7$ | oxides and silicates | | | silicates | | |
| C | $7.08\times10^6$ | — | | | titanium carbide TiC | √ | √ |
| | | | | | graphite C | √ | √ |
| | | | | | silicon carbide SiC | √ | √ |
| | | | | | diamond? | √(?) | √ |
| N | $1.95\times10^6$ | — | | | osbornite TiN | | |
| | | | | | aluminum nitride AlN | | |
| Mg | $1.02\times10^6$ | spinel $MgAl_2O_4$ | √ | √ | niningerite MgS | √ | |
| | | forsterite $Mg_2 SiO_4$ | √ | √ | spinel $MgAl_2O_4$ | √ | √ |
| | | enstatite $MgSiO_3$ | √ | √ | forsterite $Mg_2SiO_4$ | √ | √ |
| | | | | | enstatite $MgSiO_3$ | √ | √ |
| Si | $1.00\times10^6$ | gehlenite $Ca_2Al_2SiO_7$ | | | silicon carbide SiC | √ | √ |
| | | forsterite $Mg_2SiO_4$ | √ | √ | iron silicide FeSi | | |
| | | enstatite $MgSiO_3$ | √ | √ | forsterite $Mg_2SiO_4$ | √ | √ |
| | | | | | enstatite $MgSiO_3$ | √ | √ |
| Fe | $8.38\times10^5$ | iron alloy FeNi | | | iron silicide FeSi | | |
| | | schreibersite $(Fe,Ni)_3P$ | | | iron alloy FeNi | | |
| S | $4.45\times10^5$ | troilite FeS | | | oldhamite CaS | | |
| | | | | | niningerite MgS | √ | |
| | | | | | troilite FeS | √ | |
| Al | $8.41\times10^4$ | corundum $Al_2O_3$ | √ | √ | aluminum nitride AlN | | |
| | | hibonite $CaAl_{12}O_{19}$ | | √ | corundum $Al_2O_3$ | √ | √ |
| | | grossite $CaAl_4O_7$ | | | spinel $MgAl_2O_4$ | √ | √ |
| | | gehlenite $Ca_2Al_2SiO_7$ | | | anorthite $CaAl_2Si_2O_8$ | | |
| | | spinel $MgAl_2O_4$ | √ | √ | | | |
| | | anorthite $CaAl_2Si_2O_8$ | | | | | |
| Ca | $6.29\times10^4$ | hibonite $CaAl_{12}O_{19}$ | | √ | oldhamite CaS | | |
| | | grossite $CaAl_4O_7$ | | | anorthite $CaAl_2Si_2O_8$ | | |
| | | gehlenite $Ca_2Al_2SiO_7$ | | | | | |
| | | anorthite $CaAl_2Si_2O_8$ | | | | | |
| Na | $5.75\times10^4$ | albite $NaAlSi_3O_8$ | | | albite $NaAlSi_3O_8$ | | |
| | | | | | halite NaCl (?) | | |
| Ni | $4.78\times10^4$ | kamacite & taenite | √ | | kamacite & taenite | | √ |
| | | schreibersite $(Fe,Ni)_3P$ | | | schreibersite $(Fe,Ni)_3P$ | | |
| Cr | $1.29\times10^4$ | Cr in FeNi alloy | | | daubréelite $FeCr_2S_4$ | | |
| Mn | $9.17\times10^3$ | $Mn_2SiO_4$ in olivine | | | alabandite (Mn,Fe)S | | |
| P | $8.37\times10^3$ | schreibersite $(Fe,Ni)_3P$ | | | schreibersite $(Fe,Ni)_3P$ | | |
| Cl | $5.24\times10^3$ | sodalite $Na_4[AlSiO_4]_3Cl$ | | | halite NaCl (?) | | |
| K | $3.69\times10^3$ | orthoclase $KAlSi_3O_8$ | | | orthoclase $KAlSi_3O_8$ | | |
| | | | | | sylvite KCl (?) | | |
| Ti | $2.42\times10^3$ | perovskite $CaTiO_3$ | | | titanium carbide TiC | √ (?) | √ |
| | | | | | osbornite TiN | | |

[a] Adapted from Lodders and Fegley (1999). Elements are in order of decreasing solar abundance. Condensates are in order of appearance with decreasing temperature for each element. A "?" means that entry is uncertain. [b] solar photospheric abundances where Si = $1\times10^6$ atoms (Lodders 2003).

Overall, thermodynamic equilibrium calculations are very useful for modeling the chemical composition of dust from stars. We also refer the reader to the review by Sedlmayr and Krüger (1997) for non-equilibrium and kinetic considerations.

## 6.4. Synthesis: Presolar grains from RGB and AGB stars

The known presolar minerals from giant stars are as follows. Corundum, spinel, and silicate grains of the oxide groups I and III come from RGB stars. AGB stars contributed corundum, spinel, and hibonite grains to the oxide groups I, II and III; mainstream, Y, and Z SiC grains; probably graphite grains with $^{12}C/^{13}C$ ratios larger than solar, which comprise more than 70% of the high-density KFC1 graphite fraction; and possibly some diamonds.

### 6.4.1. Grains from red giant branch (RGB) stars

Dust from RGB stars should consist of oxides and silicates. After the first dredge-up, diagnostic changes in isotopic compositions relative to solar are expected in C, N, and O from processing by the CNO-cycle. Oxidized condensates should not contain much C and N (possibly as impurity, if any), so only the O-isotopes are useful to sort out the presolar corundum, spinel, hibonite, and silicate grains from RGB stars. After the first (and second) dredge-up, $^{17}O/^{16}O$ should increase over the solar ratio, but the $^{18}O/^{16}O$ ratio should remain near the solar ratio in M stars on the RGB and early AGB (Table 8). This indicates that grains from RGB stars are to be found in group I and III in Fig. 14. However, oxide grains from RGB and early AGB stars should not have evidence of $^{26}Al$. The oxide grains measured for Al-Mg systematics (Fig. 15) show that $^{26}Al$ is absent in 64% of the grains from group III and in 30% of the grains from group I. Apparently group III is dominated by oxides from RGB stars.

### 6.4.2. Presolar grains from asymptotic giant branch (AGB) stars

A star on the AGB produces either oxidized or reduced condensates, depending upon how many third dredge-up episodes have occurred to change the C/O ratio from the near-solar RGB ratio to ratios above unity. Major changes on the AGB with respect to the RGB stage are increases in C and *s*-process element abundances and rising $^{12}C/^{13}C$ ratios from the RGB values below ~20 (e.g., Smith and Lambert 1985, 1986). Nitrogen-rich in $^{14}N$ is mainly inherited from the RGB stage. Another difference between TP-AGB and RGB (and early AGB) stars is that $^{26}Al$ made during H-shell burning is brought to the surface during the third dredge-up episodes. Neutron capture reactions in AGB stars also modify the Ca and Ti isotopes, and increases in the isotopes $^{47}Ti$, $^{49}Ti$, and $^{50}Ti$ relative to $^{48}Ti$ and $^{42}Ca$, $^{43}Ca$, $^{44}Ca$, and $^{46}Ca$ relative to $^{40}Ca$ are expected (e.g., Hoppe et al. 1994, Choi et al. 1998, Amari et al. 2000a) but presolar oxides from RGB stars should contain the unaltered initial Ca and Ti isotopes of the stellar source.

**Dust from O-rich** AGB stars includes the expected condensates corundum and spinel, however, the presolar oxide grains of groups I and III plotted in the O-isotope diagram (Figs. 14, 15) are a mix of grains from RGB and AGB stars. The inferred presence of $^{26}$Al shows that group I has more grains from AGB stars (70%) than group III (36%). However, these percentages do not reflect the relative distribution of RGB and ABG stars contributing to presolar oxides because one AGB star produces much more dust than one RGB star.

Measured O-isotope ratios in RGB and AGB stars and planetary nebulae are compared to oxide grain data in Fig. 14. RGB stars have about normal $^{18}O/^{16}O$ and higher $^{17}O/^{16}O$ and plot together with group I grains. Data for AGB stars only include stars that show Tc, which ensures that they are truly on the AGB. Several AGB stars and PN plot with group I grains; however, there is a notable shift to lower $^{18}O/^{16}O$ for AGB stars (but not PN) towards group II grains.

The origin of oxide grains in groups II and III (Figs. 14, 15) was also ascribed to AGB stars (Nittler et al. 1997). Group II grains are slightly enriched in $^{17}$O and heavily depleted in $^{18}$O relative to solar, and many of these grains have the highest inferred $^{26}$Al content among oxide grains. The depletion of $^{18}$O and the large $^{26}$Al abundances in group II oxide grains is plausibly due to cool bottom processing in intermediate mass giants ($<3M_{\odot}$). However, there is no detection of $^{26}$Al in O-rich AGB stars and it is odd that the O-isotopic compositions in AGB stars do not overlap more with the field spanned by the group II grains (Figs. 14, 15). None of the stellar O-isotopic compositions overlap with grains of group III and IV, and the overlap is marginal at best for group II grains. This lack of overlap with data for current AGB stars could favor the interpretation that grains of group II and III reflect chemical galactic evolution processes (Nittler et al. 1997).

The O-isotopic compositions of the two hibonite grains found by Choi et al. (1999) favor an origin from a low mass ($\sim1.7M_{\odot}$) AGB star of solar metallicity and a $\sim1.2M_{\odot}$ RGB or early AGB star. The inferred initial $^{26}Al/^{27}Al$ ratios indicate that H-shell burning and CBP took place in the grains' stellar sources.

**Dust from C-rich ABG stars** includes most SiC grains, some yet unkown fraction of the graphite grains, and possibly – if any - some of the nanodiamonds.

A little more than 90% of all presolar SiC originated from AGB stars, mainly the **mainstream SiC and related Y and Z grains** (e.g., Hoppe et al. 1994, 1997, Hoppe and Ott 1997, Amari et al. 2001b). Their chemical and isotopic compositions are in accord with abundances in N-stars and AGB models. Within uncertainties, there is good overlap in the C and N isotopic compositions of mainstream SiC grains, N-stars and planetary nebulae (Fig. 6). The J-type C-stars overlap with data of the A+B grains (see below).

The paucity of Si-isotopic determinations in stars (Fig. 7) makes a comparison to SiC grains difficult. The two determinations for the C-star IRC+10°216 fall into the array of SiC mainstream grains. The other data are for red giants not yet on the AGB. Unfortunately, these measurements are quite uncertain. Leaving this aside, the lack of overlap between the Si-isotopic composition of SiC grains and RGB stars is expected because the ejecta from RGB stars are not reduced enough for SiC condensation. In this context we note that future measurements of Si isotopes in presolar *silicates* may show some overlap if these originate in RGB stars. However, models indicate that the Si-isotopic composition is not affected during nucleosynthesis in RGB stars and that it is probably only mildly modified in AGB stars of solar metallicity. Then the observed Si-isotopic composition is that with which a star was born and observed differences in Si-isotopic compositions point to galactic chemical evolution (Zinner et al. 2001).

The distribution in $^{12}C/^{13}C$ ratios of SiC (Fig. 22) corresponds well to that of C-stars determined by Lambert et al. (1986). The $^{12}C/^{13}C$ distributions measured in C-stars by Ohnaka and Tsuji (1996, 1999) and Schöier and Olofsson (2000) are similar but their maxima are shifted to lower $^{12}C/^{13}C$ ratios. However, the data by Lambert et al. (1986) are regarded as more reliable. The similarity in the $^{12}C/^{13}C$ distribution of SiC grains with that of N-stars is consistent with the conclusion that most of the SiC grains (93%) are from evolved AGB stars.

The trace element abundances in mainstream SiC grains provide a close link to N-stars because these are also enriched in *s*-process elements. The different abundance patterns for mainstream SiC grains (Fig. 9) and the fractionations among the elements can be understood by considering fractional condensation. Lodders and Fegley (1995) calculated the trace element abundance patterns observed in SiC grains (Amari et al. 1995c) with the goal to find the elemental abundances in the stellar sources, which are necessary to explain the patterns by fractional condensation. This then constrains the types of C-stars that could have produced the SiC grains. The patterns in Fig. 9 require source abundances of *s*-process elements ranging from solar up to 10×solar, and in some cases, the removal of ultrarefractory trace element carbides prior to SiC condensation. The Y grains also show enrichments in *s*-process elements, and their abundance patterns require condensation from a gas enriched in *s*-process element by a factor of 10 (Lodders and Fegley 1995). Such a large enrichment is consistent with the suggested origin of Y grains from low-metallicity AGB stars, where higher *s*-process yields are expected (Amari et al. 2001b).

Several of the noble gas isotope anomalies (e.g., the AGB “G” components for the noble gases in Fig. 5) can be accounted for if SiC grains formed around AGB stars. Early on, the G component of Kr was found to be strikingly similar in composition to theoretical expectations from *s*-process nucleosynthesis (Gallino et al. 1990, 1997). However, the Kr isotopes still pose some challenges to nucleosynthesis and stellar models. For example, why do the abundances of nucleosynthetic products, which are determined by the conditions in the hot interior of AGB stars, correlate with grain size, which reflects the conditions in the cool, expanding circumstellar shell outside? Lewis et al. (1990, 1994) observed a positive correlation of the *s*-process $^{86}Kr/^{84}Kr$ ratio of the G component with increasing average grain size for the different SiC aggregates. In contrast, the $^{80}Kr/^{84}Kr$ ratios decrease with average aggregate grain size. Verchovsky et al. (2004) proposed that the grain size correlation for Kr isotope abundance ratios can be explained by assuming that one low-energy Kr component was implanted into the grains during the AGB stage and a second, high-energy component was incorporated at the post-AGB stage during planetary nebula formation, which was too energetic to be captured by small grains.

The larger $^{86}Kr/^{84}Kr$ ratios in the larger-size grains fractions required larger neutron exposures for the *s*-process. Theoretical models (Gallino et al. 1990, 1997) suggest that higher neutron exposures are achieved in low-metallicity AGB stars, which then may suggests that the larger grains formed in low-metallicity stars at higher neutron exposure than the finer grains. However, this does not yet explain why low-metallicity stars produce larger grains in their outflows, and the effects of stellar mass and metallicity on grain size distributions are yet to be explored.

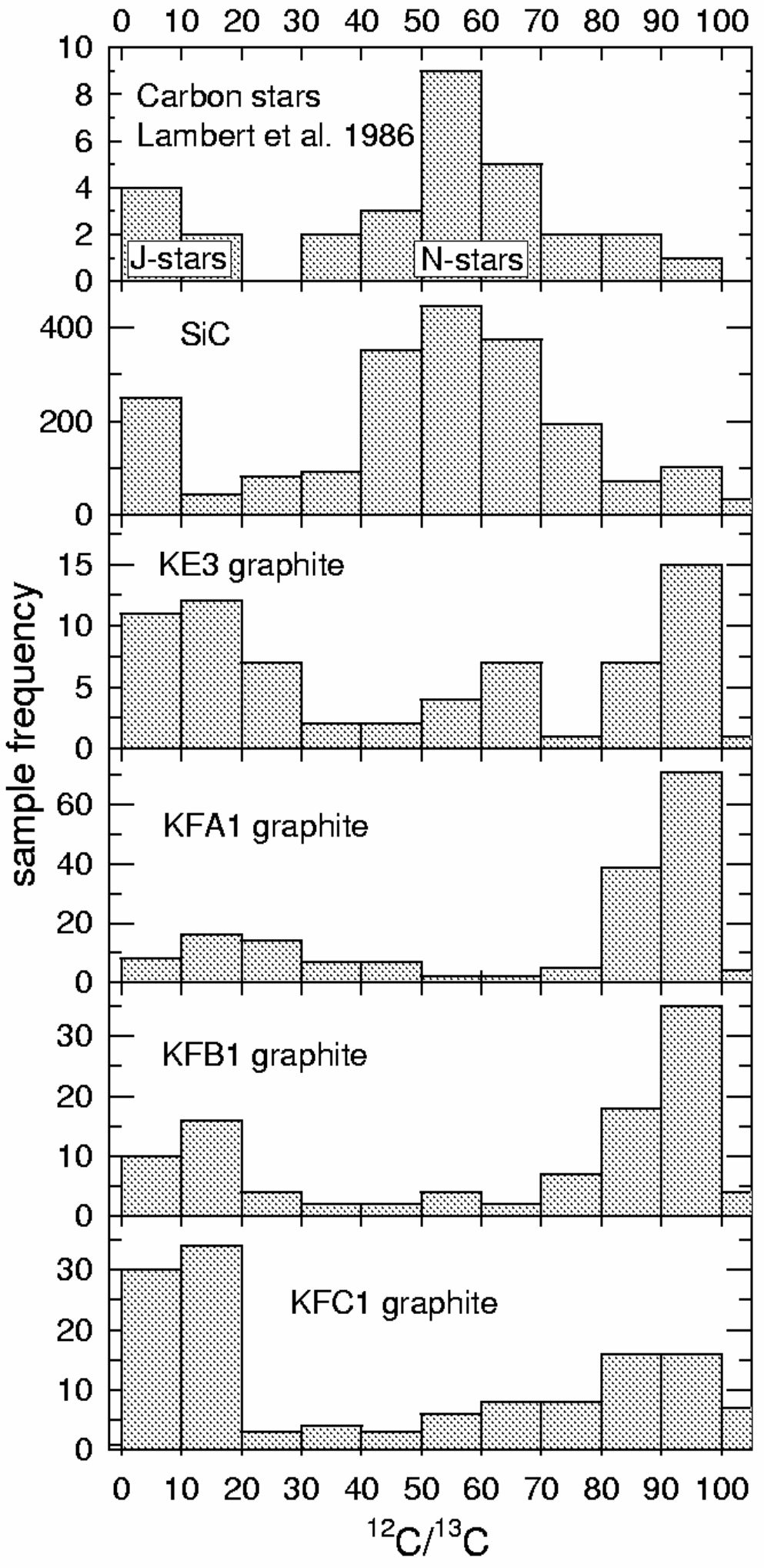

**Fig. 22.** Distribution of $^{12}C/^{13}C$ ratios in presolar SiC (Hoppe et al. 1994), presolar graphite (Hoppe et al. 1995, Travaglio et al. 1999, Amari, unpublished), and in C-stars (Lambert et al. 1986). Graphite and SiC data extend to $^{12}C/^{13}C$ of several thousand (Figs. 6, 8). Only the observed range for C-stars is covered here.

There is also a correlation between the Sr and Ba isotopes made by the *s*-process and average SiC grain size. However, the trend is *opposite* to that observed for *s*-process Kr: The $^{88}Sr/^{86}Sr$ and $^{138}Ba/^{136}Ba$ ratios decrease with increasing average aggregate grain size, which implies that neutron exposure decreased with grain size. A probable explanation is that grains carrying the *s*-pro cess Kr are not the same grains that carry the *s*-process Sr and/or Ba. Nichols et al. (2005) analyzed He and Ne in single SiC grains and found only four percent of the grains are gas-rich. Therefore, it is not unreasonable to assume that Kr-rich grains may be in a different population than the Sr- and Ba enriched grains. However, more correlated measurements for Sr, Ba, and noble gases in individual grains are required to sort out all these observations.

The link of mainstream SiC grains to carbon-rich AGB stars also keeps growing from isotopic measurements of heavy elements, such as Mo, Zr, and Ba, in *individual* grains (Nicolussi et al. 1997, 1998a, Savina et al. 2003b). One interesting recent result is the measurement of Ru isotopes which indicate that Tc, the herald of the *s*-process in AGB stars, condensed into presolar mainstream SiC grains (Savina et al. 2004).

The identification of **presolar graphite from AGB stars** is more difficult than for presolar SiC. The excellent agreement in the C-isotopic distributions (Fig. 22) among C-stars and SiC grains (mostly mainstream grains) strongly supports that mainstream grains formed around AGB stars very similar to most N-type C-stars we presently observe. If there is a major contribution of graphite grains from N-stars, we expect that the distributions for graphite grains show a "peak" that coincides with that of N-stars, as is seen for SiC grains. However, the distributions for graphites from the density fractions KFA1, KFB1, and KFC1 do not correlate with that of present-day N-stars. On the other hand, many graphite grains in the KFC1 fraction may have originated from low-metallicity AGB stars for which another $^{12}C/^{13}C$ distribution is expected. The high inferred $^{86}Kr/^{83}Kr$ = 4.80 (Amari et al. 1995a) and the high $^{12}C/^{13}C$ ratios (a few hundred or even above a thousand, Fig. 11) in many KFC1 graphites are consistent with the expected values for low-metallicity AGB stars (Gallino, private communication). In addition, many graphite grains from the high-density separate KFC1 contain refractory carbides with compositions ranging from nearly pure TiC to nearly pure Zr-Mo carbides (Bernatowicz et al. 1996). The abundances of Zr and Mo relative to Ti are larger than the solar ratio and indicate enrichments of the *s*-process elements Zr and Mo in the stellar sources. However, it is currently difficult to pin down the fractional abundance of "AGB" graphite grains in the high-density fraction.

In Fig. 22, the distributions for graphites from the density fractions KFA1, KFB1, and KFC1 show a spike near $^{12}C/^{13}C = 89$, the solar value. Such graphite grains could have a solar system origin or an origin in sources that coincided with, or better, dominated the solar system's C-isotopic composition. The KFB1 and KFC1 fractions show another major peak for $^{12}C/^{13}C < 20$, similar to A+B SiC grains and J-stars. Only the low-density fraction KE3 has a distribution that may correspond to N-stars (Fig. 22). However, Si- and O-isotopes relate many of the KE3 grains to supernovae. This illustrates that more than one criterion is required to clearly relate presolar grains to their parent stars.

The presence of **diamonds from C-rich AGB stars** among presolar diamonds cannot be excluded but currently also cannot be firmly concluded. If the spectroscopic assignments to diamond are correct, nanodiamonds seem to be present in some post-AGB stars and planetary nebulae. In that case, the known SiC and graphite from AGB stars should be accompanied by AGB nanodiamonds. However, more chemical and isotopic information on presolar nanodiamonds is required to establish a link to AGB sources.

### 6.5. Presolar grains with signatures akin to AGB stars but an imperfect match

The origin of the second largest SiC population, the A+B grains, remains enigmatic (Amari et al. 2001a). However, some types of C-stars must be responsible for their production. More than one type of stellar source is necessary to explain the spread in $^{14}N/^{15}N$ ratios (40–$10^4$) and the two types of A+B grains distinguished by their trace element concentrations. The A+B grains have either abundance patterns consistent with relative solar abundances of *s*-process elements (Fig. 9), or patterns like mainstream grains that require enhancement of *s*-process elements at the stellar sources (Amari et al. 1995c, 2001a, Lodders and Fegley 1995, 1998).

A comparison of the C- and N-isotopic composition of A+B grains and C-stars (Fig. 6) shows that J-type carbon stars have similar isotopic compositions. The C-isotope histograms (Fig. 22) also point in this direction, and the lack of *s*-process element enrichments (Utsumi 1970, 1985, Kilston 1975, Abia and Isern 2000) makes J-stars good candidates for the A+B grains.

The J-stars have C, N, and O elemental abundances similar to N-stars, but lower $^{12}C/^{13}C$ isotopic ratios (Lambert et al. 1986, Abia and Isern 1997, Ohnaka and Tsuji 1999). The low $^{12}C/^{13}C$ ratio and the absence of *s*-process enrichments in J-stars is a long- standing mystery, which now extends to the A+B grains. The evolutionary status of J-stars is unclear and they are unlikely to be on the AGB (e.g., Lloyd-Evans 1991, Abia and Isern 2000). According to nucleosynthesis models, $^{12}C$, which is necessary to make a stellar atmosphere rich in carbon, is produced during the AGB stage of low to intermediate mass stars. Without any modifications, this also increases the $^{12}C/^{13}C$ ratio, but J-stars have the lowest values among C-stars. Furthermore, the *s*-process operates during the AGB stage, but there are no s-process element enrichments in the J-stars (e.g., Utsumi 1970, 1985, Abia and Isern 2000). The situation to explain the A+B grains *with s*-process enrichments is also far from being settled (Amari et al. 2001a). Rare types of stars that can produce reduced dust are possible sources for these grains. These include low-metallicity CH stars that evolve in binary systems, and stars such as Sakurai's object, which appears to be a "born-again" AGB-like star due to external mass accretion (Amari et al. 2001a, Lodders and Fegley 1997, 1998).

# 7. Massive stars and supernovae (SNe)

Astronomical observations of supernovae and their remnants reveal the presence of dust, so they are potential sources of presolar grains. For a long time, searches for infrared excesses indicative of dust only showed small amounts of dust, but recent sub-millimeter observations indicate large amounts of very cold dust in supernova remnants (Dunne et al. 2003). Supernovae now seem to be confirmed dust producers on a similar scale as giant stars (e.g., Dunne et al. 2003, Morgan et al. 2003, Kemper et al. 2004). One would suspect that SNe produce large amounts of dust because they supply the major elements (e.g., Mg, Al, Si, Ca, Fe) to the ISM and increase the galactic metal content with time. The major elements will eventually condense, so there must be dust from supernova ejecta, and, by implication, dust grains from SNe among presolar grains.

There are different types of supernovae. Ejecta of SNe type I do not show atomic H lines in their optical spectra, whereas SNe of type II do. This indicates that a star that turns into a SN type I must have become very H-poor relative to the H-rich composition of normal main sequence and giant stars. Type I SNe fall into several subtypes (e.g., Ia, Ib, Ic), and type Ia is special among all SNe because it involves a binary system.

## 7.1. Evolution of massive stars to supernovae

The evolution of massive stars (> 8-9 $M_\odot$) is described in detail by Maeder (1990), Woosley and Weaver (1995), Arnett (1996), Pagel (1997), Wallerstein et al. (1997), Rauscher et al. (2002), Woosley et al. (2002), Meynet and Maeder (2003) and Truran and Heger (2004). The initial mass of a star and its mass-loss history determines how the star ends its life. The AGB stars end up as white dwarfs because mass-loss does not leave enough mass (<1.4 $M_\odot$) to sustain nuclear reactions in the C-O-rich core left from He-core burning. The H- and He-shell burning also stops because the envelope is lost. The evolution of more massive stars (>8-9 $M_\odot$) initially proceeds similar to that of low- and intermediate stars but occurs more quickly, and nucleosynthesis continues in several subsequent steps.

Before they go supernovae, massive stars develop a concentric shell structure in composition (Fig. 23) as a result from the succession of the major nuclear burning stages (see reviews noted above). Each major burning stage leads to ashes that provide the fuel for the next stage. For example, the first major reactions in the stellar core are H-burning to He, and then He-burning to $^{12}C$ and

$^{16}O$, which also happens in less massive AGB stars. After this, carbon burns in the core:

$$^{12}C\ (^{12}C,\alpha)\ ^{20}Ne$$

followed by Ne-burning which starts by photodisintegration of $^{20}Ne$ and subsequent $\alpha$–capture by $^{20}Ne$:

$$^{20}Ne\ (\gamma,\alpha)^{16}O$$
$$^{20}Ne(\alpha,\gamma)^{24}Mg(\alpha,\gamma)^{28}Si$$

followed by O-burning, which produces e.g., $^{28}Si$, $^{32}S$, $^{36}Ar$, $^{40}Ca$:

$$^{16}O(^{16}O,\alpha)^{28}Si$$

followed by Si-burning initiated by photodisintegration of $^{28}Si$:

$$^{28}Si(\gamma,\alpha)^{24}Mg$$

This is followed by $\alpha$-, neutron- and proton-capture reactions on $^{28}Si$, which build iron peak elements up to $^{56}Ni$.

All core-burning stages are accompanied by burning in shells of a similar succession, e.g., H-shell burning and He-core burning, He-shell burning and C-core burning; C-shell burning and C-O core burning, etc. Compared to the timescales of H- and He-core burning, the next core-burning stages are greatly accelerated: carbon-burning lasts for a few hundred years, and Si-burning around a day. After Si is exhausted as fuel in the core, the core collapses, and a resulting shockwave drives the supernova explosion. At this stage, explosive nucleosynthesis takes place. In the innermost zone, nuclear statistical equilibrium produces radioactive $^{44}Ti$, $^{48}Cr$, $^{49}V$, $^{51}Mn$, $^{52}Fe$, $^{55}Co$, $^{56}Ni$, $^{59}Cu$, and $^{60}Zn$, which decay to $^{44}Ca$, $^{48}Ti$, $^{49}Ti$, $^{51}V$, $^{52}Cr$, $^{55}Mn$, $^{56}Fe$, $^{59}Co$, and $^{60}Fe$, respectively. These isotopes are very characteristic of SNe. Aside from $^{56}Co$ and $^{57}Co$ in SN 1987A, $\gamma$-lines from $^{44}Ti$ are reported in the Cas A SN remnant (Iyudin et al. 1994), and observed excesses of $^{44}Ca$ (the decay product of $^{44}Ti$) provide a direct link of some presolar grains to supernovae.

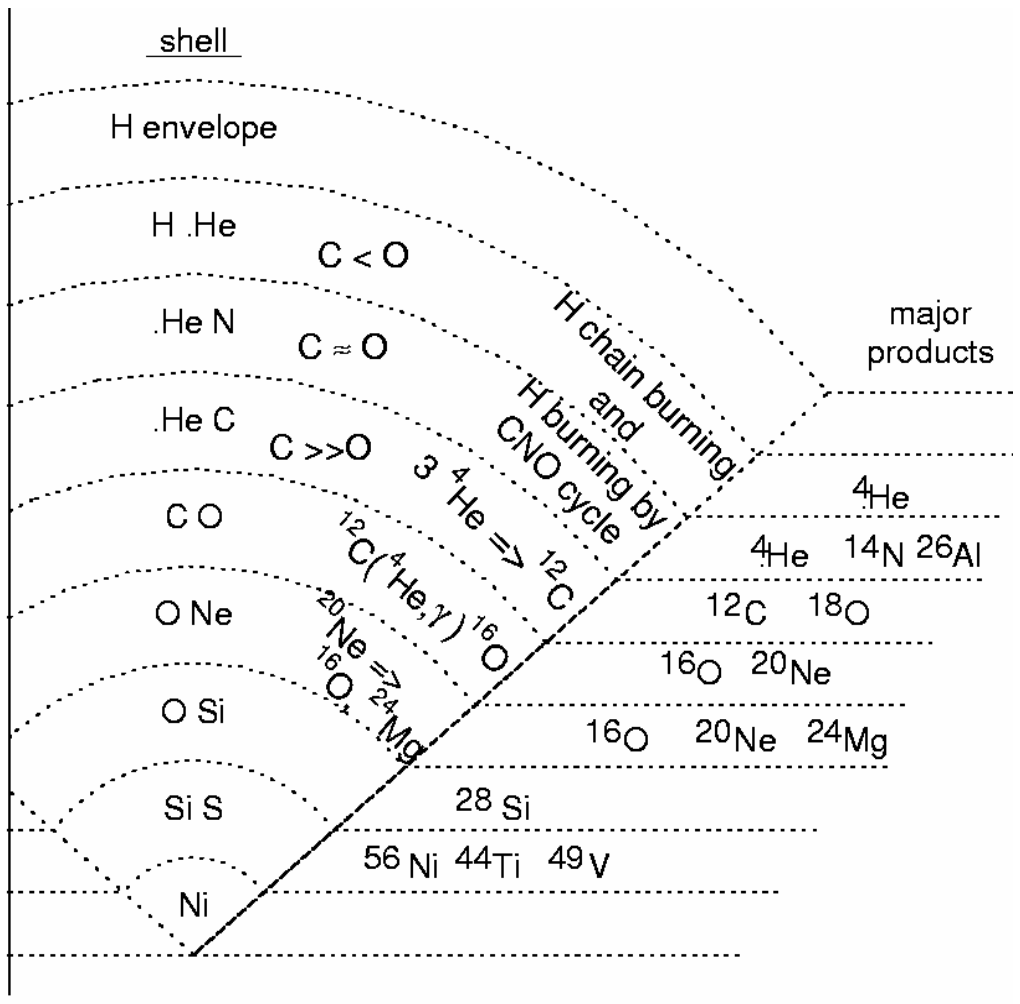

**Fig. 23.** The compositional zones resulting from nucleosynthesis in massive stars before they go supernovae are conveniently visualized in an "onion shell" diagram. Zones are labeled with the elements that make the major nuclear fuel and products. Some nuclear reactions and products of interest to presolar grain studies are indicated.

Heavy element nucleosynthesis by the ***r*- and *p*-processes** is generally associated with SN explosions, but locating the sites for these processes is still problematic (e.g., Wallerstein et al. 1997, Woosley et al. 2002). The rapid neutron capture process (*r*-process) requires high neutron densities and high temperatures so that neutron capture on seed nuclei can proceed on a faster time scale than beta decay of intermediately produced radioactive nuclei. The *p*-process is responsible for building proton-rich nuclei that cannot be produced from nuclei made by the *s*- or *r*-processes and subsequent beta decay. In the context here, the anomalies in the *p*- and *r*-process Xe isotopes (Xe-HL) were most diagnostic of supernova contributions to presolar grains.

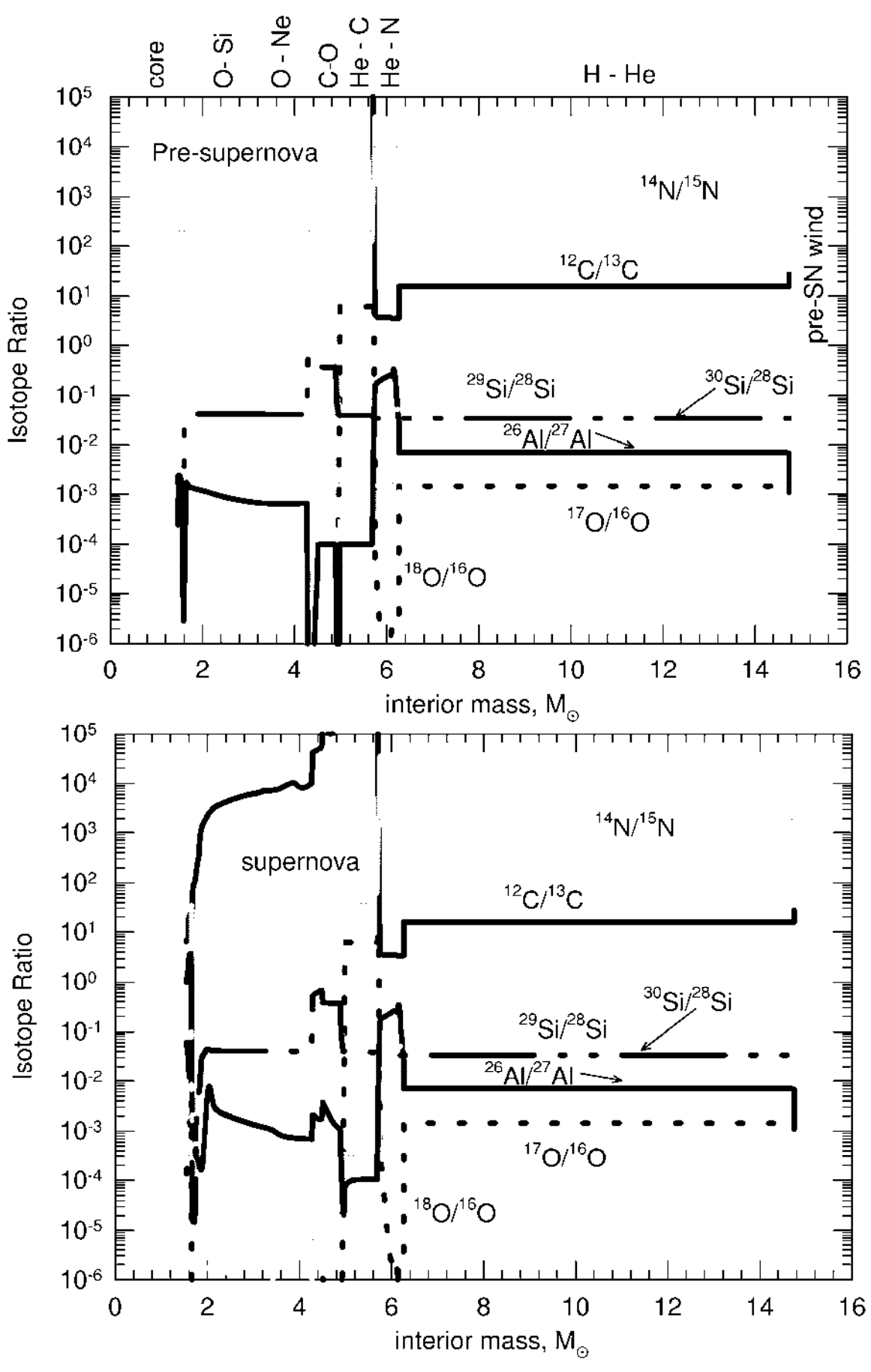

**Fig. 24.** The isotopic ratios for C, N, O, Al, and Si within a 20 solar mass star directly before the supernova explosion (top) and after the passage of the supernova shockwave (bottom). The shell structure (Fig. 23) is indicated at the top and jumps in the isotopic ratios clearly separate some of the individual zones. The mass plotted on the x-axis gives the total mass of the star that lies under a given zone. The data are from calculations by Rauscher et al. (2002).

The compositions of pre-supernova stars and supernovae have been calculated by Woosley and Weaver (1995), Thielemann et al. (1996) and Rauscher et al. (2002). Fig. 24 shows the compositional "cross-section" through a 20 $M_\odot$ star (Rauscher et al. 2002) for the major isotope ratios that are of interest for presolar grain studies. The individual zones are indicated at the top, and the zone "boundaries" are characterized by sharp changes in isotopic compositions. The important issue here is that some different zones must mix to some extent during the supernova explosions in order to produce the observed isotopic ratios in presolar grains. In particular, products from the He/N, He/C, and the innermost layers must mix, but the mixing cannot be complete because the C/O ratio must be sufficiently high to allow formation of reduced dust. The innermost zones are all rich in O so C/O < 1, and only the He/C and He/N zones contain enough C to achieve C/O $\geq$ 1, which leads to SiC and graphite condensation.

Stars with masses around 30 - 40 $M_\odot$ lose their H-rich zones in a slow wind during their red super-giant phase. Stars above ~40 $M_\odot$ shed their H- and He-rich zones in a fast wind during the main sequence stage, from where they evolve to luminous blue variables and then to **Wolf-Rayet** (WR) stars, and these very massive stars may skip the red (super) giant phase (Maeder 1990, Meynet and Maeder 2003). The substantial mass-loss produces extensive gas and dust shells around WR stars and they are therefore candidates for producing presolar grains.

Once the H-rich zone is lost, the products from H-burning by the CNO-cycle become visible as a N-rich "WN" Wolf-Rayet star. In the more massive carbon-rich Wolf-Rayet (WC) stars, loss of the H- and He-rich zones opens the view to the $^{12}C$-rich layer generated from triple α- burning. In O-rich WO stars (likely more massive than WN and WC-stars), α-capture on carbon has increased the O abundance. The circumstellar shells of WR stars may contain O-rich dust (in WN, WO), and reduced dust (in WC), depending on how much of the H- and He-zones were lost. Relative to solar, presolar grains from ejected envelope material of Wolf-Rayet stars should show a few of the following characteristics: enrichments in $^{14}N$ and $^{22}Ne$, evidence for $^{26}Al$ from the CNO-, NeNa- and MgAl-cycles during the WN-stage, $^{41}Ca$ from neutron capture, and $^{12}C$ enrichments from He-burning (e.g., Arnould et al. 1997).

### 7.2. Dust from supernovae

There is not much information yet about of the mineralogy of dust from observations of supernova ejecta and remnants. An emission feature at 22 μm seen in two supernova remnants is interpreted as "Mg-protosilicate" (Arendt et al. 1999, Chan and Onaka 2000), and a small feature at 13 μm may be attributed to corundum. However, the expected dust composition can be modeled if elemental abundances for these dust producers are available. In absence of observational data, results from nucleosynthetic network computations and hydrodynamic codes (e.g., Woosley and Weaver 1995, Meyer et al. 1995, Thielemann et al. 1996, Rauscher et al. 2002) are good resources for elemental abundances in SNe of different masses and in individual zones.

There are three possibilities to model supernova condensates. First, one can investigate the types of condensates that are expected from each individual zone composition of a supernova. The second is to take an average, overall ejecta composition to calculate the types of condensates. The third is to assume more or less selective mixing of matter between different zones. The reason for these different approaches is that the extent of mixing in supernova ejecta is not very well known. Observations and hydrodynamic models (e.g., Ebisuzaki and Shibazaki 1988, Herant et al. 1994, Hughes et al. 2000, Kifonidis et al. 2003) indicate that the ejecta are relatively well mixed. On the other hand, limited mixing between the various zones of a supernova is required to explain the isotope data of presolar grains, if nucleosynthesis predictions for the zone compositions are correct (e.g., Travaglio et al. 1999).

The condensates expected from the individual supernova shells and from the overall ejecta composition was computed by Lattimer et al. (1978). Despite the wide range in individual zone compositions, the condensate chemistry in SNe is not that much different than described above for giant stars and is mainly governed by the C/O ratio so that Table 9 also can be used as an approximate guide to the major expected supernova condensates.

Kozasa et al. (1989a,b, 1991) utilized the extensive observations that followed the explosion of SN 1987A and compared them to their thermochemical and kinetic calculations. Depending on the degree of mixing of material from different zones, they inferred that graphite, corundum, enstatite, and magnetite are expected in the ejecta. Ebel and Grossman (2001) address condensation in supernova ejecta but focus on the more specific question of whether or not reduced condensates can form at $C/O < 1$ if formation of CO is suppressed, as suggested by Clayton et al. (1999, 2001).

### 7.3. Synthesis: Presolar grains from supernovae

When searching for presolar grains from SNe, it is useful to keep in mind the following isotopic characteristics compared to the solar isotope composition. Grains from supernova (type SN II) ejecta should show strong enrichments in the stable isotopes $^{12}C$, $^{15}N$, $^{28}Si$, and excesses in $^{26}Mg$ due to radioactive decay of $^{26}Al$, $^{41}K$ from $^{41}Ca$, $^{44}Ca$ from $^{44}Ti$, and $^{49}Ti$ from $^{49}V$ (Amari et al. 1996, Nittler et al. 1996, Travaglio et al. 1999, Hoppe et al. 2000, Hoppe and Besmehn 2002). Among these isotopes, only a few (i.e., $^{28}Si$, $^{44}Ti$, $^{49}V$) are unique products of and tracers for SNe (including SN Ia for $^{28}$Si and $^{44}$Ti), whereas production of some of the other isotopes can also happen in other stellar sources (e.g., $^{26}$Al and $^{41}$Ca in AGB and WR stars).

### 7.3.1. SiC from supernovae: X grains

Only 1% of all SiC grains are from supernovae (Amari et al. 1992, Nittler et al. 1996, Travaglio et al. 1999, Hoppe et al. 2000). The X grains have large $^{28}$Si excesses (i.e., low $^{29}$Si/$^{28}$Si and $^{30}$Si/$^{28}$Si ratios), and SNe are the main source of $^{28}$Si. Further evidence for its supernova origin comes from grains that carry $^{44}$Ca excesses from decay of $^{44}$Ti ($t_{1/2}$= 60 a), which is only produced during explosive nucleosynthesis in SNe. Nittler et al. (1996) found that SiC X grains (as well as some low-density graphite grains) show evidence for $^{44}$Ti in the form of $^{44}$Ca excesses (up to 138 × solar) and that all grains (except one graphite grain) show correlated $^{28}$Si (up to 2 × solar) and $^{44}$Ca excesses. Besmehn and Hoppe (2003) found elevated $^{44}$Ca/$^{40}$Ca ratios and $^{28}$Si excesses in ~20% of all SiC type X grains that they analyzed. The high $^{12}$C/$^{13}$C ratios (>100) and relatively low $^{14}$N/$^{15}$N ratios (~20 to ~200) in X grains (Fig. 6) are consistent with nucleosynthesis in the He/C and He/N zones (Figs. 23,24), although not enough $^{15}$N is predicted in the SN models. The $^{26}$Al/$^{27}$Al ratios in X grains of up to 0.6 indicate contributions from the He/N zone to the supernova mixture from which the SiC grains condensed.

Trace element abundances in two X grains are very low (Fig. 9). Assuming relative solar abundances for the elements condensing into the SiC X grains, the observed abundance patterns can be explained if 98% of all trace elements were removed into other condensates such as refractory trace element carbides before the SiC condensed (Lodders and Fegley 1995). This seems plausible because graphite grains ascribed to a supernova origin contain Ti-carbide and metal subgrains (e.g., Croat et al. 2003). However, there is also the possibility that elemental abundances in the gas from which SiC condensed were modified by *r*- and *p*-process nucleosynthesis. Future trace element analyses of supernova SiC grains are required to sort out the different possibilities.

Isotopic ratios of Zr, Mo, and Ba in X grains have been analyzed by RIMS (Pellin et al. 2000a,b). Since the *r*-process takes place in SNe, one would expect that isotopes of heavy elements in X grains show signatures of the *r*-process. However, the expected *r*-process signature was not found for Mo in X grains. Four out of 6 X grains have excesses in $^{95}$Mo and $^{97}$Mo whereas excesses in $^{100}$Mo are expected from *r*-process models. Inspired by the new data and revisiting predictions by Howard et al. (1992), Meyer et al. (2000) reconstructed the neutron burst model. It postulates a rapid release of neutrons (on a time scale of seconds) in He-rich matter heated by the passage of the shock wave which gives a neutron flux lower than that of the classical r-process. This model successfully explained the $^{95}$Mo and $^{97}$Mo excesses of the grains. Zirconium and Ba isotopic ratios of the X grains can be also explained by the neutron burst model, assuming the grains condensed in a time scale of a few years.

### 7.3.2. Silicon nitride grains from supernovae

The few $Si_3N_4$ grains that have been discovered share many properties with the SiC X grains and are therefore also related to supernovae (Nittler et al. 1995, Hoppe et al. 2000, Lin et al. 2002, Besmehn and Hoppe 2003)

### 7.3.3. Graphite grains from supernovae

Many of the low-density graphite grains are believed to have formed in SNe. They have similar isotopic compositions as SiC type X and $Si_3N_4$ grains: high $^{12}C/^{13}C$ (>100) (Fig. 11), high $^{26}Al/^{27}Al$ up to 0.2 (Fig. 8), and Si-isotopic anomalies. Unlike SiC and $Si_3N_4$ of SN origin, graphite grains contain measurable amounts of O and often have high $^{18}O/^{16}O$ ratios - up to 184×solar (Fig. 14). The high $^{26}Al/^{27}Al$ ratios (Fig. 8) are consistent with nucleosynthesis in the He-N zone (Fig. 24), and the high $^{18}O/^{16}O$ ratios in the low-density graphites (Fig. 14) are a signature of nucleosynthesis in the He-C zone (Fig.24).

The observed isotopic composition of graphite grains requires mixing of different compositional SN zones because no single zone can explain the observations. Travaglio et al. (1999) performed mixing calculations using the different compositions of zones from Woosley and Weaver (1995) to quantitatively reproduce isotopic data for the low-density graphite grains. These models reproduced the observed $^{12}C/^{13}C$, $^{18}O/^{16}O$, and $^{30}Si/^{28}Si$ ratios and the inferred $^{41}Ca/^{40}Ca$ and $^{44}Ti/^{48}Ti$ ratios if jets of material from the inner Si-rich zone penetrated the intermediate O-rich zones and mixed with material from the outer C-rich zones. However, the major problems are that the models do not produce enough $^{15}N$ and $^{29}Si$ (e.g., Nittler et al. 1995, Travaglio et al. 1999, Hoppe et al. 2000).

### 7.3.4. Diamonds from supernovae

The essentially solar C- and N-isotopic compositions of meteoritic nanodiamonds are not helpful to relate them to any particular stellar source. On the other hand, the noble gas components with signatures of the *r*- and *p*- process leave little doubt that SNe contributed to the presolar diamonds, which leaves the diamonds' C- and N-isotopic compositions somewhat of a puzzle. If all nanodiamonds were from SNe, it would mean that the highly variable C-isotopic compositions of the different burning zones mix in such a manner as to give a $^{12}C/^{13}C$ ratio close to the solar ratio. This would also have to happen in many SNe because likely more than one contributed to the diamonds.

We noted before that Xe-HL in presolar diamonds is enriched in *p*-process $^{124}Xe$ and $^{126}Xe$ and *r*-process $^{134}Xe$ and $^{136}Xe$ (Fig. 16). Hence the conclusion that SNe are the places where the Xe-HL was implanted into the carriers. The mid-IR spectra of SN 1987A show a broad feature at 3.40 and 3.53 μm (Meikle et al. 1989), consistent with the identification of surface hydrogenated diamond (Guillois et al. 1999). This also supports the SN origin of some of the presolar diamond. Tielens et al. (1987) proposed that diamonds form by transformation of preexisting graphite in supernova shock-waves. Huss and Lewis (1994b) and Russell et al. (1996) argue that this may not work because the diamonds should retain the noble gases that are already in the graphite grains; however, graphite grains with supernova signatures do not contain Xe-HL, so diamonds could not inherit it. However, it may not be necessary for the graphites to already contain the Xe-HL component. Very massive stars go through the Wolf-Rayet (WR) stage, and in C-rich circumstellar shells of WR stars, graphite may condense. All this happens before the star goes supernova, and it is only *during* the supernova when the *r*-and *p*-processes responsible for Xe-HL are operating. Consequently, graphite formed earlier cannot contain Xe-HL because the *r*- and *p*-processes had not yet happened. However, once the products of the *r*-and *p*-processes ejected with the SN shock wave hit the older WR ejecta, the Xe-HL and other *r*-and *p*-process products may be incorporated at the same time as the preexisting graphite grains are transformed into diamonds.

Several questions remain about the details of the production of the *p*- and *r*-process nuclides of Xe seen in the nanodiamonds. Standard *p*- and *r*-process models produce equal excesses over solar for the *p*-process $^{124}$Xe and $^{126}$Xe, as well as for *r*-process $^{134}$Xe and $^{136}$Xe, but in Xe-HL the excesses are not equal for each pair (Fig. 16). In order to explain the heavy Xe isotopes ("Xe-H"), Howard et al. (1992) proposed a neutron-burst model. This model assumes that a neutron burst occurs in the He-rich zone of the supernova when this zone is heated (to about $10^9$K) by the passage of the shock wave (Meyer et al. 2000). This may lead to lower neutron-densities than in the classical *r*-process, so that $^{134}$Xe and $^{136}$Xe are produced with different efficiencies. However, the predicted $^{134}$Xe/$^{136}$Xe ratios from this model are different from those observed in Xe-H.

Remaining in the framework of the standard *r*-process, Ott (1996) suggested that a separation of the Xe isotopes and their radioactive precursors may occur before decay of their precursors is completed. All precursors of $^{136}$Xe have half-lives on the order of one minute or less, while $^{134}$Te and $^{134}$I (which decay to $^{134}$Xe) have half-lives of 42 and 52 minutes, respectively. In order to explain the $^{134}$Xe/$^{136}$Xe ratio, Ott (1996) estimated the timescale of the separation to be on the order of hours. This scenario qualitatively accounts for the $^{124}$Xe/$^{126}$Xe of Xe-L as well. The light isotopes $^{124}$Xe and $^{126}$Xe are produced from the precursors $^{124}$Ba and $^{126}$Ba, with half-lives of 12 and 100 minutes, respectively. If the separation between the Xe isotopes and their Ba precursors took place within hours after the *p*-process, $^{124}$Xe would be more enriched than $^{126}$Xe. Analyses of Te isotopes in diamond (Richter et al. 1998, Maas et al. 2001) show that the ratio of the two *r*-process isotopes, $^{128}$Te/$^{130}$Te, is also consistent with the "rapid" separation model proposed by Ott (1996). However, this requires a lot of fine tuning for the timing of the parent-daughter nuclide separation during a very turbulent SN mixing event. It remains to be seen if this can be a realistic model.

### 7.3.5. Oxide grains from supernovae

It is puzzling why presolar oxide grains of SN origin are so few if SNe are efficient dust producers. One possible explanation is that supernova oxide grains are too small (<<0.1 μm) to be recovered from meteorites during the presolar grain separation procedures. Since $^{16}O$ is the third most abundant isotope ejected from SNe, and the overall C/O ratio of the ejecta has C/O < 1 (Woosley and Weaver 1995), we would expect that oxides condense, and, for the same reason, that supernova oxides show larger excesses in $^{16}O$. However, there is only one corundum grain (shown at the bottom of group III in Fig. 14) with a significant $^{16}O$ excess, which may have a supernova origin (Nittler et al. 1998). One spinel grain, for which Choi et al. (1998) suggest a possible supernova origin, exhibits an $^{18}O$ excess (shown in Fig. 14 under the label for group IV). Otherwise, its O-isotopic composition is not very different from the presolar oxides of group IV, and its origin in a supernova appears uncertain. However, the $^{18}O/^{16}O$ composition of graphite grains with SN signatures are quite similar to the $^{18}O/^{16}O$ ratios of group IV oxide grains of uncertain origin (Fig. 14). Measurements of other isotopes for this spinel do not provide better clues. The $^{26}Al/^{27}Al$ is only $6\times10^{-4}$, far below the ratios (of up to ~0.1 $^{26}Al/^{27}Al$) seen in graphite and SiC grains of clear supernova origin (Fig. 8). Except for an excess in the neutron-rich isotope $^{50}Ti$, no isotopic anomalies in Ti and Ca were found by Choi et al. (1998), again different from what other presolar minerals from SNe show. On the other hand, such comparisons are somewhat complicated because it is not entirely clear to what extent the different zones, and hence the elements and isotopes in the supernova are mixed, and, from which "supernova mixtures" the different presolar supernova grains are coming.

# 8. Binary stars as presolar dust sources

Novae and supernovae of type Ia are dust sources involving stellar binaries. These sources may not be unimportant because ~20% of stars occur in binary or higher multiple systems. Most stars, including those in binaries, are of lower mass (< 9 $M_{\odot}$) and therefore evolve from the main sequence to giant stars and end up as white dwarfs. Of two stars, born at the same time but with somewhat different masses, the more massive one evolves faster, and, depending on their relative age and the distance from each other, there are different possibilities for mass transfer between them.

In the case where one star is still a main sequence star while the companion star is already a white dwarf, envelope mass from the main sequence star may accrete onto the hot white dwarf, which leads to a **nova.** A few presolar grains seem to come from such systems. The transferred mass can be assumed to be of normal solar composition, and the white dwarf is either C and O-rich (a CO-WD) or O and Ne -rich (a ONe-WD) with masses ranging between 0.65 to 1.35 $M_{\odot}$. When the white dwarf has accreted enough matter to exceed the Chandrasekhar mass limit (1.4 $M_{\odot}$) for thermonuclear explosions, a nova occurs.

The accreted material undergoes nuclear processing before it is ejected from the white dwarf. Dust condensation then incorporates the products of nova nucleosynthesis. Models predict that isotopes of light elements up to Ca are processed. In particular, low $^{12}C/^{13}C$ (<<10) and high $^{26}Al/^{27}Al$ ratios are expected in ejecta of CO and ONe novae (Starrfield et al. 1997, 1998, José and Hernanz 1998, José et al. 2001, 2004). In CO novae, the peak temperatures do not get high enough to significantly modify Si isotopic ratios, but enrichments in $^{30}Si$ are expected in novae of higher-mass ONe white dwarfs. A few presolar SiC and graphite grains with isotopic signatures consistent with theoretical predictions for ONe novae are known (see Figs. 6-8 and 12, Amari et al. 2001c). Presolar grains from CO novae have not yet been identified, but the absence of graphite and SiC grains from CO novae is consistent with expectations from condensation calculations for nova ejecta (José et al. 2004).

**Supernovae of type Ia (SN Ia)** seem to result from interaction of two white dwarfs. Explaining the thermonuclear explosions of SNe Ia by a merger of two white dwarfs in a close binary system can account for the absence of H in the ejecta of SNe Ia and satisfies the requirement that SNe Ia should have low- to intermediate mass progenitors that do not evolve into supernovae as single stars (e.g., Iben and Tutukov 1984, Webbink 1984). Ejecta of SN Ia explosions should also produce dust because many rock-forming elements are expelled. Clayton et al. (1997) suggested that the isotopic signatures of some SiC X grains are consistent with an origin from SNe Ia caused by He accretion onto a CO-WD. However, it appears that the origin of these grains remains best understood by an origin in mixed SN type II ejecta (Amari et al. 1998).

## 9. Conclusions and outlook

Much has been learned from presolar grains, but many questions remain and new ones must be posed. Our collection of presolar minerals is not complete; there must be other types of presolar minerals carrying abundant elements such as Fe, Cr, Mn, S, and P, just to name a few. It was about a year ago that the elusive presolar silicates were first discovered, and the future may add to the array of presolar minerals hidden in meteorites. Astronomical observations with a new generation of telescopes (e.g., the Spitzer telescope just taken into operation) already reveal amazing details of dust around giant stars and planetary nebulae. Also not long ago, observations at longer infrared and sub-millimeter wavelengths with ever-improving instruments provided the first evidence of the previously "missing" dust from supernovae, and the detection of currently not observed minerals around the different dust producing stars is awaiting. On the hand, advances in the developments of new micro- (or better nano-) analytical techniques such as the NanoSIMS and RIMS are directed towards obtaining more detailed correlated measurements of mineralogical, chemical, and isotopic properties of the known presolar components, which set the stage for probing new components yet to be found. Such measurements, in turn, provide sensitive tests to stellar evolution and nucleosynthesis models. It will not be the last time that the results from presolar grain measurements lead to revisions and updates in such models, and prompt new measurements of physical properties, such as nuclear reaction cross-sections. Overall, the understanding of presolar grains and their implications requires combined efforts from astronomy, physics, chemistry, and mineralogy and will remain an exciting field of study in the future.

*Acknowledgements*: the authors thank Klaus Keil for the invitation for this review. They thank Roberto Gallino and Laura Schaefer for careful comments on the manuscript. The authors also thank Gary Huss for comments. Work by K.L. was supported in parts by NASA grants NNG04GG13G and NNG04G157A from the NASA Astrobiology Institute. Work by S.A. was supported by NASA grants NAG5-11545 and NNG04GG13G.